\documentclass[fleqn,usenatbib,useAMS]{mnras}
\usepackage{graphicx}
\usepackage{xspace}
\usepackage{hhline} 
\usepackage{newtxtext}
\usepackage[varvw]{newtxmath} % var v
\usepackage{amsmath}
\usepackage[dvipsnames, table]{xcolor}
\everymath{\thickmuskip=2mu minus 2mu} 
\usepackage{etoolbox} % for eqs commands
\usepackage{orcidlink} % orcid links for arxiv ver
\usepackage[T1]{fontenc}
\DeclareRobustCommand{\VAN}[3]{#2}
\let\VANthebibliography\thebibliography
\def\thebibliography{\DeclareRobustCommand{\VAN}[3]{##3}\VANthebibliography}

\usepackage{defs}

\begin{document}
% TC:ignore

\title[Probing the CW with FRBs. II. Scintillation]{Probing the cosmic web with fast radio bursts: II. Scintillation}
\author[S. Lapiner]{
    Sharon Lapiner$^1$\thanks{E-mail:sharon.lapiner@mail.huji.ac.il}\orc{SL},
    Nir Mandelker$^{1,2}$\orc{NM}, 
    Paz Beniamini$^{3,4,5}$\orc{PB},
    and
    S. Peng Oh\orc{PO}$^{6}$
    \\
    $^1$Racah Institute of Physics, The Hebrew University, Jerusalem 91904 Israel\\
    $^2$University of Washington, Department of Astronomy, Seattle, WA 98195, USA\\
    $^3$Astrophysics Research Center of the Open University (ARCO), The Open University of Israel, P.O. Box 808, Ra’anana 43537, Israel\\
    $^4$Department of Natural Sciences, The Open University of Israel, P.O Box 808, Ra'anana 4353701, Israel\\
    $^5$Department of Physics, The George Washington University, 725 21st Street NW, Washington, DC 20052, USA\\
    $^6$Department of Physics, University of California, Santa Barbara, CA 93106, USA\\
}
\maketitle

% TC:endignore
\begin{abstract}
We consider the formation of multiphase gas in cosmic web objects (CWOs) through cooling and fragmentation within the shock-heated medium of cosmic sheets and filaments, and the circumgalactic medium (CGM) of haloes. This leads to dense, cool ($\sim10^{4}\,\mathrm{K}$), small-scale cloudlets in pressure equilibrium with their surroundings. 
Such ionized overdensities can affect the propagation of radio waves emitted by fast radio bursts (FRBs), allowing us to probe the small-scale structure in CWOs by examining their impact on the scintillation pattern of the observed FRB flux. 
While we find that the probability of encountering a high-temperature filament or sheet with cold cloudlets is rather small and usually confined to high redshifts, the expected high rates of FRBs may still permit detection through scintillation along rare sightlines. 
We suggest scintillation from the Milky Way (MW) interstellar medium as a promising probe: CWO plasma screens can suppress it, placing constraints on their cold-gas properties.
Filaments and sheets can suppress MW scintillation from sources at $z_\mathrm{S}\gtrsim2$ and $3$, respectively, while cold and turbulent cloudlets in the CGM can be probed by sources at redshifts as low as $z_\mathrm{S}\sim0.5$, constraining remarkably low volume-filling fractions, $f_\mathrm{v}$ (e.g. $f_\mathrm{v}\gtrsim10^{-5}$ for $z_\mathrm{S}\gtrsim2$). 
To provide testable predictions, we estimate the fraction of FRBs expected to encounter CWOs that may suppress MW scintillation below a given frequency, and express this as a function of the average extragalactic dispersion measure.
Future FRB samples could map these constraints onto bounds on the existence and properties of cloudlets in CWOs, such as their density, size, $f_\mathrm{v}$, and turbulence.
\end{abstract}
% TC:ignore

\section{Introduction}

Galaxies contain only a small fraction of the baryons in the Universe, including both their stars and the interstellar medium (ISM) \citep[e.g.][]{Peeples14, Tumlinson2017, Wechsler_Tinker18}. The majority of baryons are found in regions outside galaxies, either as gas within dark matter haloes in the \emph{circumgalactic medium} (CGM), or in the \emph{intergalactic medium} (IGM) in between haloes. By supplying galaxies with fresh gas and acting as a reservoir for enriched gas ejected from galaxies \citep[the cosmic baryon cycle, e.g.][]{Putman2012, McQuinn16, Tumlinson2017}, the CGM and IGM play an important role in the evolution of galaxies, and provide valuable constraints on the growth of large-scale structure \citep[e.g.][]{Rauch98, Viel13, Lidz_Malloy14, McQuinn16, Eilers18}.

\smallskip
On scales larger than a megaparsec, both the dark matter and the gas are structured into the elaborate network of the \emph{cosmic web} (CW), comprised of sheets and filaments. The CW was predicted by theoretical models \citep{Zeldovich1970, bbks1986}, confirmed in cosmological simulations \citep[e.g.][]{Bond96, Springel05}, and observed both in the distribution of galaxies \citep[e.g.][]{Colless2001, tegmark04, Huchra05} and in narrow-band Ly$\alpha$ emission at $z\gsim3$ \citep{Umehata19, Martin23}. Sheets and filaments of the CW comprise roughly $\sim(25-50)\%$ of the volume of the Universe, and $\sim(50-75)\%$ of its mass \citep[e.g.][and references therein]{Wang12, cautun14, Libeskind18}; where sheets dominate the volume, and filaments dominate the mass.
The intersections of filaments, at the nodes of the CW, are occupied by the most massive galaxies at any given epoch and are commonly fed by several filaments, while typical galaxies at each epoch reside along filaments. The filaments themselves are embedded and fed by sheets (or `walls'), which border cosmic voids.
At redshifts $z>2$ in the vicinity of the CW nodes, CW filaments are predicted to be the main source of gas accretion onto galaxies, manifesting as streams of cold ($T\sim10^4\Kel$) and dense ($n\sim10^{-2}\cmc$) gas, penetrating galactic haloes \citep[e.g.][]{Dekel.Birnboim.06, Dekel09a, Aung24}. This is thought to be true even for massive haloes where the CGM is hot ($T\gsim10^{6}\Kel$) and the cooling time is very long \citep[][]{Rees77, White78, Birnboim.Dekel.03, Fielding17, Stern21}.

\smallskip
The diffuse nature of the gas in the CW and CGM poses challenges for direct observations. It is most commonly studied through absorption line spectroscopy along lines of sight toward distant quasars (QSOs) or galaxies \citep[e.g.][]{Bergeron86, Hennawi06, Steidel10, Lehner22}, and through narrow-band emission line observations, for example, Ly$\alpha$ emission at $z\gsim3$ \citep{Steidel00, Cantalupo14, Martin14a, Martin14b, Martin23, Umehata19, tornotti25a, tornotti25b}. These reveal a complex multiphase structure to the gas outside galaxies, with cool clouds embedded in a hotter ambient medium (see \citealp{Tumlinson2017} for a recent CGM review). The sizes of these cool clouds are uncertain and can range from sub-parsec to a few hundred parsecs in the CGM. They generally show area-covering fractions of order unity, small volume-filling fractions ($\fv\sim 10^{-3}$), and high densities compared to the hot phase, $\chi=\rho_{\rm c}/\rho_{\rm h}\sim (100-1000)$ \citep{Tumlinson2017, Cantalupo19, FG_Oh2023}. Modern cosmological simulations that resolve the CGM with sub-kiloparsec precision similarly find a complex multiphase structure \citep{vandeVoort19, Hummels19, Suresh19, Peeples19}.

\smallskip
Recently, cosmological simulations that `zoom-in' on a region of the CW containing massive sheets and filaments at high-$z$ revealed that accretion shocks surround these structures, much like those seen around massive haloes. The post-shock regions in these CWOs display a similar multiphase structure to the CGM (\citealp[][hereafter \citeta{m19} and \citeta{m21}]{m19,m21}; \citealp{Pasha23,lu23}). 
Such cold substructure in cosmic sheets and filaments may explain puzzling observations of strong HI absorbers, $N_\mr{HI}>10^{17.2}\cms$, with remarkably low metallicities of $Z\lsim10^{-3}\zsol$ observed far from any galaxies \citep[e.g.][]{Robert19, Lehner22}. 
However, the amount and distribution of cold, dense gas in the CGM and IGM grows with increased resolution and has not converged in simulations with $\sim300\pc$ resolution in the CGM and $\sim1\kpc$ in the CW.

\smallskip
A theoretical model describing the formation of multiphase gas in the form of cold cloudlets embedded in a hot medium was proposed by \citep[][]{mccourt18}. They argued that when the cooling time of a gas cloud is shorter than its sound-crossing time, the cloud does not cool monolithically, as previously thought \citep[][]{Field65, Burkert_Lin00}, nor can it cool isobarically. Instead, it first `shatters' into small clouds that lose sonic contact, and then each fragment proceeds to contract independently \citep{mccourt18, Gronke_Oh20}. The characteristic size of these cloudlets was proposed to be of the order of the \emph{minimal cooling length}, $\lcmin \sim \min(\cs\tcool)$, where $\cs$ is the sound speed and $\tcool$ is the cooling time, with the smallest size reached at temperatures of $\sim10^4\Kel$.
At $z\sim 2-3$, $\lcmin$ typically ranges from $\sim0.1\pc$ in the CGM to $\sim 1\kpc$ in cosmic web sheets \citepa{m19,m21}. 
This model can explain large differences between area-covering and volume-filling factors \citep{FG_Oh2023}, as well as several other phenomena observed in the CGM, high-velocity clouds, quasar broad-line regions, and the ISM \citep{Gronke17, mccourt18, Stanimirovic.Zweibel.2018, FG_Oh2023, Sameer24}.

\smallskip
Recently, fast radio bursts (FRBs) have emerged as a valuable tool for probing the foreground ionized gas along their line of sight (los). FRBs are compact astrophysical sources emitting brief radio pulses, ranging from a few tens of microseconds to a few milliseconds in duration, and covering frequencies from roughly $0.1$ to $8\ghz$ \citep{Lorimer07, Lorimer24, Gajjar18, Pleunis21}. They are notable for their high occurrence rates, around $10^{3}$ per day, and strong peak fluxes exceeding $1\jy$. Although the astrophysical sources of FRBs are still uncertain, magnetars are a leading candidate \citep{Bochenek20, Good20, Kumar+17, Wadiasingh2019, beniamini25}. This supports their possible presence out to high redshifts, potentially back to the epoch of reionization \citep{beniamini21, Hashimoto_2021, Heimersheim_2022, ziegler25, shaw25}, and positions them as an exceptional probe of intergalactic gas over cosmic distances and time-scales.

\smallskip
As an FRB radio signal passes through a plasma, it suffers a reduced group velocity due to the plasma refractive index, resulting in a frequency-dependent time delay, with lower-frequency photons arriving later than higher-frequency ones. The integrated electron density along a los can be derived from a measurement of the time delay as a function of frequency, resulting in the dispersion measure (DM), $\dm=\int {\nt }/{\opzo[]} ds$, where $\nt$ is the proper electron density and $ds$ is the proper distance element. For FRBs at large distances, the DM is dominated by the IGM, enabling it to serve as an approximate indicator of the source redshift by assuming the cosmically averaged electron density along the sightline. To date, more than a hundred FRBs have been localised to their host galaxies, reaching up to a source redshift of $\zs\sim2$ \citep{Lorimer24, connor25, gordon24, caleb25}; although additional events may originate from even higher redshifts, as suggested by high DM measurements \citep[e.g.][]{zhang18}.

\smallskip
In addition to the DM, FRB scattering and scintillation can serve as a sensitive diagnostic of small-scale and turbulent structure in the CGM and IGM, similar to the usage of pulsar observations to study the MW ISM \citep[e.g.][]{rickett77, rickett90, cordes91, Armstrong95, cordes_lazio02}. 
This is enabled by FRBs' intrinsic compactness, their brief duration, the sensitivity of radio waves to fluctuations in electron density, and the growing catalogue of FRB detections \citep{Lorimer24}. Since radio wave propagation is affected by ionized density inhomogeneities on small scales, the presence of small cloudlets can substantially affect FRB signals by broadening the observed pulse \citep[e.g.][in the CGM]{vedantham19, ocker22-horizons, jow24, ocker25, mas-ribas25}, as well as causing fluctuations in the observed flux as a function of frequency, known as scintillation. This phenomenon offers a potential means to probe structure in the CGM and CW on scales far smaller than those accessible through absorption spectroscopy or simulations. Current observational evidence is inconclusive, with some FRBs showing no indication of excess scattering in the CGM \citep{prochaska19-sci, Connor20M33, Faber2024, shin25}, while a population synthesis study, based on the first CHIME/FRB catalogue \citep{chime21}, shows results that are marginally consistent with scattering from the CGM of intervening haloes \citep{chawla22}.

\smallskip
Several studies have modelled how multiphase gas in a foreground CGM could influence the observed FRB flux through scintillation. \citet{jow24} and \citet{mas-ribas25} used refractive scattering by shattered cloudlets in the CGM of intervening haloes, and constrained their expected impact on subsequent scintillation by a MW screen. \citet{ocker25} developed an empirically based model using CGM turbulence measurements, and predicted the effect of diffractive scintillation and scattering on FRB signals. 
Here, we aim to constrain the existence and properties of density fluctuations in CW sheets and filaments, as well as in the CGM of massive haloes. In particular, our focus here is to probe individual (and possibly rare) CWOs through diffractive scintillation. 

\smallskip
In our companion paper (\scattpaperl), we explore the effect of radio wave scattering caused by a CW plasma screen in the foreground of an FRB.
Diffractive scattering and scintillation are closely linked; however, as their characteristic scales are inversely proportional to one another, they tend to be dominant in different regimes. 
While scattering may be useful to explore multiple CW plasma screens (see \scattpaper), detection of an individual CW screen could be challenging and can be easily confused with scattering by the FRB host galaxy. 
To better explore individual CW screens, we consider here the impact of such screens on scintillation -- either by the scintillation pattern it imprints on the observed flux, or by its impact on the scintillation of a subsequent plasma screen in the MW-ISM. As in \scattpaper, we here adopt the shattering framework \citep{mccourt18} as a physically motivated fiducial model for multiphase gas. 
Our results are not intended to assess the validity of the shattering process itself; instead, we use it as a benchmark to explore the existence of turbulent density fluctuations in the CW.

\smallskip
Our current study also builds, in part, on findings presented in a second companion paper (\dmpaperl), where we develop an analytic framework to derive the virial properties of CW sheets as a function of redshift and halo mass of the neighbouring node.
This is analogous to models derived in \citet[hereafter \citeta{m18}]{m18} and \citet{lu23} for filaments. They assume a hierarchy of CWOs, where the virial properties of filaments are derived from the total accretion rate onto massive haloes, and the properties of sheets are constrained by the accretion rate onto the filaments embedded within them. 
We then combine the virial properties of all CWOs with the excursion set formalism of \citet{shen06} for ellipsoidal collapse in the CW, to estimate the number of CWOs along an average sightline and predict the contribution of CWOs to the total DM as a function of the source redshift ($\zs$).

\smallskip
The remainder of this paper is organized as follows. 
In \se{shatter}, we summarise our fiducial model for the formation of cold small-scale clouds in CWOs (following \scattpaper). 
In \se{dff}, we present the basic model of radio-wave scintillation. 
In \se{cosmo}, we briefly outline our model, which accounts both for the probability that an average los intercepts a CWO (\dmpaper) and for our estimates of the number of clouds within different CWOs as a function of redshift (\scattpaper). 
In \se{mdff_mw}, we use the expected flux modulation due to a plasma screen in the MW ISM to constrain the presence of a CW screen along the FRB los. 
In \se{mdff_cw}, we estimate the conditions to directly detect flux modulation caused by multiple clouds within a CW plasma screen across cosmic time. 
Finally, in \se{dc_Sc} we discuss our results and conclude. 
Throughout, we assume a flat $\Lambda$CDM cosmology, with $H_0=70\kms\Mpc^{-1}$, $\omm=1-\oml=0.3$, and a universal baryon fraction of $\fb=0.17$.

\section{Small-scale cold structure in CWOs}
\label{se:shatter}

\begin{figure}
    \centering
    \includegraphics[width=0.85\linewidth]{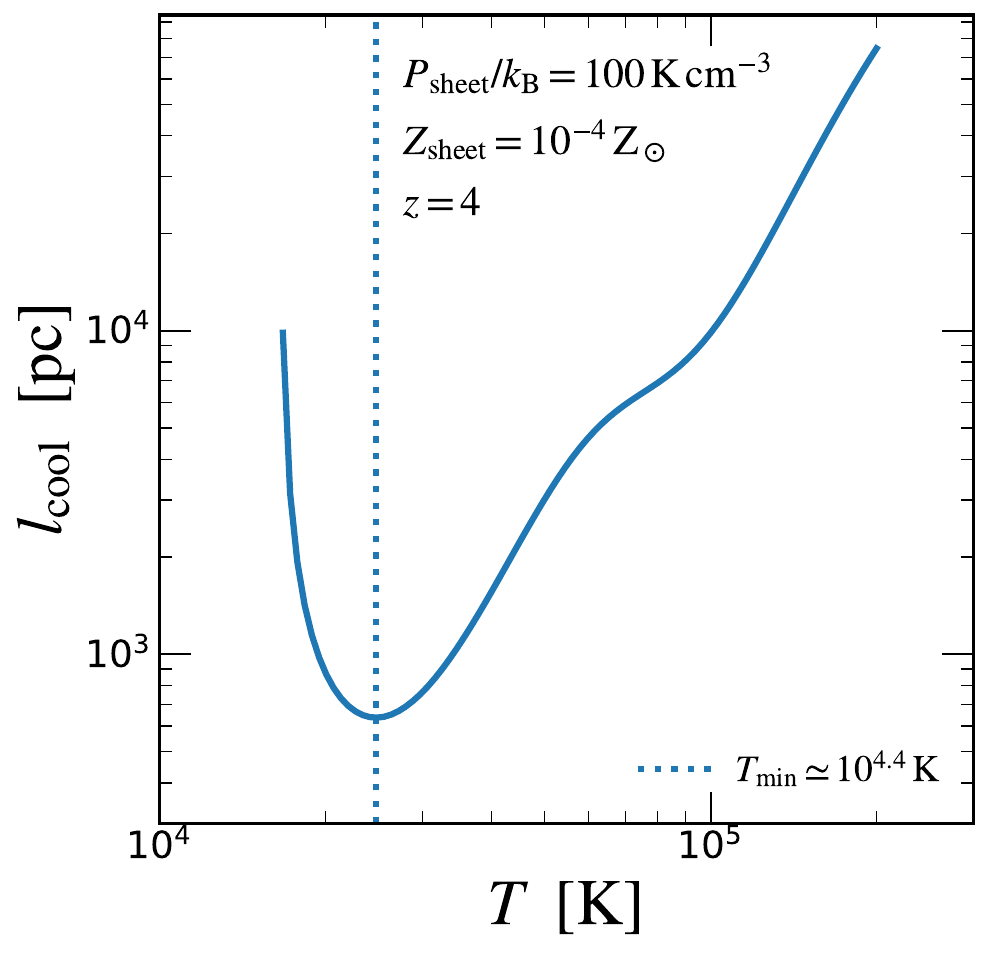}
    \caption{Illustrative example of the cooling length as a function of temperature calculated using \eq{lcool}. Shown here for a typical sheet at $z=4$, with a pressure of $100\punits$, and a metallicity of $10^{-4}\zsol$ \citepa{m21}. }
    \label{fig:lcool}
\end{figure}

\begin{figure*}
    \centering
    \includegraphics[width=0.8\linewidth]{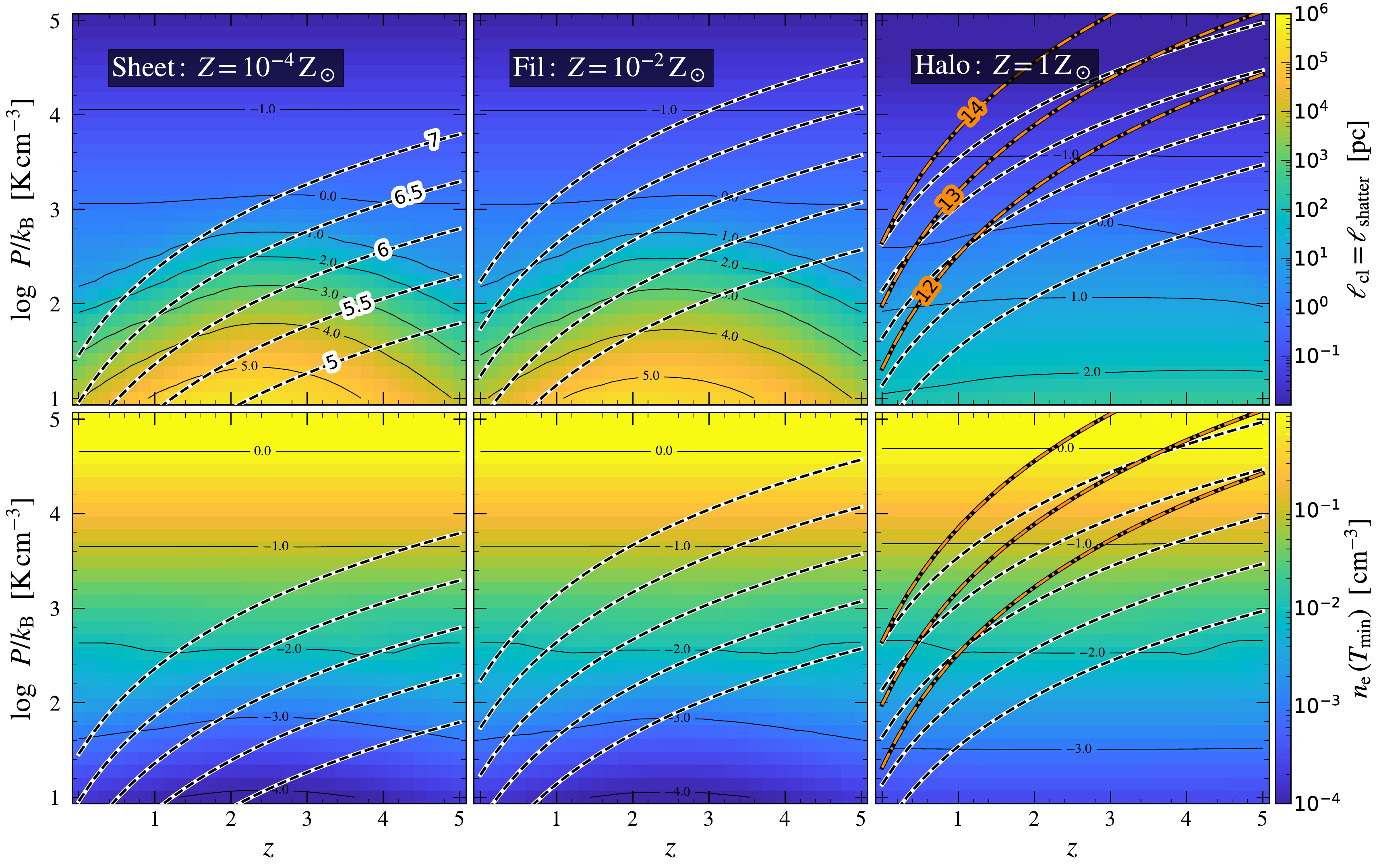}
    \caption{
    Properties of shattered cloudlets as a function of thermal pressure (y-axis), redshift (x-axis), and metallicity (columns). A metallicity of $Z=10^{-4}\zsol$ represents sheets (left column), $Z=10^{-2}\zsol$ represents filaments (middle), and $Z=1\zsol$ representing halo CGM (right). 
    The top row displays the typical size of cloudlets, defined as the minimum cooling length $\lc=\lcmin$, and the bottom row displays the cloud electron density ($\n$). 
    These assume isobaric cooling at the ambient pressure, incorporating the effects of a redshift-dependent UVB from \citet{haardt_madau96} and self-shielding \citep{rahmati13}. 
    For a given $(P,z)$, metallicity has a negligible impact for values below $10^{-2}\zsol$. 
    In each panel, dashed lines correspond to pressures at the outskirts of a CWO, with a post-shock virial temperature of $\Tv=10^{5}$, $10^{5.5}$, $10^{6}$, $10^{6.5}$, or $10^{7}\Kel$, from bottom to top, according to our model (\sea{P}, \scattpaper). Orange dash-dotted lines in the right column indicate haloes of constant mass of $\Mh=10^{12}$, $10^{13}$, and $10^{14}\msun$, from bottom to top. 
    The black contours trace the quantities depicted by the colours, with labels in log scale. 
    For a fixed $\Tv$, sheets exhibit the lowest pressures, largest cloud sizes, and lowest densities, while those in the CGM demonstrate the opposite extreme.
    }
    \label{fig:clump_prop}
\end{figure*}

In our companion paper (\scattpaper), we present in detail our fiducial model for the properties of cold cloudlets in the shock-heated medium of massive CWOs. For completeness, we provide a brief summary of the model in this section. 
Several physical processes can generate multiphase gas, with cold cloudlets embedded in an ambient hot gas (e.g. thermal instabilities, radiative shocks, turbulent mixing layers). 
We adopt as our fiducial model the `shattering' framework \citep{mccourt18}, which allows us to model the properties of cold clouds, relate them to the hot environment, and make concrete predictions. 
However, we stress that our fundamental predictions regarding FRB scintillation are not dependent on this specific scenario. The shattering model mainly affects the cloud size, $\lc$, and one can easily adjust this to account for other processes. We discuss variations from our fiducial model in \se{model_var} below, and in \S5.3 in \scattpaper.

\smallskip
In the shattering picture, a thermally unstable large gas cloud with a cooling time much shorter than its sound-crossing time shatters into small cloudlets of size $\lcool=\cs\tcool$ that can cool isobarically, thus maintaining pressure equilibrium with the hot background throughout the cooling process. For a fixed pressure, $\lcool$ strongly depends on temperature (see \fig{lcool}), and as a result, the fragmentation proceeds to smaller and smaller scales, until the minimum cooling length $\lcmin=\min(\lcool)$ is reached at $T\gsim10^{4}\Kel$.

\smallskip
$\lcmin$ mainly depends on the surrounding pressure, the gas metallicity, and the ionizing UV background radiation. 
The cooling length as a function of temperature can be expressed as
\be
    \label{eq:lcool}
    \lcool = \cs \tcool = \lt[ \frac{\gamma \kb^5}{\mu^3 \mpr (\gamma-1)^2 \xh^2} \rt]^{\half} \frac{T^{\frac{5}{2}}}{P\,\Lamtz(T,Z,z)},
\ee
where $\mu$ is the mean molecular weight, $\kb$ is Boltzmann's constant, $\mpr$ is the proton mass, $\xh$ is the hydrogen mass fraction ($\sim0.76$), and $P=n\kb T$ is the pressure. The net cooling-minus-heating rate per unit volume (accounting for the cosmic UV background) is $\nh^2\Lamtz(T,Z,z)$, with the hydrogen number density $\nh=\mu\xh n$. 
In \fig{lcool}, we show $\lcool$ as a function of temperature, for a constant pressure of $P/\kb=100\punits$, a metallicity of $Z=10^{-4}\zsol$, and a $z=4$ \citet{haardt_madau96} UV background. This is typical of high-$z$ cosmic web sheets \citepa{m19, m21}. 
The minimum cooling length can be expressed as,
\be
    \label{eq:lshatter}
    \lcmin \approx 15\pc \frac{\Tmin[,2e4]^{5/2}}{P_3\,\Lamtz[min,-23]},
\ee
where $\Tmin[,2e4]=\Tmin/(2\tm10^4\Kel)$ is the temperature where the cooling length reaches its minimum at fixed pressure, and $\Lamtz[min,-23]= \Lamtz(\Tmin)/(10^{-23},\mathrm{erg,s^{-1},cm^{3}})$.

\subsection{Cloud properties in CWOs}
\label{se:CW_clumps}

In this section, we briefly discuss the properties of cloudlets in CWOs, and refer the reader to \scattpaper for a more in-depth discussion. 
Using \eq{lcool}, we estimate the cloud size, $\lc$, by the minimum cooling length, $\lcmin=\min(\lcool)\equiv\lcool(\Tmin)$, for a given metallicity, as a function of pressure and redshift. We assume throughout that each CWO is characterized by a fixed metallicity, $Z=(10^{-4}, 10^{-2}, 1)\zsol$ for sheets, filaments and the CGM in haloes, respectively \citepa[see][]{m21}. We then compute the associated mean molecular weight $\mu(\Tmin)$, the electron density, $\n(\Tmin)$, and the net cooling, $\Lamtz(\Tmin,Z,z)$, using the \texttt{RAMSES} cooling module \citep{Teyssier.02}, which accounts for fine-structure and atomic cooling. We include a \citet{haardt_madau96} redshift-dependent UV background (UVB) and approximate self-shielding of dense gas \citep{rahmati13}. To associate our results with specific CWOs, we require a model for the ambient pressure of the hot, virialized gas in CWOs as a function of virial temperature or mass, and redshift. We model the post-shock pressure of CWOs in our companion paper (\scattpaper), and provide a brief review of the main results in \sea{P}. We refer the reader to \scattpaper for further details.

\smallskip
In \fig{lcool}, we illustrate an example of $\lcool$ as a function of temperature, for a sheet at $z=4$ with a pressure of $P/\kb=100\punits$. We find $\lcmin\sim700\pc$ at $\Tmin\simeq10^{4.4}\Kel$. 
More generally, in the top row of \fig{clump_prop}, we show the cloud size $\lc$ ($=\lcmin$) as a function of pressure and redshift ($z\leq5$). 
From left to right, the three columns show three metallicities, $Z=10^{-4}, 10^{-2},$ and $1\zsol$, crudely representative of cosmic sheets, filaments, and the CGM, respectively.
Adopting $0.1\zsol$ for the CGM yields very similar results. Comparing the panels, $\lc$ is generally very similar for sheets and filaments at a fixed $(P,z)$. At low pressures, $\lc$ in the CGM can be up to three orders of magnitude smaller than $\lc$ in sheets and filaments, while at higher pressures ($P\gsim10^3\punits$) the difference is only $\sim0.5\dex$.

\smallskip
The black-and-white dashed lines in each panel indicate the post-shock pressure near the outskirts of virialized CWOs as a function of redshift (\eqsitinp{Ph_Tv}{Psh}), assuming a fixed $\Tv$ ranging between $10^{5}$ and $10^{7}\Kel$ from bottom to top with intervals of $0.5\dex$. 
For the CGM, we also show the pressure as a function of redshift at a fixed halo mass of $\Mh=10^{12}, 10^{13},$ and $10^{14}\msun$, from bottom to top (orange and black lines). 
Note that for a fixed $\Tv$, $P\propto\opzo[]^3$, while for a fixed $\Mh$, $P\propto\opzo[]^4$.
Comparing the different panels, for a given $\Tv$, the pressure rises from sheets to filaments to the CGM, resulting in a decrease in $\lc$.
One can see that while $\lc$ can be similar for sheets and filaments at fixed $(P,z)$, realistic CWOs occupy different regions of this plane.

\smallskip
In the bottom row, we present the electron density of cold clouds, $\n$, assuming ionization equilibrium at $\Tmin$ at the external pressure. 
At fixed $(P,z)$, $\n$ shows little metallicity dependence. For $z\sim4$ and $\Tv\sim3\tm10^5\Kel$, sheets have $\n\gsim10^{-3}\cmc$ and filaments $\gsim10^{-2}\cmc$, similar to cold gas densities reported in \citeta{m21} and \citet{lu23}. In the CGM, $\n\gsim0.05\cmc$ for the same temperature. The column density in the CGM remains $\N\sim10^{17}\cms$ regardless of pressure or redshift, as suggested by \citet{mccourt18} (not shown here; see fig.~2 in \scattpaper).

\section{Diffractive scintillation - General considerations}
\label{se:dff}

When a short-duration radio burst from a compact source passes through a turbulent plasma screen, ionized overdensities can alter ray propagation and cause random wave-front distortions, which can manifest on the observed spectrum as temporal broadening of the signal (scattering) and an interference pattern on a typical spectral scale (scintillation). The degree of observed scattering and scintillation can place valuable constraints on inhomogeneities in the intervening plasma screen. Particularly for our purposes, we can use this to study the presence and properties of ionized overdensities in the hot medium of different CWOs. 
In our companion paper, \scattpaper, we studied the effect of FRB scattering due to cold cloudlets within CWOs, whereas in the current work, we focus on the observed scintillation. 
The physics of diffractive scattering and scintillation are well known and have been studied extensively using pulsars in the MW for several decades \citep[see][for a review]{rickett90}.
In the current section, we outline some fundamental concepts,\footnote{The basic quantities associated with diffractive scattering, outlined in \se{lpi} and \se{tau}, are reviewed in greater detail in \scattpaper.} 
aimed at readers unfamiliar with diffractive scintillation. Later, in \se{mdff_mw} and \se{mdff_cw}, we apply the general case presented here to study the effect of cool cloudlets in CWOs on the observed scintillation pattern of FRBs.

\subsection{Coherence scale}
\label{se:lpi}

Overdensities in a plasma screen induce fluctuations in the refractive index, $\mu_i=[1-(\nu_\mr{p}/\nuz)^2]^{1/2}$, where the plasma frequency is $\nu_\mr{p}=[e^2\n/(\pi\me)]^{1/2}$, with $\me$ and $e$ the electron mass and charge. The frequency and wavelength of the passing radio wave in the rest frame of the observer are represented by $\nuo$ and $\lamo$. Hereafter, we use a prime notation ($\square'$) when expressing these in the rest frame of the screen; $\nuz=\nuo\opzo[]=c/\lamz$. 
We assume a Kolmogorov spectrum for turbulent density fluctuations,\footnote{We assume that at the outer scale $\dn(\lo)\sim\n$; if this is not true, the results will have an extra factor related to the variance of density fluctuations within cloudlets, $\epsilon^2=\mean{\dn^2}/\n^2$, with $\n$ replaced by $\epsilon\n$ in the scintillation equations. Such a deviation from the model assumptions can be easily explored by considering a deviation from $\epsilon\sim 1$ (see \se{model_var}, where we consider similar deviations from the model).}
$\dn\sim\n(\ell/\lo)^{1/3}$
within the inertial subrange of $\li<\ell<\lo$, where $\ell$ is the eddy size, $\li$ is the inner dissipation scale, and $\lo$ is the outer injection scale.

\smallskip
In the thin screen approximation, the number of eddies with scale $\ell$ in the screen is $\dL/\ell$, where $\dL$ is the effective width of the screen occupied by eddies of size $\ell$, which may be smaller than the total width of the screen $\dL\leq\DL$. 
The cumulative RMS phase shift caused by eddies of size $\ell$ is $\Dphi\equiv(\ell/\lpi)^{5/6}$, where $\lpi$ is the \emph{coherence} or \emph{diffractive} length scale, defined to be the scale where $\Dphi(\lpi)=1\rad$. 
Diffractive scattering and scintillation are dominated by eddies of size $\ell=\lpi$, which induce both a large scattering angle and large flux modulation. 
The coherence scale is given by \citep[see][]{rickett90, macquart13} 
\be
    \lpi \sim
    \begin{cases}
        \Big(\frac{\re \lamo}{1+z}\Big)^{-\frac{6}{5}} \fa^{-\frac{3}{5}} \lo^{-\frac{1}{5}} \n^{-\frac{6}{5}} 
        &  \li<\lpi<\lo, 
        \\
          \Big(\frac{\re \lamo}{1+z}\Big)^{-1} \li^{\frac{1}{6}} \fa^{-\frac{1}{2}} \lo^{-\frac{1}{6}} \n^{-1}
        &   \lpi<\li,
    \end{cases}
    \label{eq:lpi_0}
\ee
where $\n$ is the mean electron density in the clouds and $\re=e^2/(c^2\me)$. 
The area-covering fraction of clouds, $\fa$, can also be thought of as the mean number of clouds encountered along a given los through the screen. For a screen of width $\DL$ containing cloudlets of size $\lo$ with a volume-filling fraction $\fv$, $\fa\sim\fv\DL/\lo\sim\dL/\lo$ \citep{mccourt18, FG_Oh2023}, where $\dL=\fv\DL$ is the effective width of the screen occupied by the cloudlets. 
Due to the high density contrast between the cold clouds and the surrounding hot gas, the turbulence in each media is decoupled, and the high-density clouds are expected to dominate the scintillation and scattering. Therefore, similar to \citet{vedantham19}, we take the cloud size as the outer scale of turbulence, $\lo\sim\lc$. 
Using \eq{lpi_0} for $\li<\lpi<\lc$ with $\lo=\lc$,\footnote{As discussed in \sea{dff_cw_app}, our estimates of $\li$ \citep{beniamini20} indicate that $\li<\lpi$ for most of the CW screens we explore here. Therefore, with the exception of \se{model_var}, we focus the following derivations on the regime $\li<\lpi<\lo$.}
\be
    \lpi 
    \sim
        2.4\tm 10^{12}\cm\ \ 
        \nuf^{\frac{6}{5}} \opzo^{\frac{6}{5}}  \fa^{-\frac{3}{5}} \lc[,10]^{-\frac{1}{5}} \nf^{-\frac{6}{5}},
    \label{eq:lpi}
\ee
where $\opzo=\opzo[]/5$, $\lc[,10]=\lc/(10\pc)$, $\nf=\n/(10^{-2}\cmc)$, 
and $\nuf=\nu/1\ghz$.
In the inertial subrange, the coherence scale decreases with increasing cloud size for a given area-covering fraction, $\lpi\propto\lc^{-1/5}$ for a given $\fa$, but increases with increasing cloud size for a given volume-filling fraction, $\lpi\propto\lc^{2/5}$ for a given $\fv\sim\fa\lc/L$. 
We briefly address constraints on both $\fa$ and $\fv$ in different CWOs in \se{frag_cond} (see \scattpaper for further details). 
Regardless, it is important to note that the coherence scale (\eqsnp{lpi_0,lpi}), which dominates the scintillation and scattering, is mostly affected by the density $\n$, and has a rather small dependence on $\lc$ for a Kolmogorov spectrum.

\smallskip
In addition to scattering screens in CWOs, we will often discuss screens in either the host screen or the MW screen. For these screens, it is convenient to express $\lpi$ in terms of their DM. This can be done under the assumption that the screen's $\dms$ is dominated by the same ionized gas regions that dominate the scattering; in such a case, one can assume $\dms\simeq\dmsc=\fa\lc\n$. Observational results from pulsar scattering suggest that this is the case in the MW-ISM \citep[][and references therein]{cordes91, cordes16}, and we assume it to be true in the FRB host screen as well. 
However, it is unclear if this will hold for CWOs, as simulations tend to show significant spatial variability in which phase dominates $\Nh$ (\citeta{m21}; \citealp{lu23}; see also \S4.5 in \scattpaper).
Using \eq{lpi}, along with the effective width of the screen $\dL\sim\fa\lc$, the diffractive scale can be expressed as
\be
    \label{eq:lpi_dm}
    \lpi
    \sim
        3.9\tm 10^{9}\cm\ \ 
        \nuf^{\frac{6}{5}} (1+z)_1^{\frac{6}{5}} \dL[kpc]^{} \dmsfal[2.3]^{-\frac{6}{5}},
\ee
where $(\dmsfa)_{2.3}=\dmsfa/(200\dmunits)$ and $\dL[kpc]=\dL/(1\kpc)$.

\subsection{Temporal broadening}
\label{se:tau}

Scintillation and scattering are tightly connected. To illustrate this connection, we briefly discuss the scattering time (or temporal broadening) here, and refer the reader to \scattpaper for further details on this topic and its usage as a probe for cold ionized overdensities in CWOs.

\smallskip
When rays are deflected by a scattering screen, they travel longer path lengths, causing a time delay in their arrival that can be measured as a broadening of the signal over time. 
The scattering time can be expressed as a function of the coherence scale length, $\lpi$, by 
\be 
    \label{eq:taus}
    \taus \simeq \frac{\lamo}{2\pi c} \pfrac{\rf}{\lpi}^2,
\ee 
where $\rf$ is the Fresnel scale,
\be
    \rf 
    = \pfrac{\Deff\lamz}{2\pi}^{\half} \sim 5.4\tm 10^{13}\cm \pfrac{\Deff[,9]}{\nuf\opzo}^{\half},
    \label{eq:rF}
\ee
where $\Deff=\dsl\dlo/\dso\approx\min(\dsl,\dlo)$ is the effective angular diameter distance, $\dso$ is the angular diameter distance from the source to the observer, $\dsl$ is from the source to the scattering screen, and $\dlo$ is from the screen to the observer,\footnote{For a comoving distance of $\chi_{ij}=\chi_i-\chi_j$ between two redshifts $z_i>z_j$, the angular diameter distance between these two redshifts is $d_{ij}=\chi_{ij}/(1+z_i)$. See for example, \citet{schneider92, macquart04, macquart13}, and appendix B of \scattpaper.} and $\Deff[,9]=\Deff/(1\Gpc)$. The strength of scattering and scintillation can be quantified by $u\equiv\rf/\lpi$, such that screens with $u>1$ are in the strong regime (see \se{dffu}).

\smallskip
Similar to \eq{lpi_dm}, for a screen in the host galaxy or the MW, we express the scattering time as a function of the screen DM as,
\be
    \taus  
    \sim 1.6\tm10^{-4}\ms \ \nuf^{-\frac{22}{5}} \opzo[1]^{-\frac{17}{5}} \frac{\dsc[kpc]}{\dL[kpc]^{2}} \lt[\dms\fa^{\frac{1}{3}}\rt]_{2.3}^{\frac{12}{5}}.
    \label{eq:taus_dm}
\ee
For the MW screen the distance is $\dsc\approx\dlo[,\mmw]$, and for the FRB host galaxy it is $\dsc\approx\dsl[,\hg]$.\footnote{For screens in the host galaxy and the MW, $\dsc$ and $\dL$ are usually taken to be of the same order. Here they are both normalized to $1\kpc$.}

\subsection{Source size and flux modulation -- General arguments}
\label{se:dff_1}

Besides temporal broadening, diffractive scattering also manifests as \emph{scintillation}, namely flux modulations as a function of time and/or frequency due to constructive and destructive interference patterns. Temporal scintillation occurs when a plasma screen moves with a transverse velocity, $v$, relative to the observer and the source, yielding a scintillation time of $t_\mr{scint}=\lpi/v$. This produces a `twinkling' pattern as seen in starlight. However, for FRBs, the scintillation time is too long compared to the short duration of the burst for such twinkling to manifest. However, diffractive scintillation may still be detectable in the frequency domain, even if the temporal broadening is too small to be observed. In such a case, interfering waves arriving along different paths result in enhanced regions (scintles) in the flux density, where the size of a scintle in frequency-space is the scintillation bandwidth $\dnus$ (or the decorrelation\footnote{The \emph{decorrelation} bandwidth is the scale on which the observed flux at two frequencies is not correlated; namely, when the separation between them is $\dnus$.} 
bandwidth). In this section, we review some basic concepts related to diffractive scintillation and outline the conditions that allow detectable flux modulation. We will later apply these conditions to evaluate the impact of cloudlets within a CW screen on the observed flux of an FRB source (\se{mdff_mw} and \se{mdff_cw}).

\smallskip
In practice, the decorrelation bandwidth is usually measured by taking the half-width at half maximum of the fit to the autocorrelation function (ACF) of the observed spectrum \citep{cordes86}. 
This is defined as 
\be 
    \label{eq:ACF}
    {\rm ACF}(\dnu)=\frac{1}{N_{\nu} \bar{f}^2} \sum_{\nu}\Delta f(\nu)\Delta f(\nu+\dnu),
\ee 
where $N_{\nu}$ is the number of frequency bins across the band of the burst detection and $\Delta f(\nu)\equiv f(\nu)-\bar{f}$, with $f(\nu)$ the corrected flux after subtracting the background and $\bar{f}$ the mean of the spectrum. 
The autocorrelation function in \eq{ACF} is normalized such that the amplitude is the square of the flux modulation (or modulation index), ${\rm ACF(0)}\equiv\dff^2$, defined as 
\be
    \label{eq:dff_def}
    \dff \equiv \frac{\sigma_f}{\bar{f}},
\ee
where $\sigma_f$ is the standard deviation of the observed flux. 
The ACF is typically modelled as a Lorentzian, with the half-width at half-maximum defining the decorrelation bandwidth,
\be
    {\rm ACF_{fit}}(\dnu) 
    = \frac{\dff^2}{1+\dnu^2/\dnus^2}.
\ee

\smallskip 
The decorrelation bandwidth is related to the scattering time through the Fourier theorem,\footnote{The relation $2\pi\taus\dnus=C \sim1$ has been shown to hold for several pulsars; however, given the inverse frequency dependence, $\taus$ and $\dnus$ are measured at very different frequencies, and the final comparison at the same $\nuo$ must be done by using the known frequency dependence \citep[see][and references therein]{lang71}. If a los passes through more than one screen, the measured $\taus$ and $\dnus$ could originate from different screens; in such a case, the constant $C$ could be $\gg1$ \citep[see e.g.][]{masui15, sammons23}.}
giving
\be
    \dnus 
    \sim (2\pi \taus)^{-1}
    \sim
        2 \mhz\
        \Deff[,9]^{-1}\nuf^{\frac{22}{5}} \opzo^{\frac{17}{5}} \fa^{-\frac{6}{5}}  \lc[,10]^{-\frac{2}{5}} \nf^{-\frac{12}{5}}.
    \label{eq:dnus}
\ee
We discuss and show examples of $\dnus$ for CWOs with a single cloud ($\fa=1$) in \sea{dff_cw_app}, and for multiple clouds within a CW screen in \se{mdff_cw}.

\smallskip
For a screen in the MW or in the FRB host, we  express $\dnus$ as a function of the screen DM,
\be
    \label{eq:dnus_dm}
    \dnus 
        \sim 1\mhz\ \nuf^{\frac{22}{5}} \opzo[1]^{\frac{17}{5}}  \frac{\dL[kpc]^2}{\dsc[kpc]}  
        \lt(\dms \fa^{\frac{1}{3}} \rt)_{2.3}^{-\frac{12}{5}}.
\ee
For reference, a scintillation bandwidth of $\sim1\mhz$ is of the order of typical $\dnus$ for a MW-ISM screen at high Galactic latitudes.

\smallskip
Following \citet[][hereafter \citeta{kumar24}]{kumar24}, the flux modulation due to a scattering screen depends on three components: the scintillation regime of the screen, the coherence scale of the screen with respect to the source size, and the frequency resolution of the detector. For completeness, we briefly describe here each component and refer the reader to \citeta{kumar24} for further details.

\subsubsection{Flux modulation due to the scintillation regime}
\label{se:dffu}

As mentioned previously, strong scattering and scintillation occur when the coherence length of a screen is small compared to the Fresnel scale.\footnote{In the strong regime, where $\lpi<\rf$, the coherent regions on the screen contributing to the observed wave amplitude (i.e. the Fresnel-Kirchhoff integral) are on the scale of $\lpi$; while in the weak regime ($\lpi>\rf$) coherent contribution will arise from a disc on the scale of $\rf$ (also known as the first Fresnel zone), and regions outside this zone are cancelled out due to rapid oscillations.}
The scintillation strength is defined as
\be
    u = \frac{\rf}{\lpi},
    \label{eq:u}
\ee
where a screen is in the strong scintillation regime if $u>1$, and in the weak regime if $u<1$. Using \eq{taus} and \eq{dnus}, the scattering strength can also be expressed as $u=(\nuo/\dnus)^{1/2}$. In this way, we see that strong scattering is achieved if the scintillation bandwidth is smaller than the observed frequency. However, since the scintillation bandwidth is itself a function of frequency (\eqnp{dnus}), we define the transition frequency ($\nuou$) between the weak and strong regimes as the frequency for which $u(\nuou)=1$.
We thus have $u(\nuo)=(\nuo/\nuou)^{-17/10}$, with 
\be
    \nuou 
        = \nuo \pfrac{\rf}{\lpi}^{\frac{10}{17}}
        \sim 6.2 \ghz\ \Deff[,9]^{5/17} \opzo^{-1} \fa^{\frac{6}{17}} \lc[,10]^{\frac{2}{17}} \n[,-2]^{\frac{12}{17}}.
        \label{eq:nuou}
\ee
We discuss and show examples of the scintillation strength in CWOs with a single cloud in \sea{dff_cw_app}, finding that even for $\fa=1$ most CWOs are in the strong regime at $\nuo=1\ghz$. We discuss the general case of multiple clouds within a CW screen in \se{mdff_cw}.

\smallskip
As before, for a screen in the MW or the FRB host, we express $\nuou$ as a function of the DM,
\be
    \nuou \sim 7.6 \ghz\ \dsc[kpc]^\frac{5}{17}\dL[kpc]^{-\frac{10}{17}} (1+z)_1^{-1} (\dms\fa^{\frac{1}{3}})_{2.3}^{\frac{12}{17}},
    \label{eq:nuou_dm}
\ee
where we have used \eq{lpi_dm} and \eq{rF}. For frequencies smaller (larger) than $\nuou$, the screen is in the strong (weak) scattering regime.

\smallskip
In the transition from the weak to strong scintillation regimes, an additional scale besides the diffractive scale, $\lpi$, emerges in the diffraction pattern, namely the refractive scale,
$\rref\equiv\rf^2/\lpi$.\footnote{The refractive scale is only meaningful in the strong regime, where $\rref \sim u\rf$, and $\lpi\sim \rf/u$. In the weak regime, the scale that illuminates the observed image is of order $\sim\rf$.}
A scatter-broadened image, with an angular size of $\tho$,\footnote{For a screen at redshift $z$, with a deflection angle $\dth\propto\lamz/\lpi$ (see the cartoon in \fig{time_delay}, and appendix B in \scattpaper), $\tho=\dth\dsl/\dso$ is in the rest frame of the screen, and $\thoo=\tho\opzo[]$ is in the observer rest frame.} 
is illuminated by a patch on the screen of size $\sim\rref$. This scale represents the area on the scattering screen containing all the observed rays contributing to a given image. 
Using geometric considerations, for a deflection angle of $\dth$, the refractive scale is $\rref\seq\dth\Deff=\tho\dlo$, representing the scale of the observed image projected back onto the screen. 
A helpful way to envision strong scattering is described in \citet[][hereafter \citeta{narayan92}]{narayan92}, and summarised here as follows. When the scintillation is in the strong regime, there are $\num$ coherent patches of size $\lpi$ in a disc of size $\rref$, which illuminate the observed image. If $u\gg1$, every point on the observed image receives an immense number of rays, $\num\sim \rref^2/\lpi^2=u^4$, with a strong dependence on the scattering strength. The random phases of these rays generate an interference pattern.

\smallskip 
Note that some notation conventions vary in the literature. To clarify our notation, the refractive scale is defined in the frame of the screen as $\rref=\rf^2/\lpi\propto\lamz\Deff/\lpi$, while the scattering cone scale (equivalently the scattering disc scale or the multipath scale) is defined in the observer frame as $\lcone=\opzo[]\rref\propto\lamo\Deff/\lpi$.\footnote{The scattering cone scale (illustrated in \fig{time_delay}) can be found by $\lcone\simeq\tho\chilo=\thoo\dlo$, where $\chilo$ is the comoving distance between the screen at redshift $z$ and the observer, $\chilo=\dlo\opzo[]$ (see \fig{time_delay} and appendix B in \scattpaper).}

\smallskip
For a point source scattered off a single scattering screen with a Kolmogorov perturbation spectrum, in the weak scattering regime, one can show that 
the flux modulation scales with the scattering strength as $\dffu\approx u^{5/6}$ \citepa[see derivation in][]{narayan92}. When the scattering is in the weak regime, the modulation index is low, while the strong scattering regime results in a modulation index of $\dffu=1$. 
The flux modulation due to the strength of scintillation is thus 
\be
    \dffu 
    = \min\Big[1, u^{\frac{5}{6}}\Big]
    = \min\Bigg[1, \pfrac{\nuo}{\nuou}^{-\frac{17}{12}}\Bigg].
    \label{eq:dffu}
\ee

\subsubsection{Flux modulation due to detector resolution}
\label{se:dffob}

The effect of the detector frequency resolution ($\dnuo$) on the flux modulation ($\dffob$) was explored in \citet{beniamini22} and \citeta{kumar24} in the two limiting cases of detector resolution much smaller or larger than the scintillation bandwidth. 
As discussed previously, the characteristic frequency bandwidth over which the flux modulates is $\dnus$ (\eqnp{dnus}). If the detector does not resolve this scale, $\dnuo\gg\dnus$, flux variations of order unity could be substantially suppressed when averaged over the larger observed scales. It was shown in \citeta{kumar24} that the flux modulation in this case is suppressed by a factor of $\dffob\sim(\dnus/\dnuo)^{1/2}$ \citep[see details in][]{beniamini22}. 
A high spectral resolution can also suppress the flux modulation. 
This can be understood by considering the measurement of the flux modulation between two frequencies, separated by $\dnuo\ll\dnus$. In such a case, the two frequencies are more likely to be both sampled in or out of an enhanced scintle (with a scale of $\dnus$), resulting in a small flux amplitude difference and a low flux modulation. 
In this limit, the flux modulation is suppressed by a factor of $\dffob\sim\dnuo/\dnus$ \citepa{kumar24}.
Combining these two limits, the flux modulation due to the frequency resolution of the detector is $\dffob\sim\min\lt[{\dnuo}/{\dnus}, ({\dnus}/{\dnuo})^{1/2}\rt]$, which is of order unity when the detector resolution is comparable to the scintillation bandwidth of the screen.
The $\dnuo\ll\dnus$ limit will only suppress $\dff$ when measured at a resolution where two nearby channels fall in the same scintle; in such cases, the data can be analysed on scales that optimise $\dffob$ by degrading the resolution, and $\dffob$ can be expressed as 
\be
    \dffob 
    \sim \min\lt[1, \pfrac{\dnus}{\dnuo}^{\half}\rt].
    \label{eq:dffob_1}
\ee
Thus, to detect the flux modulation of a screen at $\nuo$, the desired $\dnuo$ should be comparable to or smaller than $\dnus(\nu)$.
In \sea{dff_cw_app}, 
we show an example for $\dnus$ assuming $\nuo=1\ghz$ for CWOs with a single cloud, $\fa=1$; and we further discuss the impact of $\dnus$ on $\dff$ in \se{mdff_cw} for multiple clouds, and a given limiting frequency of the source size, $\nuoRs$.

\smallskip
Under certain circumstances, some freedom in this restriction is possible. If $\dnuo>\dnus$, one may achieve comparable scales by a slight increase in the observed central frequency. Given the strong frequency dependence of $\dnus$ ($\propto\nuo^{22/5}$), a relatively small increase in $\nuo$ will result in a much larger increase in $\dnus$. However, this is only possible if $\nuo<\nuou$, and the increased $\nuo$ is within the detector band. 
The dependence of $\dff$ on the ratio $\dnus/\dnuo$ can be used to separate different screens along the los. For instance, \citet{nimmo25} identified two distinct scintillation scales, $\dnus$, and modulation indices in the spectrum of FRB~20221022A, suggesting they originated from two different screens along its path (likely the host and MW screen). They then utilised the dependence of $\dffob$ on this ratio to clearly separate the modulation index associated with the larger $\dnus$ scale by degrading the spectral resolution to be between the two scales, which reduced the influence of the smaller scale on $\dff$.

\subsubsection{Flux modulation due to source size}
\label{se:dffRs}

While FRBs are intrinsically compact and can generally be treated as point sources, if their los passes through a strong scattering screen, their size may be perceived as an extended source by subsequent screens. 
The flux modulation for an extended source can be suppressed relative to the case of a point source discussed thus far \citepa{narayan92}. This is particularly true if the source size as perceived by the second screen is larger than the coherence length scale. In this case, every point on the resolved source undergoes diffractive scattering and, since the signal is not coherent over the size of the image, destructive interference reduces the modulation.\footnote{The modulation index may be smaller than unity both in the weak and strong scintillation regimes when the source is extended. One can distinguish between the two by the fact that in the strong (weak) regime, $\dnus/\nuo\ll 1$ ($\dnus/\nuo\sim1$).}

\smallskip
To find the condition for strong flux modulation due to a scintillating screen (screen 2) in the presence of an extended source resulting from a previous screen (screen 1, which is closer to the actual source), we compare the diffractive scale of the second screen to the scattering cone scale of the first screen when both are projected back onto the first screen plane.\footnote{For a detailed derivation following \citeta{narayan92}, see \sea{ExtPoint}.}
In such a case, a source will appear as a point source to the second screen if $\lpib[2]/\lcone[1]\geq1$, where the diffractive scale of the second screen projected back onto the first is $\lpib[2]=\lpi[2]\dlo[1]/\dlo[2]$, and $\lcone[1]=\rref[1](1+z_1)\propto\lamo\Deff[1]/\lpi[1]$.
It is often convenient to use an equivalent forward projection, where the same condition is achieved by $\lpif[1]/\lcone[2]\ge1$. In this case, the coherence scale of the first screen is projected onto the plane of the second screen, $\lpif[1]=\lpi[1]\dsl[2]/\dsl[1]$, and compared to the scattering cone of the second screen (see \sea{ExtPoint}).

\smallskip
While the results from the backward and forward projections are equivalent, we choose the one with simpler expressions in each case. 
Namely, in \se{mdff_mw}, the proximity of the MW screen to the observer simplifies the expressions of a forward projection, since $\dsl[2]\simeq\dso$. Similarly, in \se{mdff_cw}, the proximity of the host screen to the source makes the backward projection more convenient, since $\dlo[1]\simeq\dso$ (see \sea{ExtPoint}).
The projected coherence scale of the CW screen in these two cases is
\be
    \lpisl\approx\lpi\dso/\dsl, \qquad\lpiso\approx\lpi\dso/\dlo,
    \label{eq:lpip}
\ee
which can be used with results from \fig{clump_prop} and \eq{lpi} to find the projected coherence scale of a given CW screen.

\smallskip
In \se{mdff_mw} and \se{mdff_cw}, $\lcone$ is calculated for screens in the MW, $\Rmw$, and the host galaxy, $\Rhg$, respectively.\footnote{In both backward and forward projections, $\lcone$ is the scale of the screen onto which we project.}
Using  \eq{lpi_dm} and \eq{rF}, the general expression for these cases is 
\begin{equation}
    \label{eq:Rcone}
    \lcone
    {=} \frac{\lamo\dsc}{2\pi\lpi}
    {\sim} 3.8\tm 10^{12} \cm \
    \nuf^{-\frac{11}{5}}\!\opzo[1]^{-\frac{6}{5}}
    {\frac{\dsc[kpc]}{\dL[kpc]}} 
      \Big(\dms\fa^{\frac{1}{3}} \Big)_{2.3}^{\frac{6}{5}},
\end{equation}
where $d$ represents either the distance from the MW screen to the observer or the distance from the host-galaxy screen to the source.

\begin{figure}
    \centering
    \includegraphics[width=0.85\linewidth]{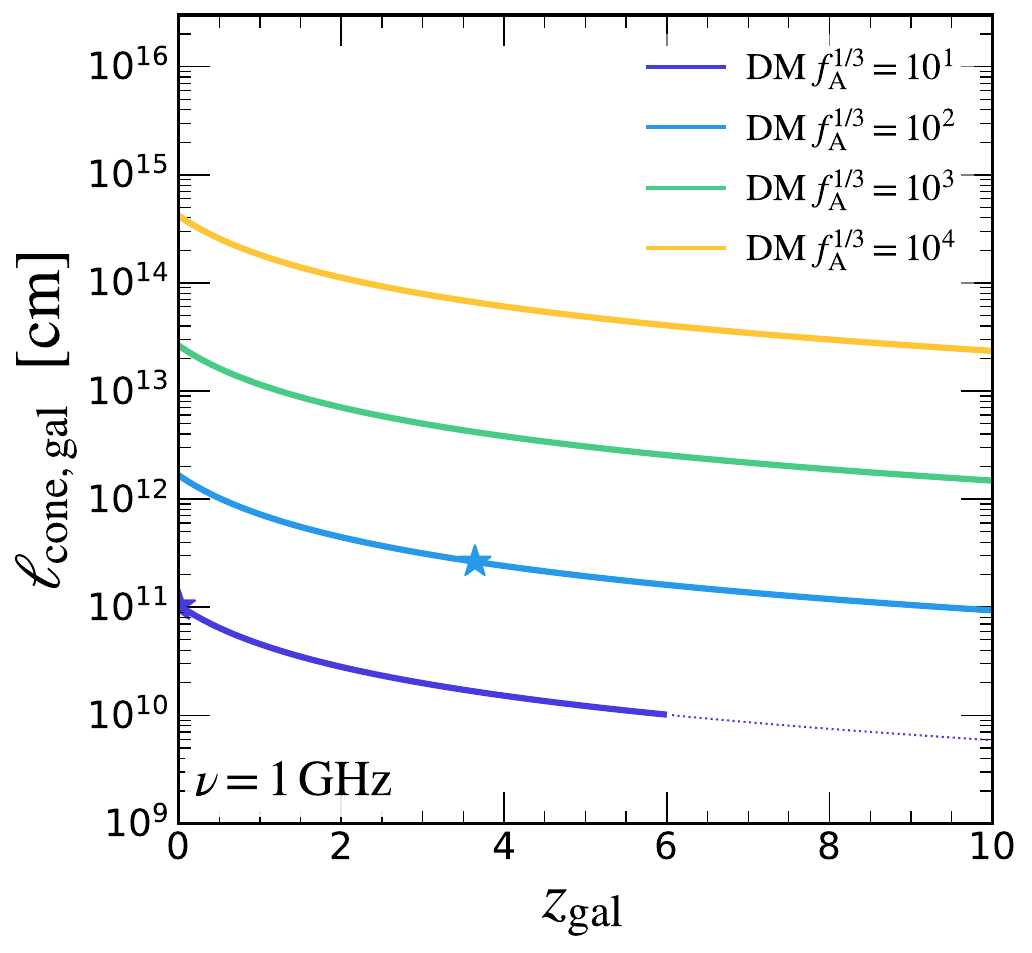}
    \caption{The scattering cone scale for a screen in the FRB host galaxy as a function of host redshift, $z_{\rm gal}$. For $z_\mr{gal}\sim0$, we can treat this as the scattering cone scale for a screen in the MW. We assume an observed frequency of $\nuo=1\ghz$. Each coloured curve represents different conditions in the screen, parametrized by $\dmsfa$ of the screen. 
    The solid lines indicate the strong scattering regime for a screen at a distance of $d=1\pc$ from the source for FRB host galaxies, or $1\pc$ from the observer for the MW. 
    From bottom (purple) to top (orange), the coloured lines correspond to an intrinsic scattering time of $\tausin[gal(1\ghz,\dsc=1\pc)]\sim10^{-4}, 10^{-1.5}, 10^{0.9}$, and $10^{3.3}\ms$, respectively.
    The intrinsic scattering time, $\tausin[gal(1\ghz,{\dsc=1\kpc})]$, is a factor of $10^3$ smaller, and the stars in the two bottom curves indicate the highest $z_\mr{gal}$ at which a screen with $d=1\kpc$ is in the strong regime.}
    \label{fig:Rcone}
\end{figure}

\smallskip
In \fig{Rcone}, we show $\lcone$ as a function of $z$ following \eq{Rcone}. This can be interpreted either as $\lcone$ of a screen in the FRB host-galaxy ($z=z_\mr{gal}>0$, $\dsc=\dsl[,\hg]$), or as $\lcone$ of a screen in the MW ($z=z_\mr{gal}\approx0$, $\dsc=\dlo[,\mmw]$). We assume $\nuo=1\ghz$, $\dsc=\dL$, and show different values of $\dmsfa$ as marked. The solid lines represent a screen in the strong regime\footnote{Since $\lcone$ is expected to be significantly smaller in the weak regime, we are only interested in cases where the screens are in the strong regime.} 
assuming $\dsc=1\pc$, appropriate for the circumburst region of an FRB.  
For a given $\dmsfa$ and $\dsc$, we can relate each line to the intrinsic scattering time of the screen, $\tausin[gal]=\taus[gal](1+z_\mr{gal})^{17/5}$ (using \eqnp{taus_dm}).
For $\dsc=1\pc$, the lines from bottom to top correspond to $\tausin[gal(1\ghz)]\sim10^{-4}, 10^{-1.5}, 10^{0.88}$, and $10^{3.3}\ms$. 
The middle two values are within the estimated range of intrinsic scattering times for the host galaxies of localised FRBs (e.g. \citealt{glowacki25, acharya25} where $\tausinhg[(1\ghz)]\sim10^{\pm2}\ms$ at $\zs<1$). 
So long as $\dsc=\dL$, $\lcone$ does not change if we assume a larger $\dsc[gal]$ (\eqnp{Rcone}); however, this does affect the scattering time (\eqnp{taus_dm}) and the transition to the weak scintillation regime at a given $\nuo$.
For example, if we assume instead $\dsc=1\kpc$, as may be relevant for a screen in the MW, then $\taus$ will be lower by a factor of $10^{3}$. 
The stars in the two bottom lines indicate the highest redshift at which the screen is in the strong regime, assuming $\dsc[gal]=1\kpc$.

\smallskip
Given the different frequency dependence of $\lpi\propto\nuo^{6/5}$ and $\lcone\propto\nuo^{-11/5}$, we can identify a minimal frequency, $\nuoRs$, above which the flux modulation (due to the second screen) is not suppressed by the broadened source (due to the first screen). This transition frequency satisfies $\lpip(\nuoRs)=\lcone(\nuoRs)$ and 
can be expressed as 
\be
    \nuoRs 
    = \nuo\pfrac{\lpif[1]}{\lcone[2]}^{-\frac{5}{17}}
    = \nuo\pfrac{\lpib[2]}{\lcone[1]}^{-\frac{5}{17}}.
    \label{eq:nuoRs_base}
\ee
The final expression for the flux modulation due to the source size is\footnote{We note that the power of $17/5$ in \eq{dffRs} is different from the results in \citeta{kumar24} for the source size. This is due to different definitions and purposes. 
In \citeta{kumar24}, the source size $\Rs$ relates to models for the radio emission region from the central engine, which has no dependence on frequency; whereas here, the scattering cone, $\lcone$, is derived from a passage through a scattering screen, and as a result, it has a dependence on $\nuo$ (and is generally much larger).}
\be
\dffRs
    \sim
    \min\Bigg[1, \pfrac{\nuo}{\nuoRs}^{\frac{17}{5}}\Bigg],
    \label{eq:dffRs}
\ee
where for any frequency $\ge\nuoRs$, $\dffRs\sim1$.

\subsubsection{Total flux modulation:}

Finally, including all contributions in \eq{dffu}, \eq{dffob_1}, and \eq{dffRs}, the total flux modulation is
\begin{equation}
{\dff 
    {\approx} 
        {\min\lt[1, \pfrac{\nuo}{\nuou}^{-\frac{17}{12}} \rt]}
        {\min\lt[1, \pfrac{\dnus}{\dnuo}^{\frac{1}{2}} \rt]}
        {\min\lt[1, \pfrac{\nuo}{\nuoRs}^{\frac{17}{5}} \rt].}}
\label{eq:dff}
\end{equation}
Assuming a detector with a resolution of $\dnuo\lsim\dnus$, if the observed frequency is in the range of $\nuoRs\!<\nuo<\nuou$ (see \eqnp{nuou} and \eqnp{nuoRs}), then the overall flux modulation is expected to be $\dff\sim 1$.

\section{Counting CWOs and cloudlets }
\label{se:cosmo}

In order to estimate how observed FRB scintillation may be affected by an encounter with cloudlets in a CWO, we must first estimate both the probability of intercepting a given CWO and the number of cloudlets within it. 
In our two companion papers, we estimate the probability of encountering a given type of CWO per unit redshift (\dmpaper), and the number of clouds expected within different CWOs (\scattpaper, Section 4). We briefly summarise these results here, and use them in \se{mdff_mw} and \se{mdff_cw} below to estimate how multiple cloudlets within CWOs affect the observed flux of an FRB.

\subsection{Average number of intercepted CWOs}
\label{se:Nobj}

To find the probability that a given los intercepts a CWO, we estimate the number density, $\nc$, and the projected area, $\Ax$, of CWOs as a function of redshift.
For simplicity, we assume the maximal projected area for sheets and filaments, such that a los is perpendicular to the sheet plane and the filament spine. These projections maximise the cross-section of the CWO on the sky, and give an upper limit to the probability of intercepting the CWO. On the other hand, such a los has a shorter path length through the CW screen, and thus represents a lower limit on its contribution to scintillation, since the los is expected to encounter fewer clouds.

\smallskip
As described in our companion papers (\scattpaper and \dmpaper), the average number of intercepted CWOs with total masses $\geq \mobj$ at redshift $z\pm dz/2$ is given by
\be
\frac{d\num(z,\geq M_\mr{obj})}{dz} 
    = \frac{c\opzo[]^2}{H(z)} \int_{M_\mr{obj}}^\infty dM\, \Ax[p](z,M)\,\frac{dn_\mr{c}(z, M)}{dM},
\label{eq:dNdz}
\ee
where $\nc$ is the comoving number density of these objects,\footnote{The proper number density is $\nc\opzo[]^3$.} and
$\Ax[p]$ is their proper projected cross-section. Here, the total CWO mass includes all the substructures it contains (e.g. a sheet of mass $M$ includes the masses of embedded filaments and haloes).

\smallskip 
For sheets and filaments, we evaluate $\dndz$ as a function of their virial temperature, $\Tv$. To do this, we combine the virial properties of CWOs as modelled in \dmpaper (sheets) and \citeta{m18} (filaments),\footnote{See also a brief summary in \sea{vir}, and \S4 in \scattpaper.}
with an excursion set model developed by \citet{shen06} that models the total mass of sheets and filaments for a given embedded halo mass, using a conditional mass function. 
A more detailed discussion is provided in \dmpaper and \S4 in \scattpaper. We discuss the implications of our results for the encounter probability in \se{mdff_mw}, and we present results for the median of $\dndz$ for different CWOs as a function of $z$ in \figsiti{dff_mw_zS05}{dff_mw_zS2}.

\subsection{The conditions for multiphase gas formation in CWOs}
\label{se:frag_cond}

In \scattpaper, we derived constraints on which CWOs might host small-scale cold gas, and on the number of cloudlets encountered along a given los through a CWO. We briefly relay the main results of this model here, and refer the reader to \scattpaper for further details.

\smallskip
In order for a given CWO to host small-scale cold cloudlets, it must satisfy two conditions. The first is that the total electron column density of the CWO is greater than the expected column density of a single cloudlet, according to our cloud model described in \se{CW_clumps}. We parametrized this using the maximal possible covering fraction, $\famax$, where we assumed that the entire column density along a sightline through a CWO, $\Nt$, consists of cold cloudlets. Namely, $\Nt[,hot]=0$ and $\Nt=\Nt[,cold]$. The maximal covering fraction, $\famax$, is then defined as the ratio of the total average column density, $\Nt=\Nt[,cold]$, to the column density of a single cloud, $\N=\lc\n$. Requiring $\famax\ge1$ restricts some CWOs to certain redshift ranges. The associated maximal possible volume-filling fraction is $\fvmax=\famax\lc/\DL=\nt/\n$, where $\DL$ is the path length through the CWO (see \scattpaper, \dmpaper).

\smallskip
The second condition is related to the ratio of the cooling time to the free-fall time, $\tcool/\tff$, which we use to estimate the susceptibility of the medium to local thermal instabilities. We adopt a threshold of $\tcool/\tff\lsim10$ \citep{mccourt12, sharma12, voit-donahue15}\footnote{A Similar threshold was found in studies using non-spherical geometries, with plane-parallel and cylindrical configurations \citep{meece15, choudhury16}.} 
for CWOs to experience condensation and fragmentation. However, we note that in the presence of strong perturbations, condensation may begin to occur at higher thresholds of $\tcool/\tff$ \citep{choudhury16, choudhury19}. 

\smallskip
In summary, we consider cold-cloud formation only in CWOs that obey the following two conditions:
\be
    \label{eq:frag_cond}
    1.\quad &\famax =\frac{\Nt}{\N}\geq 1\\
    2.\quad &\frac{\tcool}{\tff} < 10
\ee
The first condition is essential -- for a given path length through a CWO, the column density of a single cloudlet cannot exceed the total column density available for both the hot and cold phases.\footnote{We note that the first condition limits only the maximal area-covering fraction, $\famax$, while $\fa<1$ is allowed as long as $\fa<\famax$.}
The second condition, on the other hand, is taken as a crude guideline. In \scattpaper (\Ss4.2-4.4), we demonstrate how these conditions translate into redshift-dependent constraints on relevant CWOs.
In addition to these two conditions, when we examine the impact of different volume-filling fractions of cold cloudlets, we exclude CW screens with $\fv>\fvmax$ (or $\fa>\famax$).

\section{Suppression of MW scintillation by CWOs}
\label{se:mdff_mw}

\begin{figure}
    \centering
    \includegraphics[width=0.85\linewidth]{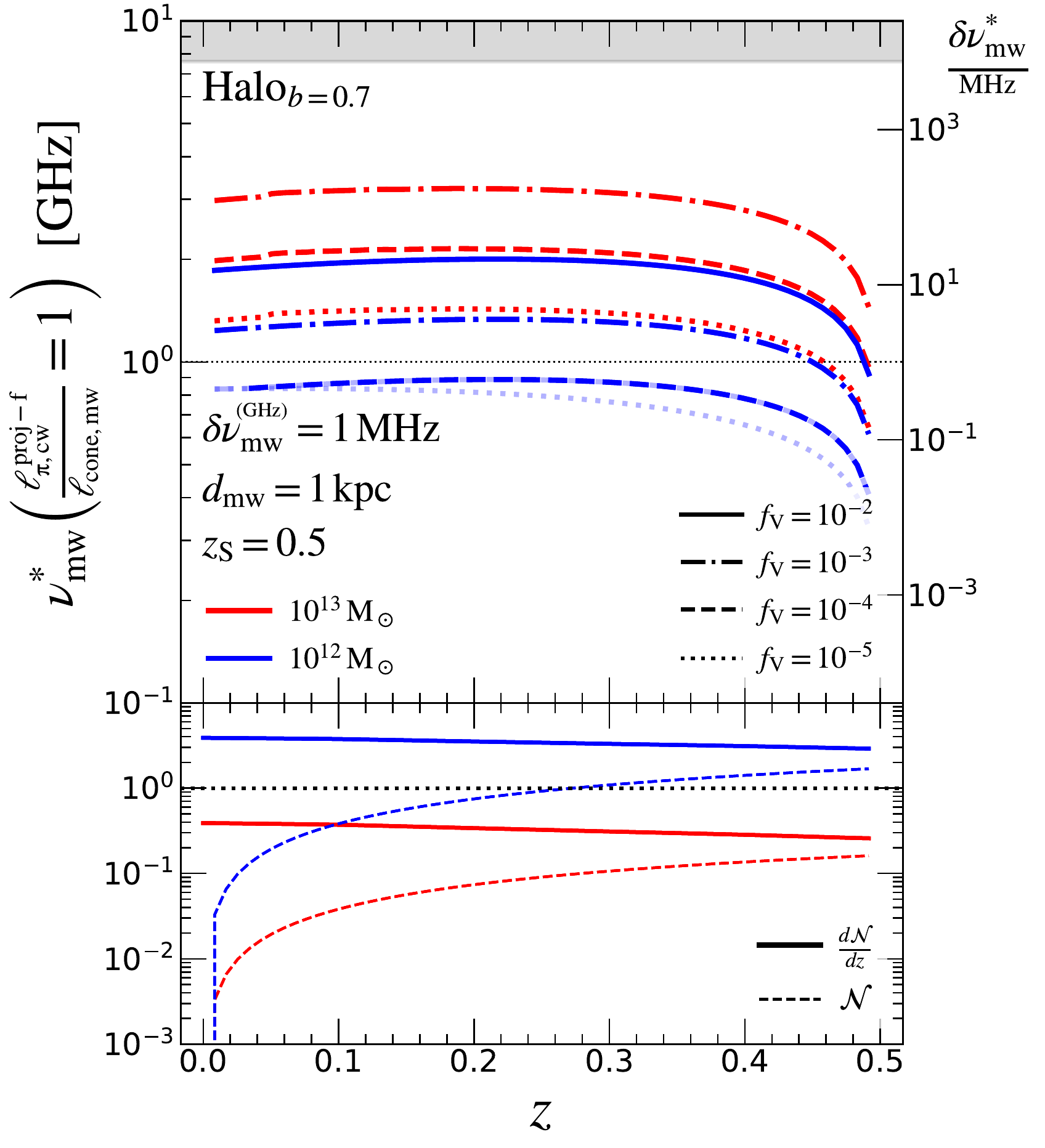}
    \caption{Suppression of MW scintillation for a los passing through an intervening halo. 
    In the top panel, we show the maximal frequency below which MW-ISM scintillation is suppressed ($\nuoRsmw$, left y-axis) by source broadening caused by a single CW screen at redshift $z$ (x-axis). 
    Blue (red) lines indicate haloes with $\Mv/\msun=10^{12}$ ($10^{13}$), with different line styles indicating different $\fv$ (see legend). We exclude redshifts and masses where clouds are not expected according to \eq{frag_cond}, and use faint lines when $\fa$ in the intervening halo is lower than one, $\fa<1$, making an interception with cloudlets less likely. The right y-axis indicates the corresponding scintillation bandwidth, $\dnusRsmw=\dnusmw(\nuoRsmw)$, and the shaded grey region at the top marks frequencies where the MW is in the weak regime for the assumed los with $\dnusmwghz$. In the bottom panel, we show the number of haloes per unit redshift interval $\dNdz$ (solid), and the cumulative number of haloes $\num(z, \Mh)$ (dashed). Blue and red lines correspond to mass ranges $\Mv/\msun=(10^{12}-10^{13})$ and $(10^{13}-10^{14})$, respectively. 
    In this example, the source is at $\zs=0.5$; its los passes through an intervening halo at an impact parameter of $b=0.7$ and a MW-ISM screen that is known to have a bandwidth of $\dnusmw=1\mhz$ at $\nuo=1\ghz$.
    Under the conditions shown here, if a halo of $10^{12}\msun$ has turbulent cloudlets with $\fv=(10^{-4}-10^{-2})$, 
    the MW flux modulation is expected to be suppressed when observed at a frequency of $\nuoRsmw\lsim 0.9, 1.3$, and $2\ghz$, respectively.}
    \label{fig:dff_mw_zS05}
\end{figure}

\begin{figure*}
    \centering
    \includegraphics[width=0.8\linewidth]{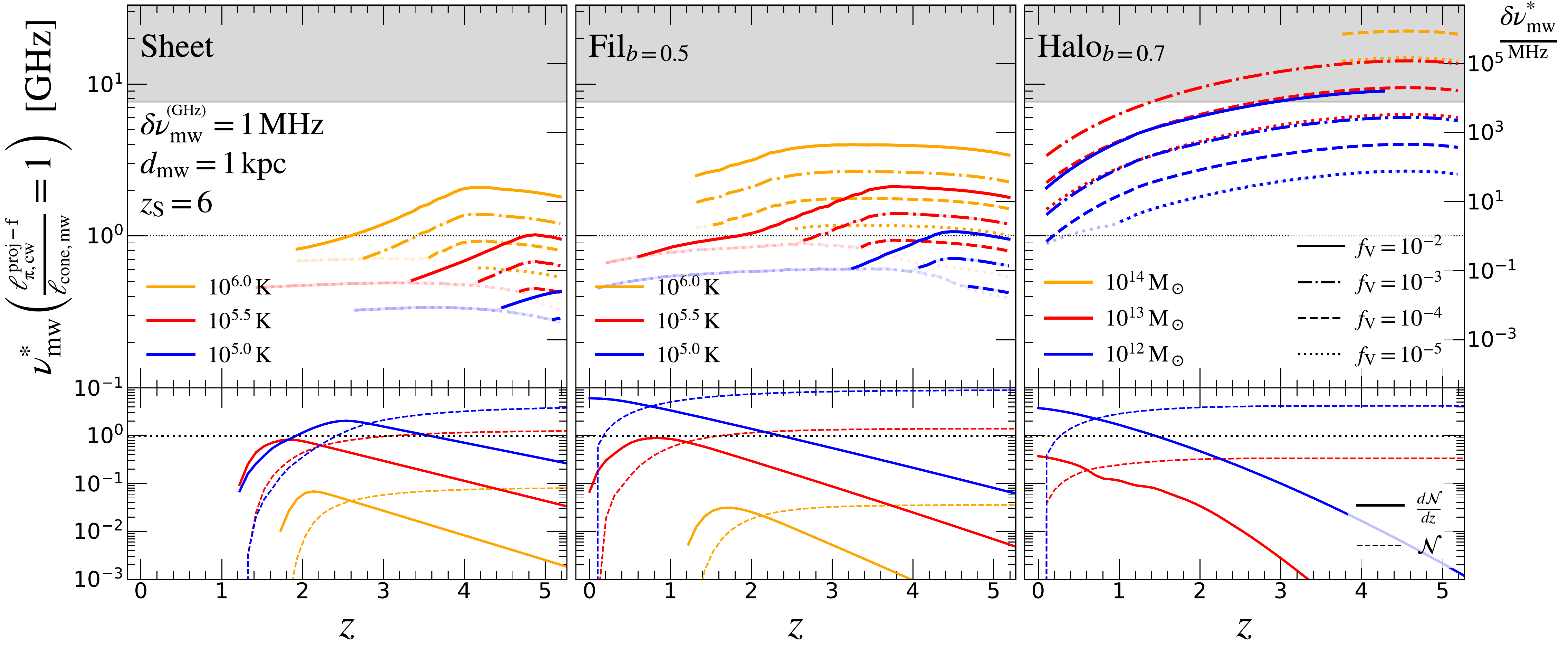}
    \caption{
    Similar to \fig{dff_mw_zS05}, suppression of MW scintillation from an FRB source at $\zs=6$ due to the los passing through an intervening CWO (sheets, filaments, or haloes, from left to right). 
    In the top row, we show the maximal frequency below which MW-ISM scintillation is suppressed ($\nuoRsmw$, left y-axis), as a function of the redshift of the intervening CWO. 
    Similar to \fig{dff_mw_zS05}, different line styles indicate different $\fv$, and different colours indicate different virial temperatures for sheets and filaments, and virial masses for haloes (see legend). The right y-axis is the MW scintillation bandwidth corresponding to the frequency $\nuoRsmw$. 
    In the bottom row, solid lines represent the median number of CWOs per unit redshift, $\dNdz$, and dashed lines represent the cumulative number of objects, $\num(z)\equiv\int_0^{z}dz\,\dNdz$.
    Different colours represent different temperature (mass) ranges for filaments and sheets (haloes), between $\Tv$ and $10^{0.5}\Tv$ ($\Mh$ and $10\Mh$). As an example, if a los from an FRB at $\zs=6$ passes through a $\gsim10^{5.5}\Kel$ filament at $z\gsim3$ or a $\gsim10^{6}\Kel$ sheet at $z\gsim3.5$, with $\fv\gsim10^{-3}$, the MW scintillation is expected to be suppressed below $\nuo\lsim1\ghz$. For the same $\zs$, $\dffmw$ is expected to be suppressed below $\nuo\sim1.3\ghz$ if the los passes a single halo of $10^{12}\msun$ at $z\gsim1.5$, with a threshold $\fv$ as low as $\gsim10^{-5}$.}
    \label{fig:dff_mw_zS6}
\end{figure*}

\begin{figure}
    \centering
    \includegraphics[width=0.99\linewidth]{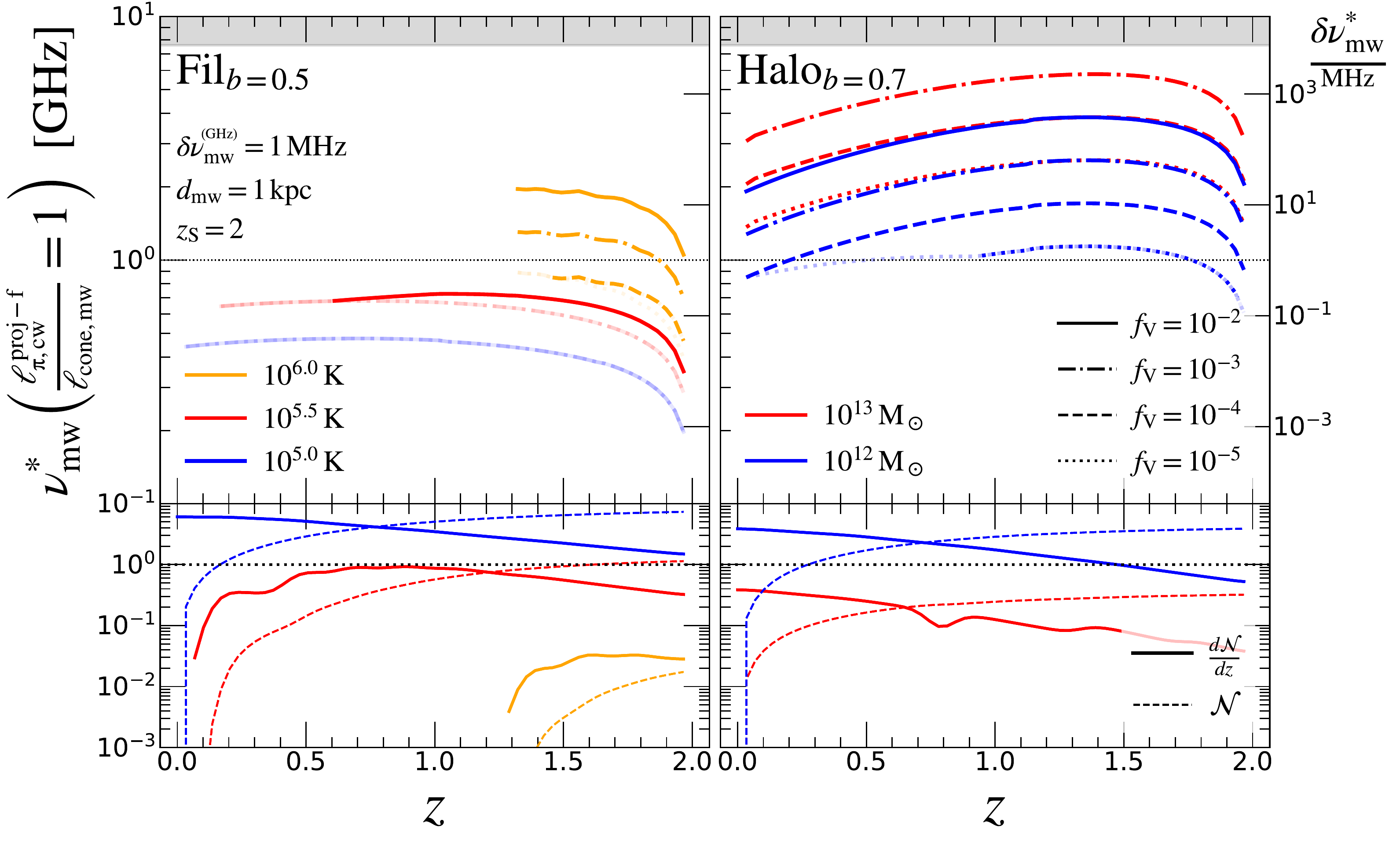}
    \caption{
    Similar to \fig{dff_mw_zS05} and \fig{dff_mw_zS6}, but for a source at $\zs=2$. On the left (right), we show a los passing through an intervening filament (halo). All assumptions about the MW-ISM screen are identical to previous figures. In this example, we see that $10^{6}\Kel$ filaments at $z\sim1.5$ with $\fv=10^{-3}$ ($10^{-2}$) are expected to suppress $\dffmw$ at $\nuo\lsim1.3\ghz$ ($\lsim1.9\ghz$). A $10^{12}\msun$ halo at $z\sim1.5$ with $\fv\sim10^{-4}-10^{-5}$ will also suppress MW scintillation below $\nuo\lsim1.3\ghz$. While distinguishing between such filaments and haloes may be difficult, some information may be drawn from the different probabilities of encountering them (bottom row).}
    \label{fig:dff_mw_zS2}
\end{figure}

\subsection{Detectable signal}
\label{se:dff_mw_sig}

We now combine the general analysis of scattering screens discussed in \se{dff} with our results from \se{cosmo}, and apply them to the case where an FRB los passes through a CW scattering screen before passing through the MW. 
In this case, the broadening caused by the CW screen (first screen, i.e. closer to the source) may suppress the flux modulation expected from the MW-ISM (second screen).\footnote{We generally assume a scintillating screen exists in both the host and the MW, in addition to at least one CW screen along the los, meaning at least three screens in total. However, the host screen typically cannot suppress the MW screen at observable frequencies due to the remarkably small ratio of $\dsl[,\hg]/\dso$ (see \se{hg_mw}). We therefore simplify the discussion by focusing on only two screens, and address the CW screen (MW screen) as the first (second) screen.}
Thus, if the expected flux modulation of the MW ($\dffmw$) is suppressed, this may indicate the existence of a scattering screen along the los. We will show below that for most relevant cases a screen in the FRB host is unlikely to effectively suppress $\dffmw$ (see \se{hg_mw}). 
Therefore, suppression of MW scintillation places constraints on other intervening scattering screens, including those that are hard to detect directly via scattering due to $\taus\ll\taus[\hg]$.
Here, we test explicitly whether diffractive scattering by cloudlets in a CWO can suppress scintillation in a MW-ISM screen, by comparing the scattering cone scale of the MW screen, $\Rmw$, to the diffractive scale of the CW screen, projected forward onto the plane of the MW, $\lpisl=\lpi\dso/\dsl$.\footnote{$\dmw/\dso\ll1$, we therefore consider a projection onto the MW screen and a projection on the observer plane as equivalent.}
A source broadened by a CW screen is perceived as a point source by the MW if $\lpisl/\Rmw>1$.

\smallskip
Using \eqsiti{lpip}{nuoRs_base}, we can express the transition frequency below which MW scintillation is suppressed by a CW screen as $\nuoRsmw=\nuo(\lpisl/\Rmw)^{-5/17}$, giving
\be
    \label{eq:nuoRsmw}
    \nuoRsmw
    &\!\!\sim\!
        0.6 \ghz 
        \frac{\fa^{\frac{3}{17}} \lc[,10]^{\frac{1}{17}} \nf^{\frac{6}{17}} 
        }{\opzo^{\frac{6}{17}}}
        \Bigg(\!\frac{\dsl\,\dmw[,kpc]}{\dso\dLmw[,kpc]} \!\Bigg)^{\!\!\frac{5}{17}} \!\!\dmsfalmw[,2.3]^{\frac{6}{17}}
    \\[5pt]
    &\!\!\sim\!
        0.6 \ghz
        \frac{\fa^{\frac{3}{17}} \lc[,10]^{\frac{1}{17}} \nf^{\frac{6}{17}} 
        }{\opzo^{\frac{6}{17}} }
        \Bigg(
           \frac{\dsl}{\dso}  
      \Bigg)^{\frac{5}{17}} 
        \lt( \frac{\dmw[,kpc]}{\dnusmwghz[,6]}\rt)^{\frac{5}{34}},
\ee
where the source is assumed to be at $\zs=6$.\footnote{$\nuoRsmw$ in \eq{nuoRsmw} depends on $\zs$ through the ratio $(\dsl/\dso)^{5/17}$, where $\dsl$ is the angular distance of the CW screen from the source. See \fig{dso/dsl} for a rough estimate of $\dso/\dsl$ as a function of $\zs$ for different $z$ of the screen.} 
For the MW screen we assume a typical screen-width and distance of $\dmw=\dL[\mmw]\sim 1\kpc$, and choose $\dmsfamw[,2.3]=\dmsfamw/(200\dmunits)$. 
These fiducial values correspond to a MW screen with $\dnusmwghz[,6]=\dnusmw(1\ghz)/(1\mhz)$, motivated by typical observed MW scintillation bandwidths at high latitudes, where the $\dnusmw$ as a function of $\dmsmw$ is given by \eq{dnus_dm}.

\smallskip
Under these conditions, an FRB source broadened by a CW screen and observed at a frequency $\nuo\geq\nuoRsmw$ is perceived as a point source by the MW screen, giving an order-unity flux modulation, $\dffmw\sim1$, on a corresponding spectral scale of $\dnusmw(\nuo)$. 
Conversely, at lower frequencies $\nuo<\nuoRsmw$ the broadened source becomes an extended source with respect to the MW screen, thus smearing the MW screen scintillation interference pattern, and inhibiting the flux modulation to low values.

\smallskip
In the upper panel of \fig{dff_mw_zS05}, we show $\nuoRsmw$ (left y-axis) as a function of the CW screen redshift. The right y-axis shows the corresponding scintillation bandwidth using \eq{dnus_dm}, $\dnusRsmw\equiv\dnusmw(\nuoRsmw)$. We assume a source at $\zs=0.5$, an impact parameter through a single intervening halo of $b=0.7$, and a los passing through a MW-ISM screen with $\dmw=\dLmw\sim 1\kpc$.
Different coloured lines represent different halo masses, $\Mh$, while different line styles represent different volume-filling fractions, $\fv$ (see legends). 
The grey shaded region marks frequencies where the MW screen is in the weak regime ($\ge\nuou[,\mmw]$, see \eqnp{nuou_dm}). If the CW screen is in the weak regime (which does not happen for the parameters we show in \fig{dff_mw_zS05}, nor in any of the other related figures that follow), the source size can usually be treated as a point source. 
In the bottom panel, solid lines show the number of haloes per unit redshift interval, $\dNdz$.\footnote{In \figsiti{dff_mw_zS05}{dff_mw_zS2}, $\dNdz$ is the integral of $d^2\num/(dzdT)$ over the range $[\Tv,10^{0.5}\Tv]$ for filaments and sheets with a given $\Tv$, and the integral of $d^2\num/(dzdM)$ over the range $[\Mh,10\Mh]$ for haloes with a given $\Mh$.}
The dashed lines indicate the cumulative $\num(z)=\int_0^z dz\, \dNdz$. In both panels, we exclude CWOs that are not expected to form cold cloudlets (\eqnp{frag_cond}). 
In the upper panel, we also exclude the redshift ranges for haloes with $\fvmax<\fv$ for a given $\fv$ (\se{frag_cond}). For example, haloes of $\Mh=10^{13}\msun$ have $\fvmax<10^{-2}$, and therefore such haloes with $\fv=10^{-2}$ are not shown (the missing red solid lines in \figsiti{dff_mw_zS05}{dff_mw_zS2}).\footnote{Similarly, $10^{12}\msun$ haloes at $z\gsim4$ with $\fv=10^{-2}$ are also excluded in \fig{dff_mw_zS6} due to their lower $\fvmax$.}
All other exclusions in this figure are due to the conditions in \eq{frag_cond}.

\smallskip
Consider, for example, a MW screen, which is known to have a scintillation bandwidth of $\dnusmwghz=1\mhz$ at $1\ghz$\footnote{One can easily adjust these fiducial values using \eqs{dnus_dm, Rcone, nuoRsmw}.} 
\citep[e.g. using][models]{cordes_lazio02, yao17}. 
The bottom panel of \fig{dff_mw_zS05} tells us that an average los passes through $\sim1$ halo of $10^{12-13}\msun$ at $0.3<z<0.5$. Therefore, if such a halo has turbulent cold cloudlets with $\fv=10^{-4},10^{-3}$, and $10^{-2}$, we predict that the MW flux modulation will be suppressed when observed at a frequency below the threshold of $\nuo\lsim\nuoRsmw = 0.9, 1.3$, and $2\ghz$, respectively. 
A MW screen with lower $\dnusmwghz$, typical for lower latitudes, will further constrain $\nuoRsmw$ to higher values. For example, if $\dnusmwghz$ is lower by a factor of $10$, it will increase $\nuoRsmw$ by a factor of $\sim1.4$ (see \eqnp{nuoRsmw}).

\smallskip
Similar to \fig{dff_mw_zS05}, in \figsii{dff_mw_zS6}{dff_mw_zS2} we show $\nuoRsmw$ for sources at $\zs=6$ and $\zs=2$, respectively, passing through different CWOs at redshift $z$ ($x$-axis). We assume an impact parameter of $b=0.7, 0.5$ for haloes and filaments, respectively. Different coloured lines represent different values of $\Tv$ for sheets and filaments, and $\Mh$ for haloes. The higher source redshifts allow us to place constraints on other CWOs using MW scintillation. 
For example, if a los from an FRB at $\zs=6$ passes through a $10^{5.5}\Kel$ filament at $z\sim2$ with $\fv=10^{-2}$, MW flux modulation will be suppressed below $\nuo\lsim\nuoRsmw\sim1\ghz$. Higher redshift sources also allow us to place constraints on remarkably low values of $\fv$. For a source at $\zs=2$, we find that a $10^{12}\msun$ halo at $z\sim1.5$ (a very likely encounter based on the bottom right panel of \fig{dff_mw_zS2}, which shows $\dndz\sim1$) with $\fv=10^{-5}$ is expected to suppress $\dffmw$ below $\nuo\sim1\ghz$.

\smallskip
The chances of detection can be greatly affected by the probability of interception, $\dndz$, depicted as solid lines in the bottom row of \figsiti{dff_mw_zS05}{dff_mw_zS2}.
For example, using \fig{dff_mw_zS6}, $10^{5.5}\Kel$ filaments at $z\sim3\pm0.5$ have $\dndz\sim10^{-1}$, while $10^{6}\Kel$ sheets at the same redshift have $\dndz\sim10^{-2}$. Rare CWOs may still be explored if the high rate of FRBs partially compensates for the relatively low probabilities for interception along an average los. 
One can evaluate the expected frequency of an FRB sightline passing through a given CWO using $\dNdz$ and the rate of FRBs, $\dotNum$. 
As discussed in \scattpaper, $\dNdz$ can be thought of as the covering fraction on the sky per unit redshift of a given type of CWO. Or alternatively, for $\dndz\le1$ it represents the fraction of FRBs with lines-of-sight that are expected to intercept a given type of CWO per unit redshift. 
\citet{beniamini21} used results from \citet{lu-piro19} to estimate the rate of FRBs above a threshold of observed specific fluence ($e_\nu$). For $\zs>6$ and $e_\nu\geq1\jy\ms$ at $\nuo=1\ghz$, they conservatively find a rate of $\dotNum(\zs>6)\sim10^{4}\yr^{-1}\sky$. For FRBs at lower $\zs$, this rate increases substantially \citep[][see their fig.~3]{beniamini21}.
These high rates of FRBs may allow the detection of rare sightlines. 
For instance, $10^{6}\Kel$ sheets at $z\sim4\pm0.5$ (with $\nuoRsmw\sim0.9-2\ghz$ for $\fv=10^{-4}-10^{-2}$) are expected to cover $\sim10^{-2}$ of the sky. 
Considering only sources with $\zs>6$ and a threshold of $e_\nu\geq1\jy\ms$, then $\dotNum\sim10^{4}\yr^{-1}\sky$ at $1\ghz$;\footnote{The rate of FRBs scales roughly linearly with frequency.}
namely, more than $10^{2}$ such systems are expected to be intercepted by an FRB los per year over the entire sky.\footnote{We address here only the geometric distribution on the sky, and note that actual numbers could be significantly smaller once observational effects are folded in.}

\smallskip
There are certain degeneracies between different CWOs. For example, for a source at $\zs=2$, if a los passes either a $10^{6}\Kel$ filament with $\fv=10^{-3}$ or a $10^{12}\msun$ halo with $\fv\sim10^{-5}$ at $z\sim1.5$, the MW scintillation will be suppressed below similar frequencies around $\nuoRsmw\sim1.3\pm0.1\ghz$. While it is hard to distinguish between two CWOs with such similar scales, some information may be drawn in this case from the very different probabilities of encountering these two CWOs, as shown in the bottom row of \fig{dff_mw_zS2}.

\smallskip 
The possibility of confusion with other screens (whether another type of CW screen, the host screen, or the MW screen) should always be considered. In \se{hg_mw}, we describe one of the important advantages of using the MW scintillation to infer the presence of a CW screen, namely, the fact that the host screen cannot effectively suppress $\dffmw$.

\subsection[Fraction of FRBs with suppressed dffmw by CWOs]{Fraction of FRBs with suppressed $\dffmw$ by CWOs}
\label{se:ffrbmw}

The current sample of FRBs with MW scintillation measurements is small, and more importantly, many of them are at low redshifts of $\zs\lsim0.1$ (see \se{other_works} below). 
A meaningful comparison would require intermediate source redshifts of $\zs\gsim0.3$ (or a statistically significant sample at $\zs<0.3$). 
However, with forthcoming observations in mind, we wish to provide observationally motivated constraints by estimating the fraction of FRBs expected to pass through a CWO that would suppress the MW scintillation at a given observed $\nuo$, $\ffrbmw$. 
We describe below the calculation for haloes, noting that a parallel calculation can be done for sheets and filaments using the temperature instead of the mass (see \sea{ffrbmw_fsh}).

\smallskip
To find $\ffrbmw$, we first use the number of intercepted haloes per unit redshift and per unit mass, $d^2\num/(dzdM)$, and multiply it by $\min(1,\fa)$; in doing so, we include the probability to intercept a cloudlet in CWOs with $\fa<1$.\footnote{For simplicity, we crudely ignore CWOs with $\fa<1$ in most of our results, given that an interception with a cloudlet is less likely. In the current calculation (\sesixi{ffrbmw}{, }{model_var}, and \sea{ffrbmw_fsh}), we combine the effects of the two covering fractions, $d^2\num/(dzdM)$ and $\fa$, to find the sky coverage of cloudlets within CWOs, making it easy to include the probability of encountering at least one cloudlet in CWOs with $\fa<1$.}
Next, we limit the systems to those that can cause $\dffmw$ suppression at or below a given frequency by including only redshift ranges where the transition frequency exceeds it, $\nuo\leq\nuoRsmw$. This results in
\be
    \frac{d^2\nmw}{dz\,dM} =
    \begin{cases}
    {\frac{d^2\num}{dzdM} \min(1, \fa)}, & \nuo\leq\nuoRsmw,
    \\[5pt]
    0, & \nuo>\nuoRsmw,
    \end{cases}
    \label{eq:dnmw_dzdm}
\ee
where $\nuoRsmw$ is calculated by setting the covering fraction of cloudlets to $\max(1,\fa)$ (as shown in faint lines in \figsiti{dff_mw_zS05}{dff_mw_zS2}), since the transition frequency is only meaningful if the los encounters at least one cloudlet. 
The cumulative number of haloes that can cause suppression at $\leq\nuo$ is given by the integration over redshift and mass in \eq{dnmw_dzdm}, 
\be
    \nmw(\zs,\nuo,\geq\Mh) =
    \int_{0}^{\zs}\int_{\Mh}
    \frac{d^2\nmw}{dz\,dM} \ dz\,dM.
    \label{eq:nmw}
\ee
Finally, the fraction of FRBs at $\zs$ that encounter a halo causing MW suppression at $\leq\nuo$ is
\be
    \ffrbmw(\zs,\nuo,\geq\Mh) = 1-\e{-\nmw},
    \label{eq:ffrbmw}
\ee
assuming Poisson statistics for the distribution of haloes along the los.

\begin{figure}
    \centering
    \includegraphics[width=0.85\linewidth]{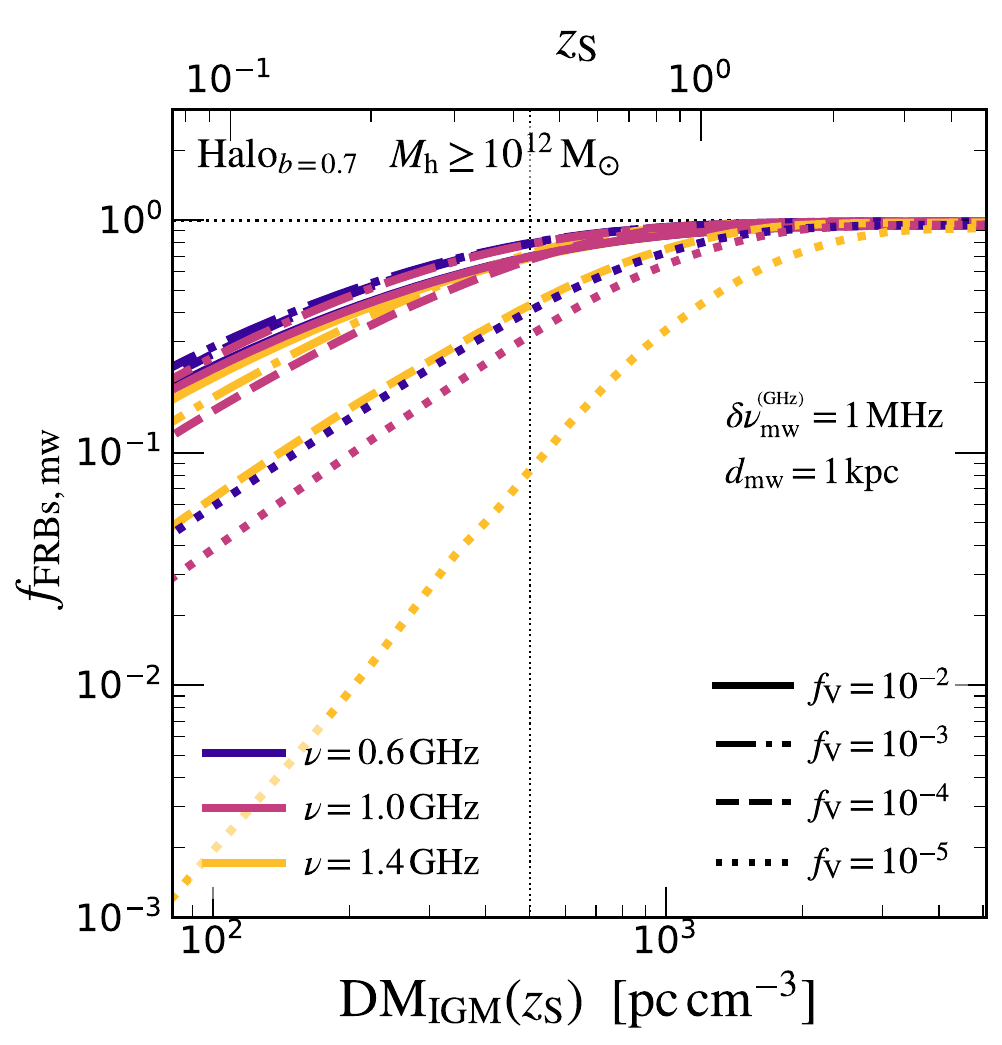}
    \caption{Fraction of FRBs whose MW scintillation is suppressed by $\Mh\geq10^{12}\msun$ haloes ($\ffrbmw$, y-axis).
    To allow comparison with future non-localised samples, we express $\ffrbmw$ as a function of the average extragalactic DM, $\dmigm$ (bottom x-axis, \eq{dmigm}). The top x-axis represents the redshift of the source.
    We assume a los passes through a halo with an impact parameter of $b=0.7$, and through a MW screen with a typical $\dnusmwghz=1\mhz$, $\dmw=1\kpc$.
    The different colours represent different observed frequencies, and the line styles indicate the volume-filling fraction of the cold cloudlets in haloes. 
    For instance, $\gsim40\%$ of FRBs with excess DM of $\gsim500\dmunits$ are expected to have suppressed MW scintillation below a frequency of $\nuo\sim1.4\ghz$ due to haloes with $\fv\gsim10^{-4}$.}
    \label{fig:ffrbmw_h}
\end{figure}

\smallskip
Most current observations with scintillation measurements focus on localised FRBs; however, future observations may include much larger samples of non-localised FRBs. The observed DM includes all contributions along the los. For FRBs at larger distances, the contribution of the IGM ($\dmigm$) becomes dominant over that of the host, the MW, and their haloes, in which case it can be used as an approximate indicator for the source redshift.\footnote{While we consider only an average $\dmigm$ for clarity, the relation between $\zs$ and $\dmigm$ includes a scatter \citep[e.g. see][]{macquart20,jaroszynski19}.}
To broaden comparison with observations, including sources without known redshifts, we express $\ffrbmw$ as a function of the average extragalactic DM expected for a source at $\zs$,
\be
    \dmigm 
    = c\int_{0}^{\zs} dz \frac{\nt(z)}{\opzo[]^2 H(z)},
    \label{eq:dmigm}
\ee
where the average electron density along the path is $\nt\sim\xie\omb\rhou(z)/\mpr$, $\rhou$ is the mean matter density of the Universe, and the free-electron fraction per nucleon is taken to be $\xie=0.88$, which assumes that both hydrogen and helium are fully ionized (for a primordial hydrogen fraction of $\xh=0.76$).
For instance, out of the thousands of FRB sources in the second CHIME catalogue \citep{chime_cat2}, a large fraction have a DM of over $500\dmunits$ 
(marked by the dotted vertical line in \figsii{ffrbmw_h}{ffrbmw_var_h} and \fig{ffrbmw_h_mbins}).
For an average los, this roughly corresponds to $\zs\sim0.5$ \citep[e.g. see][for additional support using spatial correlations from galaxy distributions]{rafiei21_dm_z_cf_chime_cat1, wang25_dm_z_cf_chime_cat2}.

\smallskip
In \fig{ffrbmw_h}, we use our fiducial model to show $\ffrbmw$ due to haloes, as a function of $\dmigm$ (bottom x-axis) and $\zs$ (top x-axis). We include halo masses of $\Mh\geq10^{12}\msun$ with an impact parameter $b=0.7$, and as before we assume a MW screen with $\dnusmwghz=1\mhz$ and $\dmw=1\kpc$. The different colours represent different observed frequencies, and the line styles represent different $\fv$. As before, we include only CW screens that obey the conditions for fragmentation (\eqnp{frag_cond}), and with $\fv\leq\fvmax$.\footnote{The $\fv=10^{-2}$ lines in \fig{ffrbmw_h} are slightly lower than $\fv=10^{-3}$. This is due to the condition $\fv\leq\fvmax$, causing the highest mass haloes with $\fvmax<10^{-2}$ to be excluded (see \se{frag_cond} and \sea{mdff_mw_app}).}
For instance, we find that most FRBs with a DM excess of $\gsim500\dmunits$ (thin vertical dotted line) are expected to have suppressed $\dffmw$ below $\nuo\sim1\ghz$ (purple lines) due to haloes with $\fv\gsim10^{-4}$, with $\ffrbmw\gsim0.7$. Only remarkably low $\fv\sim10^{-5}$ are expected to show lower fractions; even then, for the same DM threshold and below $1\ghz$, at least $20\%$ of the FRBs are expected to have a suppressed $\dffmw$.

\subsection{Caveats to the fiducial model and deviations from it}
\label{se:model_var}
 
In \scattpaper (\S5.3), we discuss the effect of deviations from our fiducial model assumptions. Similarly, here, we can examine how such deviations affect the transition frequency $\nuoRsmw$. For instance, consider a larger $\lc=10\lcmin$, caused either by fragmentation limited in some way, halting at scales large compared to the shattering scale, or by a different process altogether dominating the formation of clouds. For a fixed $\fv$, $\nuoRsmw\propto\lc^{-2/17}$, so this would result in a lower $\nuoRsmw$ by a factor of $\sim0.76$. 
Another issue discussed in \scattpaper is the possibility of significant non-thermal pressure, which affects both $\n$ and $\lc$. If the thermal pressure is a fraction of $f_{P_\mr{th}}=1/2$ of the total pressure, $\nuoRsmw$ would be $f_{P_\mr{th}}^{8/17}\sim0.72$ of the fiducial value. 
Another potential issue is that unaccounted-for damping mechanisms may interrupt the cascade and result in a substantially larger inner scale compared to our estimates. 
Using the second line in \eq{lpi_0}, for the limit where $\lpi\leq\li$, and following a similar derivation as in \se{dff}, the expression for the transition frequency is similar (with a slightly different power), 
\be
    \nuoRsmw[(\lpi\leq\li)] = \nuo \pfrac{\lpisl[(\lpi<\li)]}{\Rmw}^{-\frac{5}{16}},
    \label{eq:nuoRsmw_li}
\ee
with a dependence of $\nuoRsmw[(\lpi\leq\li)]\propto\li^{-5/96}$.
We can quantify this effect by considering different ratios of $\fli$. For example, changing this ratio by a factor of $\sim10$ changes $\nuoRsmw$ by a factor of $(\fli)^{-5/96}\sim0.9$ compared to the fiducial model.

\begin{figure}
    \centering
    \includegraphics[width=0.85\linewidth]{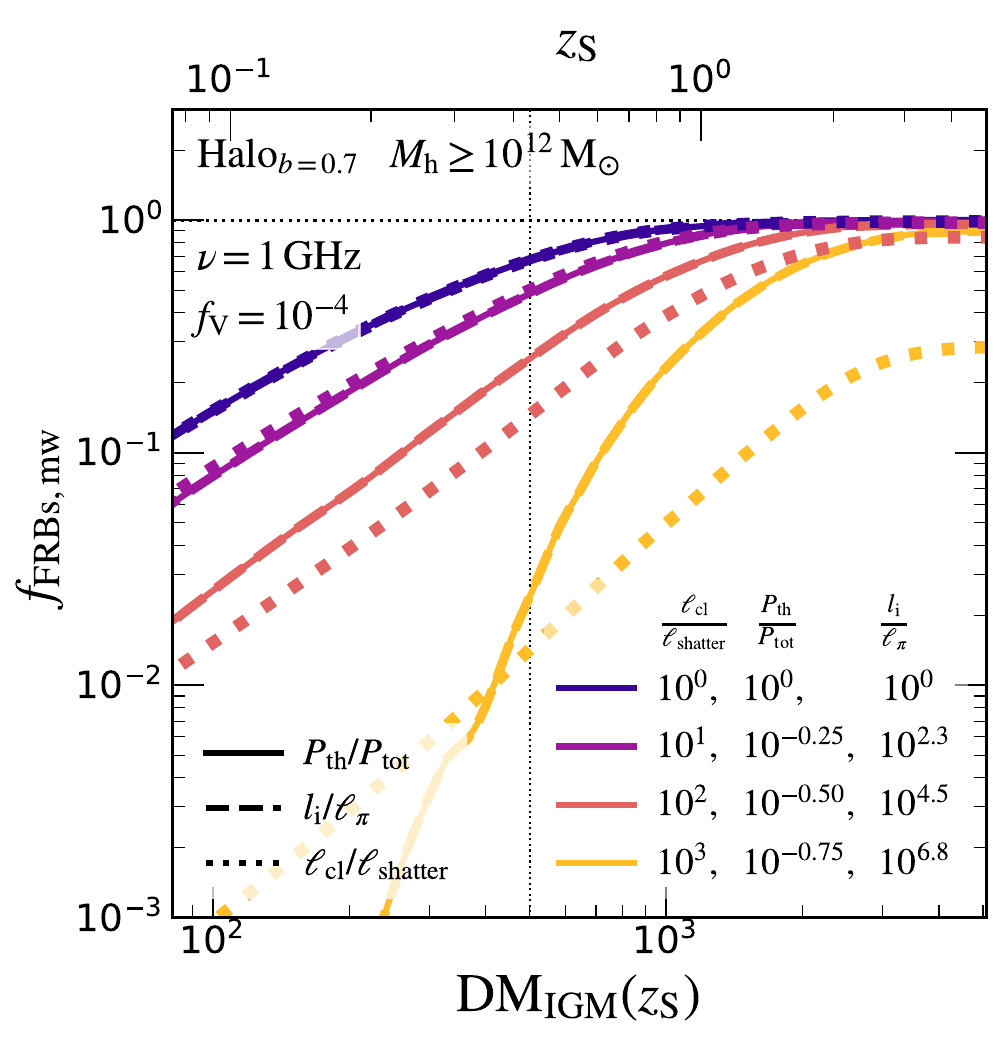}
    \caption{Similar to \fig{ffrbmw_h}, but exploring the impact of deviations from the fiducial model. We show the fractions of FRBs intercepting haloes of $\Mh\geq10^{12}\msun$ that are expected to suppress $\dffmw$ ($\ffrbmw$, y-axis), as a function of the average extragalactic DM, $\dmigm$ (bottom x-axis), and source redshift (top x-axis). 
    As before, we assume the los intercepts haloes at an impact parameter $b=0.7$, and a MW screen with $\dnusmwghz=1\mhz$ and $\dmw=1\kpc$.
    For clarity, we show here results for one observed frequency, $\nuo=1\ghz$, and one volume-filling fraction, $\fv=10^{-4}$.
    The different line styles and colours represent different deviations from the fiducial model, with the blue lines representing the fiducial model.
    For instance, for haloes with either $\fpth=10^{-0.25}$, $\fli=10^{2.3}$, or $\flc=10^1$ (purple solid, dashed, and dotted lines, respectively), we find that $\gsim 50\%$ of the FRBs with DM excess of $\gsim500\dmunits$ are expected to result in suppression at an observed frequency of $\nuo=1\ghz$, compared to $\gsim70\%$ in the fiducial model. 
    All the deviations from the model that we explore reduce the fraction of FRBs with suppressed $\dffmw$ for a given $\dmigm$ or $\zs$; however, most of the explored deviations are still expected to result in a substantial $\ffrbmw$ at $\dmigm\lsim10^3\dmunits$ ($\zs\lsim1$).
}
    \label{fig:ffrbmw_var_h}
\end{figure}

\smallskip
In \fig{ffrbmw_var_h}, we illustrate the effect of deviations from our fiducial model for haloes of $\geq10^{12}\msun$.\footnote{We explore the dominant masses affecting $\ffrbmw$ using two mass bins in \sea{mdff_mw_app}.}
As before, we assume an impact parameter $b=0.7$, and a los passing through a MW screen with $\dnusmwghz=1\mhz$ at a distance of $\dmw=1\kpc$ from the observer. For clarity, we show the results for one observed frequency, $\nuo=1\ghz$ (purple lines in \fig{ffrbmw_h}), and one $\fv$.
In \citeta{l26a}, our discussion in \S5.3 focused on high volume-filling fractions, as only $\fv\gsim10^{-3}$ were expected to cause detectable features in our fiducial scattering model. Here, the sensitivity of MW screens to intervening scattering screens allows us to probe lower $\fv$, so we show in this example $\fv=10^{-4}$ (dashed lines in \fig{ffrbmw_h}). The different line styles represent different types of deviations from our fiducial model. The solid lines represent the ratio of thermal pressure to total pressure in the CGM ($\fpth$), the dashed lines represent the ratio of the inner turbulence scale to the coherence scale ($\fli$), and the dotted lines represent the ratio of the cloudlet size to the shattering scale ($\flc$). The different colours indicate the strength of the deviation (for the values shown in the right legend, presented in a table). The blue lines indicate the fiducial model, where all ratios are one ($\fpth=\fli=\flc=1$), and the purple, orange, and yellow lines indicate increasingly stronger deviations.
Each deviation is explored separately, keeping all other parameters fixed at their fiducial value.

\smallskip
For a given colour, the different line styles (namely, the amplitudes of the individual deviations in the various model parameters) are chosen to give the same $\ffrbmw(\dmigm)$ when $\fa\geq1$. 
However, since $\nuoRsmw$ is only meaningful when $\fa\geq1$, $\nuoRsmw$ will be calculated at a fixed covering fraction when $\fa$ is smaller than one (by replacing $\fa$ with $\max(1,\fa)$ in \eqnp{nuoRsmw}) rather than at a fixed $\fv$. As a result, when a substantial contribution originates from haloes with $\fa<1$, the three lines may differ in their $\ffrbmw(\dmigm)$.
A deviation in $\fli$ does not affect $\lc$ and therefore will not reduce $\fa$. 
The overlap between the solid and dashed lines indicates that the haloes contributing to $\ffrbmw$ have $\fa>1$ for the different ratios of $\fpth$ explored here. On the other hand, for strong deviations in $\flc$, the impact on $\fa$ results in a substantial difference.
For example, the dotted orange line of $\flc=10^2$ is about a factor of two lower than the corresponding orange lines of $\fpth=10^{-0.5}$ (solid) and $\fli=10^{4.5}$ (dashed).
The dotted yellow line of $\flc=10^3$ shows an entirely different behaviour compared to the solid and dashed yellow lines.
In this case, at redshift ranges dominated by haloes with $\fa<1$, the exclusion of redshifts where $\nuo>\nuoRsmw$ (\eqsitinp{dnmw_dzdm}{ffrbmw}) is greatly diminished, leaving the main impact on $\ffrbmw$ to the capped covering fraction, $\min(1,\fa)$, in \eq{dnmw_dzdm}.
As shown, at low $\zs\lsim0.3$ the variation of $\flc=10^{3}$ can result in a higher $\ffrbmw$ than $\fpth=10^{-0.75}$. 
We note that the differences between $\flc$ and $\fpth$ (or $\fli$) are only important for low volume-filling fractions. For instance, these effects had no bearing on the results discussed in \citeta{l26a} \S5.3, where only high volume-filling fractions were discussed. For haloes with $\fv=10^{-3}$ (not shown here), the only meaningful difference between the results of deviations in different parameters is a factor of $\sim2-3$ difference between $\flc=10^{3}$ and $\fpth=10^{-0.75}$.

\smallskip
As shown in \fig{ffrbmw_var_h}, $\ffrbmw$ decreases as we increase the deviations from the fiducial model. 
For $\dmigm\gsim500\dmunits$, deviations of $\flc=10$, $\fpth=10^{-0.25}$, or $\fli=10^{2.3}$ still result in significant suppression of MW scintillation by haloes, with $\ffrbmw\gsim0.5$. For stronger deviations of $\fpth=10^{-0.5}$ or $\fli=10^{4.5}$, we have $\ffrbmw\gsim0.25$, with the corresponding $\flc=10^2$ lower by an additional factor of $\sim2$. However, for somewhat higher $\dmigm$, these still result in significant suppression fractions. The only exception is the extreme case of $\flc=10^3$ in the dotted yellow line, which reaches $\ffrbmw\sim0.3$ only at $\zs\gsim2-3$ or $\dmigm\gsim3000\dmunits$.

\smallskip
In conclusion, we find that deviations from the fiducial model can lower $\ffrbmw$, with strong deviations of $\lc$ from $\lcmin$ (assuming $\fv\lsim10^{-4}$) reducing the probability for MW suppression more efficiently than the other options considered here. However, barring the most extreme case explored here ($\flc=10^3$), the fraction of FRBs with suppressed $\dffmw$ is expected to be substantial if turbulent cloudlets exist in the CGM, with increasingly higher fractions for higher $\dmigm$ (or $\zs$).

\subsection{Can the FRB host suppress MW scintillation?}
\label{se:hg_mw}

\begin{figure}
    \centering
    \includegraphics[width=0.85\linewidth]{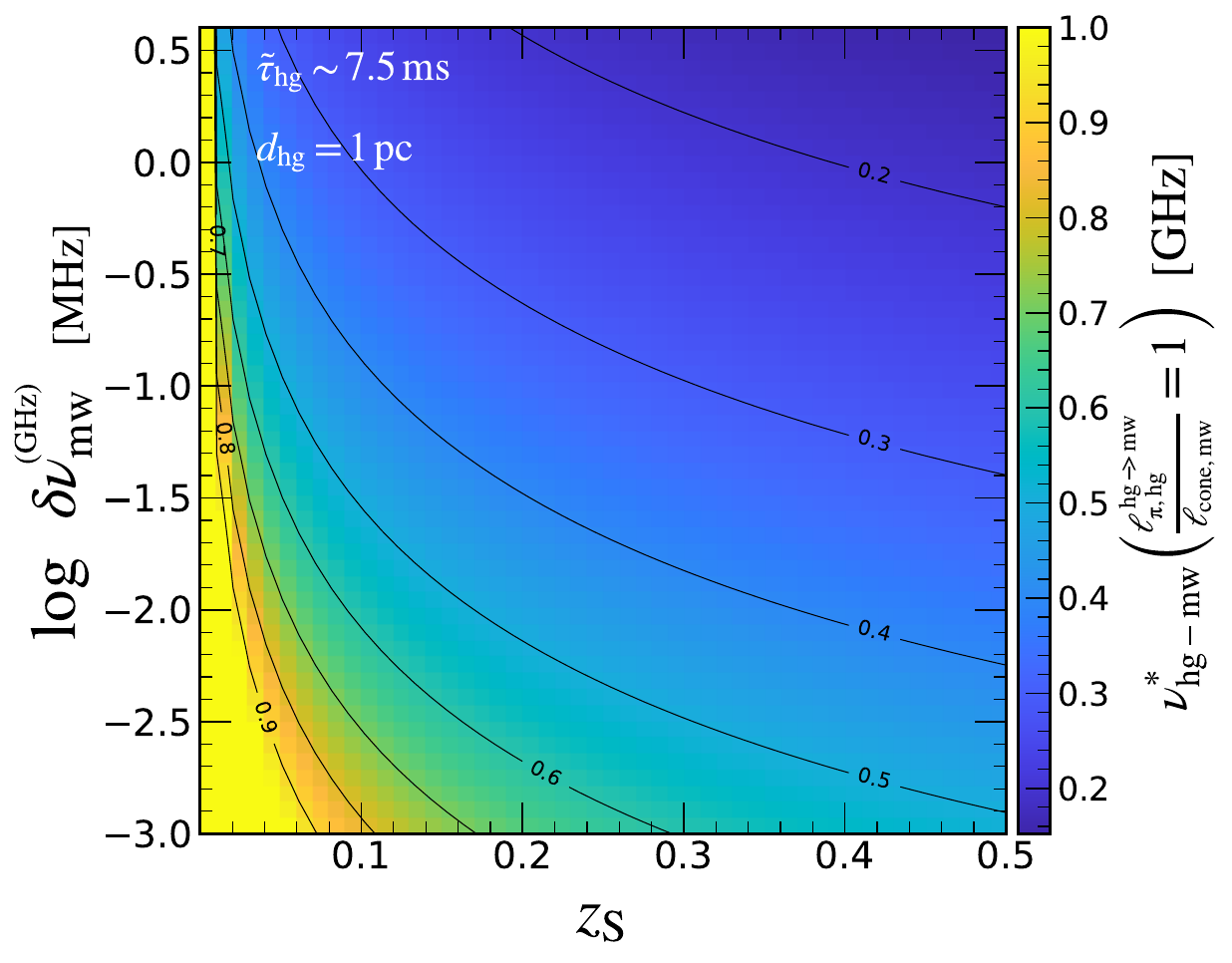}
    \caption{
    The maximal frequency (in colour) below which source size broadening by a strong scattering screen in the FRB host can suppress MW scintillation. The transition frequency $\nuoRs[\hg-\mmw]$ is shown as a function of the scintillation bandwidth of the MW screen at $1\ghz$ ($\dnusmwghz$, y-axis) and the redshift of the source ($\zs$, x-axis). The host screen is assumed to be at a distance of $\dhg=1\pc$ from the source, with an intrinsic scattering time of $\tausin[\hg]\sim7.5\ms$ at $\nuo=1\ghz$. 
    Only sources at very small $\zs$, or sightlines passing through a very small $\dnusmwghz$ (low latitudes) can raise $\nuoRs[\hg-\mmw]$ to sufficiently high values to suppress $\dffmw$ at typical observed frequencies.
    For a typical MW screen at high latitude, with a scintillation bandwidth of $\dnusmwghz\sim1\mhz$, only sources at very small $\zs$ are expected to suppress $\dffmw$.}
    \label{fig:dff_hgmw_2d}
\end{figure}

Given our results in \sesiti{dff_mw_sig}{model_var}, one may ask whether a screen in the FRB host can suppress scintillation from a screen in the MW, thus preventing us from attributing such suppression to a CW screen. 
The answer to this question illustrates one of the main advantages of using suppression of MW scintillation to detect CW screens, since it turns out that the host screen is usually incapable of suppressing MW scintillation (unless measured at low frequencies, below the typically observed range). 
For most FRBs, the projection of the diffractive scale onto the MW plane, $\lpisl[,\hg]$, is quite large due to the proximity of the host screen to the source. For sources at cosmological distances of $\dso\sim1\gpc$, and a host screen with $\dsl[,\hg]\sim(1-10^{3})\pc$, the ratio $\dso/\dsl[,\hg]\sim10^{6}-10^{9}$ results in a large $\lpisl[,\hg]\approx\lpi[,\hg]\dso/\dsl[,\hg]$, thus driving $\nuoRs[\hg-\mmw]$ to low values.
The transition frequency, $\nuoRs[\hg-\mmw]=\nuo(\lpisl[,\hg]/\Rmw)^{-5/17}$, above which $\dffmw$ is not suppressed by the FRB host screen is, 
\begin{equation}
    \nuoRs[\hg-\mmw]\!\sim 0.1\ghz\ \opzs[7]^{-\frac{6}{17}} \!\pfrac{\dsl[,\hg]}{\dso}^{\frac{5}{17}} \!\lt[ \frac{\tausinhgghz[,7.5]}{\dhg[,pc]} \frac{\dmw[,kpc]}{\dnusmwghz[,6]}\rt]^{\frac{5}{34}}.
    \label{eq:nuoRshgmw}
\end{equation}

\smallskip
In \fig{dff_hgmw_2d}, we show in colour the frequency below which the MW scintillation is suppressed, as a function of $\zs$ (x-axis), and of the MW screen scintillation bandwidth, $\dnusmw$ at $1\ghz$ (y-axis).
We assume substantial scattering in the host, with rest-frame scattering time of $\tausinhg(1\ghz)=7.5\ms$.\footnote{This corresponds, for example, to $\dhg=\dLhg=1\pc$ and $\dmsfahg\sim10^3\dmunits$. See the green line in \fig{Rcone}.}
For $\dnusmw(1\ghz)\sim \mhz$, one can consider almost every host as incapable of suppressing the MW screen scintillation below a reasonable detection frequency, where we crudely take the central frequency of the CHIME/FRB bandwidth, $600\mhz$, as our guideline for the minimum detection frequency.
The host screen can only suppress the MW scintillation under uncommon circumstances where $\zs$ is very small, or for sightlines with very low $\dnusmw(1\ghz)$ (typical for low Galactic latitudes). 
Meaning that, if the MW flux modulation is found to be suppressed below a certain frequency, $\nuoRsmw$, it may indicate an encounter with an intervening CW screen along the los.

\subsection[Ambiguity between screens]{Ambiguity between screens}
\label{se:screen_ambiguity}

As mentioned, the FRB host is expected to have a limited impact on suppression of scintillation by MW screens, eliminating the host-CW multi-screen ambiguity. Likewise, any ambiguity of MW-CW screen pairs can be alleviated by detailed modelling of MW screens from pulsar surveys \citep[e.g.][]{cordes_lazio02, yao17}. However, the CW-CW screen ambiguity remains an issue.

\smallskip
Regardless of the method used, since fragmentation into small-scale cloudlets in sheets and filaments is less likely at lower redshifts, a detection of scintillation or scattering for sources at $\zs\lsim1$, which are not attributed to the host, is less likely to be caused by these systems. 
At higher redshifts, considering two CWOs with turbulent cloudlets, CW screens could have a similar intrinsic scintillation or scattering (with different $\fv$), and although some clues can be drawn from the expected probability for encounter, in the absence of independent information \citep[e.g.][using the FLIMFLAM survey in the foreground of FRBs at low-$z$]{lee22}, it will be hard to know which class of CWO is responsible.

\subsection{Comparison to other works}
\label{se:other_works}

A small number of studies reported measurements of scintillation from localised FRBs attributed to the MW \citep{masui15, marcote20, ocker22-20190520, sammons23, nimmo25, scott25}. 
Most are at $\zs<0.3$, many are at $\zs\lsim0.1$. 
\citet{scott25} used a sample of $34$ FRBs, detected scintillation in $13$, and found significant evidence for two screens in $6$, where it is likely that one of the screens is in the MW. 
It is important to note that the non-scintillating FRBs are not necessarily due to suppression caused by a previous screen. Rather, a non-detection in the sample could be caused by a larger MW decorrelation bandwidth than that predicted using pulsar surveys \citep{sammons23}. 
When $\nuoRsmw$ is within the detector band, it may be possible to separate non-detections due to uncertainties in $\dnusmw$ from actual suppression. In such cases, one can observe a frequency-dependent scintillation quenching, where $\dffmw$ is expected to be of order unity at $\nuo\geq\nuoRsmw$, and increasingly suppressed below it.

\smallskip
The small number of FRBs in the sample greatly raises the uncertainty of any comparison, but the main limitation of the sample is the low $\zs$. 
Our model predicts that the cumulative number of $10^{12}\lsim\Mh<10^{13}\msun$ haloes along an average los is less than one at $\zs\lsim0.3$ (\fig{dff_mw_zS05}, bottom panel, dashed blue line), implying that some of the FRBs in the sample are not expected to pass through a massive halo, and the probability is considerably lower for those at $\zs\lsim0.1$. 
Therefore, even if cold cloudlets in haloes exist and are able to suppress MW scintillation, this is not expected to be common in the current low $\zs$ sample, where a substantial fraction of sightlines could reach us without intercepting a massive halo or CWO. 
Although we cannot meaningfully compare our results to current observations, forthcoming observations with scintillation measurements for sources at higher $\zs$ ($\zs>0.3$ for haloes, $\zs\gsim2$ for filaments, and $\zs>3$ for sheets) may help shed light on the effect of intervening CWOs with cold cloudlets.

\smallskip
The effect of MW scintillation suppression due to shattered CGM cloudlets was studied by \citet{jow24} and \citet{mas-ribas25} using refractive scattering. 
\citet{jow24} argued that if cloudlets exist in the CGM with $\fv=10^{-4}$, they are expected to cause tens of refractive images, which will suppress the MW scintillation. 
However, for an average los (i.e. with a large impact parameter with respect to the intervening halo), their fixed fiducial value of $\lcmin=0.1\pc$ is more adequate for higher redshifts (see \fig{clump_prop}), which are not well probed by current localised FRBs. 
Indeed, \citet{mas-ribas25} showed that once $\lcmin$ is taken to be density dependent \citep[as described in][]{mccourt18}, MW scintillation is not expected to be suppressed by refractive scattering off of shattered clouds at redshifts that can be probed by most localised FRBs, even for high $\fv=10^{-1}$ (assuming CGM densities derived in \citealt{werk14, prochaska17, zahedy19}).
However, at high-$z$ it may be interesting to leverage the different dependencies of diffractive and refractive scattering (for instance, the dependence on $\lc$, the fact that refractive scattering does not require turbulence in cloudlets), and use the two approaches as complementary probes to help partially break some of the degeneracies between the parameters involved.

\subsection{Comparison to other detection methods}
\label{se:comp_methods}

Overall, suppression of MW scintillation can be a far more useful tool for probing small-scale structures in CWOs than direct detection of either scintillation or scattering (discussed in \scattpaper):

\itemethos{Limited effect of the host screen:}
    It is often difficult to disentangle pairs of scattering screens, such as host-CW, CW-MW, or CW-CW. 
    However, an FRB host screen is not expected to impact MW scintillation in most cases (\fig{dff_hgmw_2d}); thus reducing one of the main origins of confusion.
    In contrast, if we compare with direct detection of scattering (i.e. $\taus$), a screen in the FRB host may overwhelm the scattering, making a distinction from a single CW screen challenging (\scattpaper).
\itemethos 
    {The effect of a single CW screen:} 
    Contrary to direct detection of scattering, even a single CW screen may suffice in some cases to suppress the MW scintillation and hint at the existence of a CW screen (\figsiti{dff_mw_zS05}{dff_mw_zS2}).
    Moreover, for haloes, this may be utilised with moderate source redshifts, $\zs\gsim0.3$ (\fig{dff_mw_zS05}). 
    
\itemethos{Low $\fv$:}
    Detection via MW scintillation may help constrain remarkably low values of $\fv$, especially at higher redshifts.
    For example, cloudlets in a $10^{12}\msun$ halo are expected to suppress $\dffmw$ if $\fv\gsim10^{-4}$ for a source at $\zs=0.5$; for a similar halo at $z\sim1.5$ and $\zs=2$, suppression is expected for $\fv\gsim10^{-5}$ (\fig{dff_mw_zS05}, \fig{dff_mw_zS2}). Such low values of $\fv$ are extremely difficult to detect via scattering (\scattpaper).
    
\itemethos{Suppression of $\dffmw$ vs. direct detection of $\dff[cw]$}:
    Direct detection of scintillation from a CW screen ($\dff[cw]$, see \se{mdff_cw} below) requires identifying the scintillation bandwidth scale, after a complex reduction of various noise sources. 
    However, the rich existing knowledge of MW screens, including the expected $\dnusmw$ (albeit with some uncertainties), makes the identification of MW scintillation much easier.

\section{Detection of scintillating CWOs}
\label{se:mdff_cw}

Under certain conditions, a CW screen can imprint its own pattern on the observed flux. We will show below that direct detection of a CW screen flux modulation is more challenging than the indirect effect on the MW scintillation. However, we wish to explore this possibility not only as another avenue for detection of a CW screen, but also as a complementary method to the one presented in \se{mdff_mw} that can help to increase the certainty of an encounter with a CW screen.

\smallskip
To assess the conditions for a strong modulation index due to an intervening CW screen, we assume the source is first broadened by a previous screen in the host galaxy ($\hg$), and find the transition frequency distinguishing between an extended and a point source, $\nuoRscw=(\lpiso/\Rhg)^{-5/17}$. 
The coherence scale of the CW screen projected back onto the plane of the host screen, $\lpiso$, can be found using results from \fig{clump_prop}, \eq{lpi}, and \eq{lpip}.\footnote{See also \fig{lpiso} in \sea{ExtPoint} for an example of $\lpiso$ of CWOs, assuming $\fa=1$.}
The scattering cone scale of the host, $\Rhg$, can be found using \eq{Rcone} (see also \fig{Rcone}). 
For a given $\dmsfahg$, and $\dhg\sim\dLhg$, we relate $\Rhg$ to the intrinsic scattering time of the host screen, $\tausinhg=\taushg\opzs^{17/5}$, where we use \eq{taus_dm} to find $\taushg$. We then place $\tausinhg$ in \eq{Rcone} to find $\Rhg\propto(\tausinhg\dhg)^{1/2}\opzs^{-6/5}$.

\smallskip
Using \eqsiti{Rcone}{nuoRs_base}, the minimal frequency for $\dffRs\sim1$ is $\nuoRscw\sim\nuo ({\lpiso}/{\Rhg})^{-{5}/{17}}$, giving
\be
    \label{eq:nuoRs}
    \nuoRscw
    &
    \!\!\sim\!
        1.07 \ghz 
        \frac{\fa^{\frac{3}{17}} \lc[,10]^{\frac{1}{17}} \nf^{\frac{6}{17}}}{\opzo^{\frac{6}{17}}  }
           \!\pfrac{\dlo\,\dhg[,pc]}{\dso\,\dLhg[,pc]}^{\!\!\frac{5}{17}}
        \!\!\frac{\dmsfalhg[,3]^{\frac{6}{17}}}{ \opzs[7]^{\frac{6}{17}}} 
    \\[3pt]
    &\!\!\sim \!
        1.07 \ghz
        \frac{\fa^{\frac{3}{17}} \lc[,10]^{\frac{1}{17}} \nf^{\frac{6}{17}}}{\opzo^{\frac{6}{17}}  }
        \pfrac{\dlo}{\dso}^{\frac{5}{17}}
        \frac{\Big[\dhg[,\pc]\tausinhgghz[,7.5]\Big]^{\frac{5}{34}}}{\opzs[7]^{\frac{6}{17}}}.
\ee
In the last line, we express $\nuoRscw$ as a function of the host screen intrinsic scattering time scaled to a fixed $\nuo=1\ghz$, $\tausinhgghz=\tausinhg(1\ghz)$, using \eq{taus_dm}, with $\tausinhgghz[,7.5]\sim\tausinhgghz/(7.5\ms)$.

\smallskip
While we provide here one illustrative example, our model can be easily adjusted for different host-screen properties.
For instance, our fiducial choice of $\dhg=1\pc$ is motivated by recent studies \citep[e.g.][]{glowacki25, acharya25}, which did not find a significant correlation between the observed scattering time and the host galaxy properties, such as star formation rate or inclination, suggesting that the typical scattering measured in localised FRBs may be dominated by the circumburst region rather than the ISM of the host (though this is still uncertain, as the sample sizes are small). The fiducial value of $\tausinhgghz\sim7.5\ms$ is within the wide range derived from current localised FRBs with measurements of $\tausinhgghz\sim10^{\pm2}\ms$ at $\zs<1$ \citep{glowacki25, acharya25}.

\begin{figure}
    \centering
    \includegraphics[width=1\linewidth]{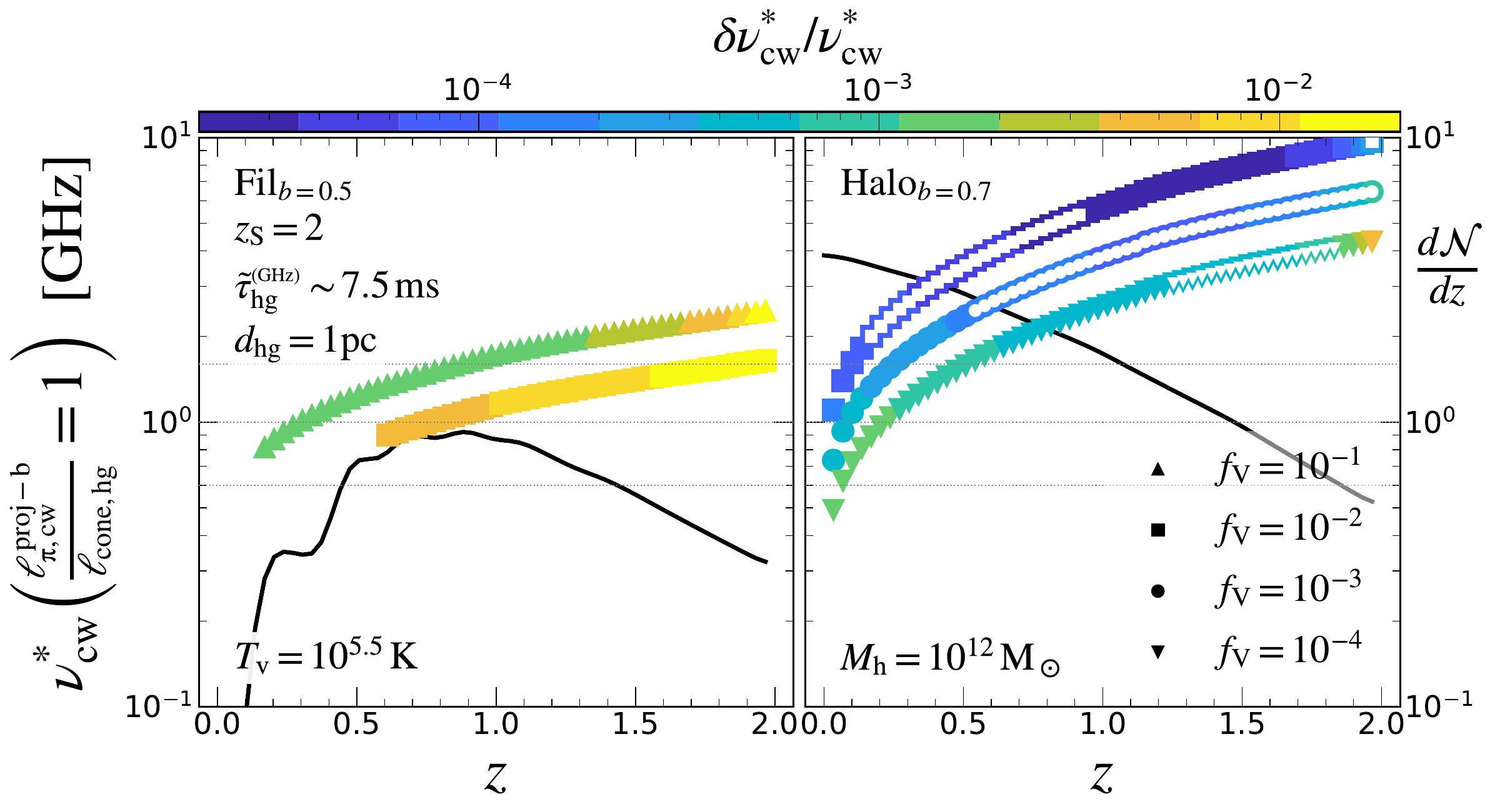}
    \caption{Detectability of flux modulation and scintillation from CWOs. Coloured symbols show, as a function of redshift, the minimal frequency above which $\dff$ of these CWOs is expected to be of order unity, $\nuoRscw$ (\eqnp{nuoRs}, left y-axis). 
    The symbols are coloured by the ratio $\dnusRscw/\nuoRscw$, where $\dnusRscw=\dnus(\nuoRs)$ is the corresponding bandwidth. We assume a source at $\zs=2$ and a host screen $\dhg=1\pc$ from the source with an intrinsic scattering time of $\tausinhg\sim7.5\ms$ at $\nuo=1\ghz$. 
    We show for example $\Tv=10^{5.5}\Kel$ filaments (left) and $\Mh=10^{12}\msun$ haloes (right).
    Triangles, squares, circles, and reverse triangles represent CWOs with $\log\fv=-1, -2, -3, -4$, respectively. The black lines show the number of these CWOs per unit redshift interval, $\dNdz$ (right y-axis), where we include only CWOs expected to form cold clouds, and in the temperature and mass ranges of $10^{5.5-6}\Kel$ and $10^{12-13}\msun$, respectively. For clarity, we exclude CWOs with $\fa<1$. 
    Open symbols indicate $\dnusRscw$ within an order of magnitude of the assumed scintillation bandwidth of the host screen.}
    \label{fig:dff_cw}
\end{figure}

\smallskip
In \fig{dff_cw}, we show using coloured symbols the minimal frequency required to achieve $\dffRs\sim1$, $\nuoRscw$ (left y-axis), as a function of redshift for CW filaments (left) and haloes (right). Different symbols (triangles, squares, circles, and reverse triangles) represent different values of $\fv$, as indicated in the legend. 
For clarity, we exclude CWOs with $\fa<1$.
We assume the line-of-sight has an impact parameter of $b=0.7, 0.5$ for haloes and filaments, respectively.\footnote{See \sea{vir} for a description of the path length through a CWO as a function of impact parameter.}
The symbol colour indicates the ratio $\dnusRscw/\nuoRscw$, where $\dnusRscw=\dnus(\nuoRscw)$ is the scintillation bandwidth at this frequency (\eqnp{dnus}). This quantity will be important for estimating the flux modulation due to detector resolution, as described below. 
The FRB source is assumed to be at $\zs=2$, with a screen at a distance of $\dhg=1\pc$ from the source, a comparable width of $\dLhg=1\pc$, and an intrinsic scattering time of $\tausinhgghz\sim7.5\ms$ at $\nuo=1\ghz$. This corresponds to $\dms\fa^{1/3}\sim 10^3\pc\cmc$, namely to the green line in \fig{Rcone}. 
Open symbols mark CW screens with $\dnusRscw$ within an order of magnitude of the scintillation bandwidth of the host screen assumed here, namely within the range $\dnus[,\hg]\tm10^{\pm1}$ (where $\dnus[,\hg]\sim2.4\khz$ at $1\ghz$). As can be seen in \fig{dff_cw}, for sufficiently high $\tausinhg$ as shown here, the scintillation bandwidth of the host is generally small compared to $\dnusRscw$ in the frequency range of $10^{\pm0.2}\ghz$, and we can normally avoid confusion in such cases.
The black line in each panel shows the number of CWOs per redshift interval along a los, $\dNdz$ (right y-axis). Similar to the solid lines of $\dndz$ in the bottom row of \fig{dff_mw_zS2}, $\dndz$ is integrated here over the range of $10^{5.5-6}\Kel$ for the filaments shown here, and over $10^{12-13}\msun$ for the haloes. As in \fig{dff_mw_zS2}, we exclude CWOs that do not meet the conditions of \eq{frag_cond}.

\smallskip
Following the derivation by \citeta{kumar24}, we showed in \se{dff} that $\dff\sim1$ can be achieved when the observed frequency is in the range $\nuoRs\leq\nuo<\nuou$, and $\dnuo\lsim\dnus(\nuo)$. 
In our interpretation of \fig{dff_cw}, we take into account the above restrictions and translate them to practical capabilities of current or near-future detectors. For instance, the spectral resolution of current detectors differs substantially; however, the ratio of the resolution with respect to the central frequency of the detector bandwidth is typically $\dnuo/\nuo[0]\sim\few\tm10^{-4}$. We therefore use a rough threshold of $\dnus/\nuo\gsim10^{-4}$ to assess whether $\dnuo\lsim\dnus$, thereby avoiding the need to assume a specific detector.
As for the upper limit, although formally the upper-limit frequency that yields $\dff\sim1$ is $\nuou$,\footnote{As discussed in \sea{mdff_cw_app} (see \fig{nuou} for a general CWO with a general $\fa$, and the bottom row of \fig{dnus_u} for $\fa=1$), $\nuou$ is significantly larger than $\nuoRs$ for all CWOs, and therefore, the upper limit of $\nuou$ is usually much less restrictive.}
high-frequency detectors are not expected to detect many new FRBs due to their relatively narrow field of view.
We therefore focus our attention on CW screens within about $\sim(0.2-0.3)\dex$ around $\nuoRs\sim1\ghz$.
The possibility of confusion with MW or host screen raises additional restrictions. Confusion with the MW screen can be partially managed, given our knowledge of MW screens mapped by pulsar surveys.
Similarly, the host scintillation bandwidth should also be considered. In this case, the observed scattering time of the FRB may help place upper limits on the scattering time of the host and therefore on its scintillation bandwidth.\footnote{As discussed in \scattpaper (\S5), if the cumulative contribution of massive haloes to the observed scattering time ($\taut$) exceeds that of a typical host screen, their abundance along an average los at $z\lsim1$ would result in a correlation between $\taut$ and $\zs$. So far, observations do not indicate such a correlation. We therefore assume that $\taushg$ is typically the main contributor to $\taut$.}

\smallskip
For the example shown in \fig{dff_cw}, we can identify a few CWOs that may yield $\dff\sim1$. 
For instance, a $10^{12}\msun$ halo with $\fv=10^{-4}$ at $z\sim0.2-0.5$ has $\nuoRscw\sim(0.6-1.6)\ghz$ and $\dnusRscw/\nuoRscw\sim\few\tm10^{-4}-10^{-3}$, yielding $\dnus\sim1\mhz$. 
Haloes with $\fv=10^{-3}$ and $10^{5.5}\Kel$ filaments with $\fv=10^{-1}$ (which is roughly equal to the maximal possible volume-filling fraction for such filaments; see \se{frag_cond}) are in the same frequency range at $z\lsim0.3$ and $z\lsim0.7$, respectively.
These values may allow detection; however, care should be taken to avoid confusion with the MW screen. For example, if the MW scintillation bandwidth along the los is known to be $\dnusmw(1\ghz)\sim1\mhz$, a halo with $\fv=10^{-4}$ at $z\sim0.2$ or a filament with $\fv=10^{-1}$ at $z\sim0.3$ can be confused with the MW.

\smallskip
We note that if more than one CW screen is encountered along the los, $\dff$ of the second one is likely to be suppressed by the first due to the typical $\Deff/\dL$ ratio. Namely, contrary to the case where the suppression is caused by the host screen (where $\dhg\sim\dLhg$), for typical $\Deff\sim\gpc$ distances of CW screens and $\dL\sim10^{1-3}\kpc$, their ratio can have a substantial impact and increase $\nuoRs(\lpib[,cw2]/\lcone[,cw1]=1)$ to very high values. Therefore, the CW screen that is likely to imprint its pattern on the flux is the first\footnote{The \emph{first} screen does not refer to the halo of the FRB host galaxy or to the filament or sheet feeding it, but rather the first intervening CW screen between the source and the observer.}
one following the host screen.

\smallskip
We conclude that while direct detection of the imprinted CW screen flux modulation is possible in some cases, this method is challenging in many respects and, at best, provides a limited range of possibilities for a given set of conditions in the host and in the CW screen.
However, despite the challenges, it could be valuable in specific cases, especially when combined with the method presented in \se{mdff_mw}. 
Consider, for example, an FRB source at $\zs\sim1$, first encountering a host screen with $\tausinhgghz(\dhg=1\pc)\sim1\ms$, followed by a $10^{12}\msun$ halo with $\fv\sim10^{-3}$ at $z\sim0.3$, and finally a MW screen with $\dnusmwghz\sim1\mhz$ at $\dmw=1\kpc$.\footnote{These conditions can be estimated for a given CW screen and redshift using \fig{dff_cw}, rescaling to $\zs=1$ and $\tausinhgghz=1\ms$ according to \eq{nuoRs}, and taking into account the different $\dlo/\dso$ using \fig{dso/dsl}. For the example discussed here, the difference in $\dlo/\dso$ is very small and can be ignored for a rough estimate.}
Both $\nuoRscw$ and $\nuoRsmw$ are expected to be $\sim1.5\ghz$. In this case, the expected scintillation bandwidths of the host, MW, and CW screen are well separated, $\sim10^{1}, 10^{2.5}, 10^{3.8}\khz$, respectively. Namely, this CW screen is expected to imprint its pattern on the observed flux at $\sim1.5\ghz$ and suppress $\dffmw$ below this frequency.

\section{Conclusions}
\label{se:dc_Sc}

In this work, we explored the emergence of multiphase gas within the shock-heated medium of massive cosmic web sheets and filaments, as well as in the circumgalactic medium (CGM) of massive dark matter haloes, which are collectively referred to as cosmic web objects (CWOs). 
We focus on the formation of cold, dense cloudlets, and evaluate how these small-scale ionized overdensities could influence the propagation of radio waves emitted from fast radio bursts (FRBs).
Specifically, we explore their effects on the scintillation pattern imprinted on the FRB flux and whether these can constrain the properties of the small-scale structure in the multiphase gas of CWOs.

%%%
\itemconc{Fiducial model for small-scale cold structure in CWOs (\se{shatter}):}

\itembul
    To relate the hot environment of massive CWOs to the properties of small-scale cold clouds, we utilise the `shattering' model \citep{mccourt18} as a fiducial reference. 
    We find the cloudlets' size, $\lc$, for a given pressure, metallicity, and redshift, by identifying the minimum cooling length $\lc=\lcmin$ (\eqnp{lcool}, \fig{lcool}), the associated temperature, $\Tmin$, and the electron density of the clouds, $\n$ (\fig{clump_prop}, see also fig.~2 in \scattpaper).
    We stress that our work is not meant to assess the validity of the shattering model itself, but rather to use it as a physically motivated starting point. This allows us to first make predictions within the fiducial model framework and then address the effects of deviations from its assumptions.

%%%
\itemconc{Scintillation -- general case (\se{dff}):}

\itembul
    When radio waves pass through an inhomogeneous, turbulent plasma screen, they undergo a phase shift and are deflected. This results in both scattering (temporal broadening, $\taus$) and scintillation. The relevant length scale of perturbations in the plasma screen for both processes is the \emph{coherence scale} (or \emph{diffractive scale}) length ($\lpi$, \eqnp{lpi}). 
    For FRBs, \emph{Scintillation} manifests as flux modulations with frequency ($\dff$), 
    which occurs when multipath propagation of rays results in an interference pattern in the FRB spectrum, with enhanced regions having a typical spectral scale of the \emph{scintillation bandwidth} ($\dnus$, \eqnp{dnus}, \fig{dnus_u}). 
    We assume throughout that clouds in CW screens are turbulent with a Kolmogorov power spectrum, and an outer scale of turbulence $\lo=\lc$.

\itembul
    In our companion paper (\scattpaper), we employed a population-level analysis to constrain scattering from $z\lsim1$ haloes, and showed that probing a single CWO via scattering is expected to be challenging, especially if $\n$ ($\lc$) is relatively low (large). 
    However, the relation $\dnus\sim(2\pi\taus)^{-1}$ implies that when $\taus$ is too small for detection, scintillation may be a better-suited approach.
    
\itembul
    A significant flux modulation\footnote{In diffractive scintillation, the flux modulation of a screen is saturated at $\dff=1$ \citepa{narayan92}.} ($\dff\sim1$) depends on three components \citepa{kumar24}: the scintillation regime (strong or weak) of the screen ($\dffu$), the impact of an extended source size ($\dffRs$), and the effect of the detector spectral resolution ($\dffob$).

\itembul
    $\dffu$: We find the transition frequency between the weak and strong regimes, $\nuou$, for different CWOs (\eqsnp{nuou,nuou_dm}). This is defined such that $\dffu\sim1$ if the observed central frequency is $\nuo<\nuou$ (\eqnp{dffu}).

\itembul
    $\dffob$: To achieve $\dffob\sim1$ (\eqnp{dffob_1}), the detector spectral resolution, $\dnuo$, should be comparable to or smaller than $\dnus$ (\eqsnp{dnus,dnus_dm}) of the screen we wish to detect.
    Under certain conditions, this restriction allows some freedom (\se{dffob}).

\itembul
    $\dffRs$: 
    A compact source broadened by a scattering screen can smear the interference pattern of a subsequent screen and reduce the observed flux modulation, $\dffRs$ (\eqnp{dffRs}). 
    We define the transition frequency, $\nuoRs$, above which the broadened source size is unresolved by the following screen.

\itembul
    $\dff$: The total flux modulation is given by $\dff=\dffu\,\dffRs\,\dffob$ (\eqnp{dff}). If $\dnuo\lsim\dnus$, then $\dff\sim1$ if the central observed frequency obeys $\nuoRs<\nuo<\nuou$ (\eqsnp{nuou,nuou_dm,nuoRs_base}). 

%%%
\itemconc{Counting CWOs and cloudlets (\se{cosmo}):}

\itembul
    We use results from our two companion papers to account for the probability of encountering a CWO, and the number of cloudlets intercepted within it.
    In \dmpaper, we estimate the probability of a los intercepting a CWO (see also fig.~4 in \scattpaper).
    In \scattpaper, we constrain the number of cloudlets within a CWO and identify CWOs that are more likely to form cold cloudlets (see \eqnp{frag_cond} and Figs.~5,~6 in \scattpaper).

%%%
\itemconc{Suppression of MW scintillation by CW screens (\se{mdff_mw}):}

\itembul
    We find that one of the most promising methods for detection of multiphase CWOs is to use the effect of a CW screen on scintillation from a plasma screen in the MW-ISM, following the considerations in \se{dff}. 
    We identify the transition frequency, $\nuoRsmw$ (\eqnp{nuoRsmw}), below which the MW flux modulation, $\dffmw$, is suppressed. In other words, the absence of expected MW scintillation below $\nuoRsmw$ hints at the existence of a CW screen along the FRB los.

\itembul
    For instance, for a los from a source at $\zs=0.5$ passing through a halo at $z\sim0.3$, and then through a MW screen with a known $\dnusmw(\nuo=1\ghz)\sim1\mhz$, $\dffmw$ is expected to be suppressed below $\nuo\lsim\nuoRsmw\sim0.9, 1.3$ and $2\ghz$ for a $10^{12}\msun$ halo with $\fv\sim10^{-4},10^{-3}$, and $10^{-2}$, respectively (\fig{dff_mw_zS05}).
    Passing a $10^{13}\msun$ halo at these redshifts with $\fv=10^{-5}, 10^{-4}$, and $10^{-3}$,\footnote{Our estimates for the maximal volume-filling fraction of cloudlets in $10^{13}\msun$ haloes is $\fvmax<10^{-2}$ for the entire redshift range explored here (see \se{frag_cond}, and \scattpaper \S4.4 for further details).}
    the constraints are even stronger, with suppression below $\nuo\lsim\nuoRsmw\sim 1.3, 2$, and $3$, respectively.

\itembul
    The expected high rates of FRBs may improve the chances of detecting rare systems at high-$z$. 
    \citet{beniamini21} estimated conservatively an FRB rate of $\dotNum(\zs\geq6,\nuf)\sim10^{4}\yr^{-1}\sky$ with a fluence threshold of $e_{\nuo}\geq1\jy\ms$.
    For instance, the sky covering fraction of $10^{6}\Kel$ sheets at $z\sim3.5\pm0.5$ is $\sim10^{-2}$. Assuming $\nuo\sim1\ghz$ (\fig{dff_cw}), $\sim 100$ FRBs at $\zs\gsim6$ are expected to pass through such systems per year over the entire sky.
    For $\zs=6$ we find that a $10^{5.5}\Kel$ filament or a $10^{6}\Kel$ sheet at $z\sim4$, is expected to suppress $\dffmw$ below $\nuo\sim0.9, 1.4, 2\ghz$ for $\log\fv=-4, -3, -2$, respectively (\fig{dff_mw_zS6}).
    Similar values are expected for a source at $\zs=2$ and a $10^{6}\Kel$ filament at $z\sim1.5$.

\itembul
    The current FRB sample with scintillation measurements is too small for direct comparison, with most FRBs at $\zs<0.3$. 
    To provide testable predictions to compare with forthcoming observations, we estimate the fraction of FRBs ($\ffrbmw$) that are expected to have $\dffmw$ suppressed by CWOs, as a function of the average extragalactic DM, $\dmigm(\zs)$ (\eqnp{dmigm}). 
    Our fiducial model suggests that most FRBs with $\dmigm\gsim500\dmunits$ are expected to have $\dffmw$ suppressed below $\nuo\sim1\ghz$ if turbulent shattered cloudlets exist in haloes of mass $\geq10^{12}\msun$ with $\fv\gsim10^{-4}$ (\fig{ffrbmw_h}).

\itembul
    After exploring the results within the framework of our fiducial shattering model, we use it as a baseline and study the impact of different deviations from it, separately altering one parameter at a time to examine the effect of: 
    (i) the assumption that the cloudlet sizes are of the order of the shattering scale, where $\flc>1$;
    (ii) the assumption that the non-thermal pressure in CWOs is negligible, using the ratio between the thermal pressure and the total pressure, $\fpth$;
    (iii) the assumption that there are no damping mechanisms that are unaccounted for in our estimates for the inner scale of turbulence, which may halt the cascade and result in an inner-scale of turbulence larger than the coherence scale, using the ratio of $\fli$.

\itembul 
    We find that for cloudlets in the CGM, small or moderate deviations from the fiducial model still show a significant $\ffrbmw$ for FRBs with $\dmigm\gsim500\dmunits$, even with volume-filling fraction as low as $\fv\gsim10^{-4}$ (\fig{ffrbmw_h}). For a significantly larger $\lc$, with $\flc\gsim10^{2}$, the effect of the deviation can reduce the probability of suppressing $\dffmw$ more efficiently than the corresponding $\fpth$ and $\fli$ due to its strong impact on the covering fraction.

\itembul
    Several advantages highlight this method as a particularly useful approach to explore CW screens (\se{comp_methods}). 
    For instance, in scattering measurements, the FRB host galaxy can be a major source of ambiguity. One crucial advantage of utilising the suppression of $\dffmw$ as an indicator is the fact that it is usually unaffected by the FRB host galaxy screen (\fig{dff_hgmw_2d}).

%%%
\itemconc{Flux modulation of CW screen (\se{mdff_cw}):}

\itembul
    Combining the general theory in \se{dff} with our results from \se{cosmo}, we investigate the possibility of direct detection of $\dff[cw]$ due to small-scale density fluctuations in CWOs. We assume the source size is first broadened by a scattering screen in the FRB host, and estimate the expected transition frequency ($\nuoRs$, \eqnp{nuoRs}).
    We find that while direct detection of the scintillation pattern of a CW screen is possible for some parameter sets (\fig{dff_cw}), the challenges remain substantial. However, in some cases, this method may be valuable as a complementary approach to detection via suppression of MW scintillation (\se{mdff_mw}).

\section*{Acknowledgements}
  We thank Yakov Faerman, Wenbin Lu, Matt McQuinn, Stella Ocker, Xavier Prochaska, and Nadav Shoval for very helpful discussions. 
  SL and NM acknowledge support from BSF grant 2022281 and NSF-BSF grant 2022736. 
  PB's work was funded by a grant (no. 2024788) from the United States-Israel Binational Science Foundation (BSF), Jerusalem, Israel, by a grant (no. 1649/23) from the Israel Science Foundation and by a NASA grant (80NSSC24K0770). 
  SPO acknowledges NSF grant AST240752 for support.

\section*{Data Availability}
The data produced in this study will be available on
reasonable request to the corresponding author.

\bibliographystyle{mnras}
\bibliography{bib.bib}

@ARTICLE{acharya25,
       author = {{Acharya}, Sandeep Kumar and {Beniamini}, Paz},
        title = "{Utilizing localized fast radio bursts to constrain their progenitors and the expansion history of the Universe}",
      journal = {\jcap},
         year = 2025,
        month = oct,
       volume = {2025},
       number = {10},
          eid = {073},
        pages = {073},
          doi = {10.1088/1475-7516/2025/10/073},
archivePrefix = {arXiv},
       eprint = {2503.08441},
 primaryClass = {astro-ph.CO},
       adsurl = {https://ui.adsabs.harvard.edu/abs/2025JCAP...10..073A}
}

@ARTICLE{Armstrong95,
       author = {{Armstrong}, J.~W. and {Rickett}, B.~J. and {Spangler}, S.~R.},
        title = "{Electron Density Power Spectrum in the Local Interstellar Medium}",
      journal = {\apj},
         year = 1995,
        month = apr,
       volume = {443},
        pages = {209},
          doi = {10.1086/175515},
       adsurl = {https://ui.adsabs.harvard.edu/abs/1995ApJ...443..209A}
}

@ARTICLE{aung24,
       author = {{Aung}, Han and {Mandelker}, Nir and {Dekel}, Avishai and {Nagai}, Daisuke and {Semenov}, Vadim and {van den Bosch}, Frank C.},
        title = "{Entrainment of hot gas into cold streams: the origin of excessive star formation rates at cosmic noon}",
      journal = {\mnras},
         year = 2024,
        month = aug,
       volume = {532},
       number = {3},
        pages = {2965-2987},
          doi = {10.1093/mnras/stae1673},
archivePrefix = {arXiv},
       eprint = {2403.00912},
 primaryClass = {astro-ph.GA},
       adsurl = {https://ui.adsabs.harvard.edu/abs/2024MNRAS.532.2965A}
}

@ARTICLE{bbks1986,
       author = {{Bardeen}, J.~M. and {Bond}, J.~R. and {Kaiser}, N. and {Szalay}, A.~S.},
        title = "{The Statistics of Peaks of Gaussian Random Fields}",
      journal = {\apj},
         year = 1986,
        month = may,
       volume = {304},
        pages = {15},
          doi = {10.1086/164143},
       adsurl = {https://ui.adsabs.harvard.edu/abs/1986ApJ...304...15B}
}

@ARTICLE{beniamini20,
       author = {{Beniamini}, Paz and {Kumar}, Pawan},
        title = "{What does FRB light-curve variability tell us about the emission mechanism?}",
      journal = {\mnras},
         year = 2020,
        month = oct,
       volume = {498},
       number = {1},
        pages = {651-664},
          doi = {10.1093/mnras/staa2489},
archivePrefix = {arXiv},
       eprint = {2007.07265},
 primaryClass = {astro-ph.HE},
       adsurl = {https://ui.adsabs.harvard.edu/abs/2020MNRAS.498..651B}
}

@ARTICLE{beniamini21,
       author = {{Beniamini}, Paz and {Kumar}, Pawan and {Ma}, Xiangcheng and {Quataert}, Eliot},
        title = "{Exploring the epoch of hydrogen reionization using FRBs}",
      journal = {\mnras},
         year = 2021,
        month = apr,
       volume = {502},
       number = {4},
        pages = {5134-5146},
          doi = {10.1093/mnras/stab309},
archivePrefix = {arXiv},
       eprint = {2011.11643},
 primaryClass = {astro-ph.CO},
       adsurl = {https://ui.adsabs.harvard.edu/abs/2021MNRAS.502.5134B}
}

@ARTICLE{beniamini22,
       author = {{Beniamini}, Paz and {Kumar}, Pawan and {Narayan}, Ramesh},
        title = "{Faraday depolarization and induced circular polarization by multipath propagation with application to FRBs}",
      journal = {\mnras},
         year = 2022,
        month = mar,
       volume = {510},
       number = {3},
        pages = {4654-4668},
          doi = {10.1093/mnras/stab3730},
archivePrefix = {arXiv},
       eprint = {2110.00028},
 primaryClass = {astro-ph.HE},
       adsurl = {https://ui.adsabs.harvard.edu/abs/2022MNRAS.510.4654B}
}

@ARTICLE{beniamini25,
       author = {{Beniamini}, Paz and {Kumar}, Pawan},
        title = "{The Role of Magnetic and Rotation Axis Alignment in Driving Fast Radio Burst Phenomenology}",
      journal = {\apj},
         year = 2025,
        month = mar,
       volume = {982},
       number = {1},
          eid = {45},
        pages = {45},
          doi = {10.3847/1538-4357/adb8e6},
archivePrefix = {arXiv},
       eprint = {2410.19043},
 primaryClass = {astro-ph.HE},
       adsurl = {https://ui.adsabs.harvard.edu/abs/2025ApJ...982...45B}
}

@ARTICLE{Bergeron86,
       author = {{Bergeron}, J.},
        title = "{The MG II absorption system in the QSO PKS 2128-12 : a galaxy disc/halo with a radius of 65 kpc.}",
      journal = {\aap},
         year = 1986,
        month = jan,
       volume = {155},
        pages = {L8-L11},
       adsurl = {https://ui.adsabs.harvard.edu/abs/1986A&A...155L...8B}
}

@ARTICLE{Birnboim.Dekel.03,
       author = {{Birnboim}, Yuval and {Dekel}, Avishai},
        title = "{Virial shocks in galactic haloes?}",
      journal = {\mnras},
         year = 2003,
        month = oct,
       volume = {345},
       number = {1},
        pages = {349-364},
          doi = {10.1046/j.1365-8711.2003.06955.x},
archivePrefix = {arXiv},
       eprint = {astro-ph/0302161},
 primaryClass = {astro-ph},
       adsurl = {https://ui.adsabs.harvard.edu/abs/2003MNRAS.345..349B}
}

@ARTICLE{Bochenek20,
       author = {{Bochenek}, C.~D. and {Ravi}, V. and {Belov}, K.~V. and {Hallinan}, G. and {Kocz}, J. and {Kulkarni}, S.~R. and {McKenna}, D.~L.},
        title = "{A fast radio burst associated with a Galactic magnetar}",
      journal = {\nat},
         year = 2020,
        month = nov,
       volume = {587},
       number = {7832},
        pages = {59-62},
          doi = {10.1038/s41586-020-2872-x},
archivePrefix = {arXiv},
       eprint = {2005.10828},
 primaryClass = {astro-ph.HE},
       adsurl = {https://ui.adsabs.harvard.edu/abs/2020Natur.587...59B}
}

@ARTICLE{Bond96,
       author = {{Bond}, J. Richard and {Kofman}, Lev and {Pogosyan}, Dmitry},
        title = "{How filaments of galaxies are woven into the cosmic web}",
      journal = {\nat},
         year = 1996,
        month = apr,
       volume = {380},
       number = {6575},
        pages = {603-606},
          doi = {10.1038/380603a0},
archivePrefix = {arXiv},
       eprint = {astro-ph/9512141},
 primaryClass = {astro-ph},
       adsurl = {https://ui.adsabs.harvard.edu/abs/1996Natur.380..603B}
}

@ARTICLE{Burkert_Lin00,
       author = {{Burkert}, A. and {Lin}, D.~N.~C.},
        title = "{Thermal Instability and the Formation of Clumpy Gas Clouds}",
      journal = {\apj},
         year = 2000,
        month = jul,
       volume = {537},
       number = {1},
        pages = {270-282},
          doi = {10.1086/308989},
archivePrefix = {arXiv},
       eprint = {astro-ph/0002106},
 primaryClass = {astro-ph},
       adsurl = {https://ui.adsabs.harvard.edu/abs/2000ApJ...537..270B}
}

@ARTICLE{chime_cat2,
       author = {{Chime/Frb Collaboration} and {Abbott}, Thomas and {Andersen}, Bridget C. and {Andrew}, Shion and {Bandura}, Kevin and {Bhardwaj}, Mohit and {Bhusare}, Yash and {Brar}, Charanjot and {Cassanelli}, Tomas and {Chatterjee}, Shami and {Cliche}, Jean-Francois and {Cook}, Amanda M. and {Curtin}, Alice and {Dobbs}, Matt and {Dong}, Fengqiu Adam and {Eadie}, Gwendolyn and {Eftekhari}, Tarraneh and {Fonseca}, Emmanuel and {Gaensler}, B.~M. and {Good}, Deborah and {Halpern}, Mark and {Hessels}, Jason W.~T. and {Ibik}, Adaeze and {Jain}, Naman and {Joseph}, Ronniy C. and {Kader}, Zarif and {Kaspi}, Victoria M. and {Khan}, Afrokk and {Kharel}, Bikash and {Kumar}, Ajay and {Landecker}, T.~L. and {Lang}, Dustin and {Lanman}, Adam E. and {L'Argent}, Magnus and {Lazda}, Mattias and {Leung}, Calvin and {Li}, Dong Zi and {Lintott}, Chris J. and {Main}, Robert and {Masui}, Kiyoshi W. and {Mate}, Sujay and {McGregor}, Kyle and {McKinven}, Ryan and {Mena-Parra}, Juan and {Meyers}, Bradley W. and {Michilli}, Daniele and {Ng}, Cherry and {Ng}, Mason and {Nimmo}, Kenzie and {Noble}, Gavin and {Pandhi}, Ayush and {Patil}, Swarali S. and {Pearlman}, Aaron B. and {Pen}, Ue-Li and {Pleunis}, Ziggy and {Prochaska}, J. Xavier and {Rafiei-Ravandi}, Masoud and {Ransom}, Scott and {Renard}, Andre and {Sammons}, Mawson W. and {Sand}, Ketan R. and {Scholz}, Paul and {Shah}, Vishwangi and {Shin}, Kaitlyn and {Siegel}, Seth R. and {Sirota}, Sloane and {Smith}, Kendrick and {Stairs}, Ingrid and {Stenning}, David C. and {Tendulkar}, Shriharsh P. and {Vanderlinde}, Keith and {Walmsley}, Mike and {Wang}, Haochen and {Wulf}, Dallas},
        title = "{The Second CHIME/FRB Catalog of Fast Radio Bursts}",
      journal = {\apjs},
         year = 2026,
        month = mar,
       volume = {283},
       number = {1},
          eid = {34},
        pages = {34},
          doi = {10.3847/1538-4365/ae3828},
archivePrefix = {arXiv},
       eprint = {2601.09399},
 primaryClass = {astro-ph.HE},
       adsurl = {https://ui.adsabs.harvard.edu/abs/2026ApJS..283...34C}
}

@ARTICLE{caleb25,
       author = {{Caleb}, Manisha and {Nanayakkara}, Themiya and {Stappers}, Benjamin and {Pastor-Marazuela}, In{\'e}s and {Khrykin}, Ilya S. and {Glazebrook}, Karl and {Tejos}, Nicolas and {Prochaska}, J. Xavier and {Rajwade}, Kaustubh and {Mas-Ribas}, Lluis and {Driessen}, Laura N. and {Fong}, Wen-fai and {Gordon}, Alexa C. and {Hoffmann}, Jordan and {James}, Clancy W. and {Jankowski}, Fabian and {Kahinga}, Lordrick and {Kramer}, Michael and {Simha}, Sunil and {Barr}, Ewan D. and {Christiaan Bezuidenhout}, Mechiel and {Deng}, Xihan and {Lin}, Zeren and {Marnoch}, Lachlan and {Martin}, Christopher D. and {Nugent}, Anya and {Shaji}, Kavya and {Tian}, Jun},
        title = "{A fast radio burst from the first 3 billion years of the Universe}",
      journal = {arXiv e-prints},
         year = 2025,
        month = aug,
          eid = {arXiv:2508.01648},
        pages = {arXiv:2508.01648},
          doi = {10.48550/arXiv.2508.01648},
archivePrefix = {arXiv},
       eprint = {2508.01648},
 primaryClass = {astro-ph.HE},
       adsurl = {https://ui.adsabs.harvard.edu/abs/2025arXiv250801648C}
}

@ARTICLE{Cantalupo14,
       author = {{Cantalupo}, Sebastiano and {Arrigoni-Battaia}, Fabrizio and {Prochaska}, J. Xavier and {Hennawi}, Joseph F. and {Madau}, Piero},
        title = "{A cosmic web filament revealed in Lyman-{\ensuremath{\alpha}} emission around a luminous high-redshift quasar}",
      journal = {\nat},
         year = 2014,
        month = feb,
       volume = {506},
       number = {7486},
        pages = {63-66},
          doi = {10.1038/nature12898},
archivePrefix = {arXiv},
       eprint = {1401.4469},
 primaryClass = {astro-ph.CO},
       adsurl = {https://ui.adsabs.harvard.edu/abs/2014Natur.506...63C}
}

@ARTICLE{Cantalupo19,
       author = {{Cantalupo}, Sebastiano and {Pezzulli}, Gabriele and {Lilly}, Simon J. and {Marino}, Raffaella Anna and {Gallego}, Sofia G. and {Schaye}, Joop and {Bacon}, Roland and {Feltre}, Anna and {Kollatschny}, Wolfram and {Nanayakkara}, Themiya and {Richard}, Johan and {Wendt}, Martin and {Wisotzki}, Lutz and {Prochaska}, J. Xavier},
        title = "{The large- and small-scale properties of the intergalactic gas in the Slug Ly {\ensuremath{\alpha}} nebula revealed by MUSE He II emission observations}",
      journal = {\mnras},
         year = 2019,
        month = mar,
       volume = {483},
       number = {4},
        pages = {5188-5204},
          doi = {10.1093/mnras/sty3481},
archivePrefix = {arXiv},
       eprint = {1811.11783},
}

@ARTICLE{cautun14,
       author = {{Cautun}, Marius and {van de Weygaert}, Rien and {Jones}, Bernard J.~T. and {Frenk}, Carlos S.},
        title = "{Evolution of the cosmic web}",
      journal = {\mnras},
         year = 2014,
        month = jul,
       volume = {441},
       number = {4},
        pages = {2923-2973},
          doi = {10.1093/mnras/stu768},
archivePrefix = {arXiv},
       eprint = {1401.7866},
 primaryClass = {astro-ph.CO},
       adsurl = {https://ui.adsabs.harvard.edu/abs/2014MNRAS.441.2923C}
}

@ARTICLE{choudhury16,
       author = {{Choudhury}, Prakriti Pal and {Sharma}, Prateek},
        title = "{Cold gas in cluster cores: global stability analysis and non-linear simulations of thermal instability}",
      journal = {\mnras},
         year = 2016,
        month = apr,
       volume = {457},
       number = {3},
        pages = {2554-2568},
          doi = {10.1093/mnras/stw152},
archivePrefix = {arXiv},
       eprint = {1512.01217},
 primaryClass = {astro-ph.GA},
       adsurl = {https://ui.adsabs.harvard.edu/abs/2016MNRAS.457.2554C}
}

@ARTICLE{choudhury19,
       author = {{Choudhury}, Prakriti Pal and {Sharma}, Prateek and {Quataert}, Eliot},
        title = "{Multiphase gas in the circumgalactic medium: relative role of t$_{cool}$/t$_{ff}$ and density fluctuations}",
      journal = {\mnras},
         year = 2019,
        month = sep,
       volume = {488},
       number = {3},
        pages = {3195-3210},
          doi = {10.1093/mnras/stz1857},
archivePrefix = {arXiv},
       eprint = {1901.02903},
 primaryClass = {astro-ph.GA},
       adsurl = {https://ui.adsabs.harvard.edu/abs/2019MNRAS.488.3195C}
}

@ARTICLE{Colless2001,
       author = {{Colless}, Matthew and {Dalton}, Gavin and {Maddox}, Steve and {Sutherland}, Will and {Norberg}, Peder and {Cole}, Shaun and {Bland-Hawthorn}, Joss and {Bridges}, Terry and {Cannon}, Russell and {Collins}, Chris and {Couch}, Warrick and {Cross}, Nicholas and {Deeley}, Kathryn and {De Propris}, Roberto and {Driver}, Simon P. and {Efstathiou}, George and {Ellis}, Richard S. and {Frenk}, Carlos S. and {Glazebrook}, Karl and {Jackson}, Carole and {Lahav}, Ofer and {Lewis}, Ian and {Lumsden}, Stuart and {Madgwick}, Darren and {Peacock}, John A. and {Peterson}, Bruce A. and {Price}, Ian and {Seaborne}, Mark and {Taylor}, Keith},
        title = "{The 2dF Galaxy Redshift Survey: spectra and redshifts}",
      journal = {\mnras},
         year = 2001,
        month = dec,
       volume = {328},
       number = {4},
        pages = {1039-1063},
          doi = {10.1046/j.1365-8711.2001.04902.x},
archivePrefix = {arXiv},
       eprint = {astro-ph/0106498},
 primaryClass = {astro-ph},
       adsurl = {https://ui.adsabs.harvard.edu/abs/2001MNRAS.328.1039C}
}

@ARTICLE{cordes86,
       author = {{Cordes}, J.~M.},
        title = "{Space velocities of radio pulsars from interstellar scintillations.}",
      journal = {\apj},
         year = 1986,
        month = dec,
       volume = {311},
        pages = {183-196},
          doi = {10.1086/164764},
       adsurl = {https://ui.adsabs.harvard.edu/abs/1986ApJ...311..183C}
}

@ARTICLE{cordes91,
       author = {{Cordes}, J.~M. and {Weisberg}, J.~M. and {Frail}, D.~A. and {Spangler}, S.~R. and {Ryan}, M.},
        title = "{The galactic distribution of free electrons}",
      journal = {\nat},
         year = 1991,
        month = nov,
       volume = {354},
       number = {6349},
        pages = {121-124},
          doi = {10.1038/354121a0},
       adsurl = {https://ui.adsabs.harvard.edu/abs/1991Natur.354..121C}
}

@ARTICLE{cordes_lazio02,
       author = {{Cordes}, J.~M. and {Lazio}, T.~J.~W.},
        title = "{NE2001.I. A New Model for the Galactic Distribution of Free Electrons and its Fluctuations}",
      journal = {arXiv e-prints},
         year = 2002,
        month = jul,
          eid = {astro-ph/0207156},
        pages = {astro-ph/0207156},
          doi = {10.48550/arXiv.astro-ph/0207156},
archivePrefix = {arXiv},
       eprint = {astro-ph/0207156},
 primaryClass = {astro-ph},
       adsurl = {https://ui.adsabs.harvard.edu/abs/2002astro.ph..7156C}
}

@ARTICLE{cordes16,
       author = {{Cordes}, J.~M. and {Wharton}, R.~S. and {Spitler}, L.~G. and {Chatterjee}, S. and {Wasserman}, I.},
        title = "{Radio Wave Propagation and the Provenance of Fast Radio Bursts}",
      journal = {arXiv e-prints},
         year = 2016,
        month = may,
          eid = {arXiv:1605.05890},
        pages = {arXiv:1605.05890},
          doi = {10.48550/arXiv.1605.05890},
archivePrefix = {arXiv},
       eprint = {1605.05890},
 primaryClass = {astro-ph.HE},
       adsurl = {https://ui.adsabs.harvard.edu/abs/2016arXiv160505890C}
}

@ARTICLE{chawla22,
       author = {{Chawla}, P. and {Kaspi}, V.~M. and {Ransom}, S.~M. and {Bhardwaj}, M. and {Boyle}, P.~J. and {Breitman}, D. and {Cassanelli}, T. and {Cubranic}, D. and {Dong}, F.~Q. and {Fonseca}, E. and {Gaensler}, B.~M. and {Giri}, U. and {Josephy}, A. and {Kaczmarek}, J.~F. and {Leung}, C. and {Masui}, K.~W. and {Mena-Parra}, J. and {Merryfield}, M. and {Michilli}, D. and {M{\"u}nchmeyer}, M. and {Ng}, C. and {Patel}, C. and {Pearlman}, A.~B. and {Petroff}, E. and {Pleunis}, Z. and {Rahman}, M. and {Sanghavi}, P. and {Shin}, K. and {Smith}, K.~M. and {Stairs}, I. and {Tendulkar}, S.~P.},
        title = "{Modeling Fast Radio Burst Dispersion and Scattering Properties in the First CHIME/FRB Catalog}",
      journal = {\apj},
         year = 2022,
        month = mar,
       volume = {927},
       number = {1},
          eid = {35},
        pages = {35},
          doi = {10.3847/1538-4357/ac49e1},
archivePrefix = {arXiv},
       eprint = {2107.10858},
 primaryClass = {astro-ph.HE},
       adsurl = {https://ui.adsabs.harvard.edu/abs/2022ApJ...927...35C}
}

@ARTICLE{chime21,
       author = {{CHIME/FRB Collaboration} and {Amiri}, Mandana and {Andersen}, Bridget C. and {Bandura}, Kevin and {Berger}, Sabrina and {Bhardwaj}, Mohit and {Boyce}, Michelle M. and {Boyle}, P.~J. and {Brar}, Charanjot and {Breitman}, Daniela and {Cassanelli}, Tomas and {Chawla}, Pragya and {Chen}, Tianyue and {Cliche}, J. -F. and {Cook}, Amanda and {Cubranic}, Davor and {Curtin}, Alice P. and {Deng}, Meiling and {Dobbs}, Matt and {Dong}, Fengqiu Adam and {Eadie}, Gwendolyn and {Fandino}, Mateus and {Fonseca}, Emmanuel and {Gaensler}, B.~M. and {Giri}, Utkarsh and {Good}, Deborah C. and {Halpern}, Mark and {Hill}, Alex S. and {Hinshaw}, Gary and {Josephy}, Alexander and {Kaczmarek}, Jane F. and {Kader}, Zarif and {Kania}, Joseph W. and {Kaspi}, Victoria M. and {Landecker}, T.~L. and {Lang}, Dustin and {Leung}, Calvin and {Li}, Dongzi and {Lin}, Hsiu-Hsien and {Masui}, Kiyoshi W. and {McKinven}, Ryan and {Mena-Parra}, Juan and {Merryfield}, Marcus and {Meyers}, Bradley W. and {Michilli}, Daniele and {Milutinovic}, Nikola and {Mirhosseini}, Arash and {M{\"u}nchmeyer}, Moritz and {Naidu}, Arun and {Newburgh}, Laura and {Ng}, Cherry and {Patel}, Chitrang and {Pen}, Ue-Li and {Petroff}, Emily and {Pinsonneault-Marotte}, Tristan and {Pleunis}, Ziggy and {Rafiei-Ravandi}, Masoud and {Rahman}, Mubdi and {Ransom}, Scott M. and {Renard}, Andre and {Sanghavi}, Pranav and {Scholz}, Paul and {Shaw}, J. Richard and {Shin}, Kaitlyn and {Siegel}, Seth R. and {Sikora}, Andrew E. and {Singh}, Saurabh and {Smith}, Kendrick M. and {Stairs}, Ingrid and {Tan}, Chia Min and {Tendulkar}, S.~P. and {Vanderlinde}, Keith and {Wang}, Haochen and {Wulf}, Dallas and {Zwaniga}, A.~V.},
        title = "{The First CHIME/FRB Fast Radio Burst Catalog}",
      journal = {\apjs},
         year = 2021,
        month = dec,
       volume = {257},
       number = {2},
          eid = {59},
        pages = {59},
          doi = {10.3847/1538-4365/ac33ab},
archivePrefix = {arXiv},
       eprint = {2106.04352},
 primaryClass = {astro-ph.HE},
       adsurl = {https://ui.adsabs.harvard.edu/abs/2021ApJS..257...59C}
}

@ARTICLE{Connor20M33,
       author = {{Connor}, L. and {van Leeuwen}, J. and {Oostrum}, L.~C. and {Petroff}, E. and {Maan}, Y. and {Adams}, E.~A.~K. and {Attema}, J.~J. and {Bast}, J.~E. and {Boersma}, O.~M. and {D{\'e}nes}, H. and {Gardenier}, D.~W. and {Hargreaves}, J.~E. and {Kooistra}, E. and {Pastor-Marazuela}, I. and {Schulz}, R. and {Sclocco}, A. and {Smits}, R. and {Straal}, S.~M. and {van der Schuur}, D. and {Vohl}, D. and {Adebahr}, B. and {de Blok}, W.~J.~G. and {van Cappellen}, W.~A. and {Coolen}, A.~H.~W.~M. and {Damstra}, S. and {van Diepen}, G.~N.~J. and {Frank}, B.~S. and {Hess}, K.~M. and {Hut}, B. and {Kutkin}, A. and {Loose}, G. Marcel and {Lucero}, D.~M. and {Mika}, {\'A}. and {Moss}, V.~A. and {Mulder}, H. and {Oosterloo}, T.~A. and {Ruiter}, M. and {Vedantham}, H. and {Vermaas}, N.~J. and {Wijnholds}, S.~J. and {Ziemke}, J.},
        title = "{A bright, high rotation-measure FRB that skewers the M33 halo}",
      journal = {\mnras},
         year = 2020,
        month = dec,
       volume = {499},
       number = {4},
        pages = {4716-4724},
          doi = {10.1093/mnras/staa3009},
archivePrefix = {arXiv},
       eprint = {2002.01399},
 primaryClass = {astro-ph.HE},
       adsurl = {https://ui.adsabs.harvard.edu/abs/2020MNRAS.499.4716C}
}

@ARTICLE{connor25,
       author = {{Connor}, Liam and {Ravi}, Vikram and {Sharma}, Kritti and {Ocker}, Stella Koch and {Faber}, Jakob and {Hallinan}, Gregg and {Harnach}, Charlie and {Hellbourg}, Greg and {Hobbs}, Rick and {Hodge}, David and {Hodges}, Mark and {Kosogorov}, Nikita and {Lamb}, James and {Law}, Casey and {Rasmussen}, Paul and {Sherman}, Myles and {Somalwar}, Jean and {Weinreb}, Sander and {Woody}, David and {Konietzka}, Ralf M.},
        title = "{A gas-rich cosmic web revealed by the partitioning of the missing baryons}",
      journal = {Nature Astronomy},
         year = 2025,
        month = aug,
       volume = {9},
        pages = {1226-1239},
          doi = {10.1038/s41550-025-02566-y},
archivePrefix = {arXiv},
       eprint = {2409.16952},
 primaryClass = {astro-ph.CO},
       adsurl = {https://ui.adsabs.harvard.edu/abs/2025NatAs...9.1226C}
}

@article{danovich12,
    author = {Danovich, Mark and Dekel, Avishai and Hahn, Oliver and Teyssier, Romain},
    title = "{Coplanar streams, pancakes and angular-momentum exchange in high-z disc galaxies}",
    journal = {Monthly Notices of the Royal Astronomical Society},
    volume = {422},
    number = {2},
    pages = {1732-1749},
    year = {2012},
    month = {04},
    issn = {0035-8711},
    doi = {10.1111/j.1365-2966.2012.20751.x},
    url = {https://doi.org/10.1111/j.1365-2966.2012.20751.x},
    eprint = {https://academic.oup.com/mnras/article-pdf/422/2/1732/3516962/mnras0422-1732.pdf},
}

@ARTICLE{Dekel.Birnboim.06,
       author = {{Dekel}, A. and {Birnboim}, Yuval},
        title = "{Galaxy bimodality due to cold flows and shock heating}",
      journal = {\mnras},
         year = 2006,
        month = may,
       volume = {368},
       number = {1},
        pages = {2-20},
          doi = {10.1111/j.1365-2966.2006.10145.x},
archivePrefix = {arXiv},
       eprint = {astro-ph/0412300},
 primaryClass = {astro-ph},
       adsurl = {https://ui.adsabs.harvard.edu/abs/2006MNRAS.368....2D}
}

@ARTICLE{Dekel09a,
       author = {{Dekel}, A. and {Birnboim}, Y. and {Engel}, G. and {Freundlich}, J. and
         {Goerdt}, T. and {Mumcuoglu}, M. and {Neistein}, E. and {Pichon}, C. and
         {Teyssier}, R. and {Zinger}, E.},
        title = "{Cold streams in early massive hot haloes as the main mode of galaxy formation}",
      journal = {\nat},
         year = 2009,
        month = jan,
       volume = {457},
       number = {7228},
        pages = {451-454},
          doi = {10.1038/nature07648},
archivePrefix = {arXiv},
       eprint = {0808.0553},
 primaryClass = {astro-ph},
       adsurl = {https://ui.adsabs.harvard.edu/abs/2009Natur.457..451D}
}

@ARTICLE{dekel13,
       author = {{Dekel}, A. and {Zolotov}, A. and {Tweed}, D. and {Cacciato}, M. and {Ceverino}, D. and {Primack}, J.~R.},
        title = "{Toy models for galaxy formation versus simulations}",
      journal = {\mnras},
         year = 2013,
        month = oct,
       volume = {435},
       number = {2},
        pages = {999-1019},
          doi = {10.1093/mnras/stt1338},
archivePrefix = {arXiv},
       eprint = {1303.3009},
 primaryClass = {astro-ph.CO},
       adsurl = {https://ui.adsabs.harvard.edu/abs/2013MNRAS.435..999D}
}

@ARTICLE{Eilers18,
       author = {{Eilers}, Anna-Christina and {Davies}, Frederick B. and
         {Hennawi}, Joseph F.},
        title = "{The Opacity of the Intergalactic Medium Measured along Quasar Sightlines at z {\ensuremath{\sim}} 6}",
      journal = {\apj},
         year = 2018,
        month = sep,
       volume = {864},
       number = {1},
          eid = {53},
        pages = {53},
          doi = {10.3847/1538-4357/aad4fd},
archivePrefix = {arXiv},
       eprint = {1807.04229},
 primaryClass = {astro-ph.GA},
       adsurl = {https://ui.adsabs.harvard.edu/abs/2018ApJ...864...53E}
}

@ARTICLE{Faber2024,
       author = {{Faber}, Jakob T. and {Ravi}, Vikram and {Ocker}, Stella Koch and {Sherman}, Myles B. and {Sharma}, Kritti and {Connor}, Liam and {Law}, Casey and {Kosogorov}, Nikita and {Hallinan}, Gregg and {Harnach}, Charlie and {Hellbourg}, Greg and {Hobbs}, Rick and {Hodge}, David and {Hodges}, Mark and {Lamb}, James W. and {Rasmussen}, Paul and {Somalwar}, Jean J. and {Weinreb}, Sander and {Woody}, David P.},
        title = "{A Heavily Scattered Fast Radio Burst Is Viewed Through Multiple Galaxy Halos}",
      journal = {arXiv e-prints},
         year = 2024,
        month = may,
          eid = {arXiv:2405.14182},
        pages = {arXiv:2405.14182},
          doi = {10.48550/arXiv.2405.14182},
archivePrefix = {arXiv},
       eprint = {2405.14182},
 primaryClass = {astro-ph.HE},
       adsurl = {https://ui.adsabs.harvard.edu/abs/2024arXiv240514182F}
}

@ARTICLE{FG_Oh2023,
       author = {{Faucher-Gigu{\`e}re}, Claude-Andr{\'e} and {Oh}, S. Peng},
        title = "{Key Physical Processes in the Circumgalactic Medium}",
      journal = {\araa},
         year = 2023,
        month = aug,
       volume = {61},
        pages = {131-195},
          doi = {10.1146/annurev-astro-052920-125203},
archivePrefix = {arXiv},
       eprint = {2301.10253},
 primaryClass = {astro-ph.GA},
       adsurl = {https://ui.adsabs.harvard.edu/abs/2023ARA&A..61..131F}
}

@ARTICLE{Field65,
       author = {{Field}, George B.},
        title = "{Thermal Instability.}",
      journal = {\apj},
         year = 1965,
        month = aug,
       volume = {142},
        pages = {531},
          doi = {10.1086/148317},
       adsurl = {https://ui.adsabs.harvard.edu/abs/1965ApJ...142..531F}
}

@ARTICLE{Fielding17,
       author = {{Fielding}, Drummond and {Quataert}, Eliot and {McCourt}, Michael and {Thompson}, Todd A.},
        title = "{The impact of star formation feedback on the circumgalactic medium}",
      journal = {\mnras},
         year = 2017,
        month = apr,
       volume = {466},
       number = {4},
        pages = {3810-3826},
          doi = {10.1093/mnras/stw3326},
archivePrefix = {arXiv},
       eprint = {1606.06734},
 primaryClass = {astro-ph.GA},
       adsurl = {https://ui.adsabs.harvard.edu/abs/2017MNRAS.466.3810F}
}

@ARTICLE{Gajjar18,
       author = {{Gajjar}, V. and {Siemion}, A.~P.~V. and {Price}, D.~C. and {Law}, C.~J. and {Michilli}, D. and {Hessels}, J.~W.~T. and {Chatterjee}, S. and {Archibald}, A.~M. and {Bower}, G.~C. and {Brinkman}, C. and {Burke-Spolaor}, S. and {Cordes}, J.~M. and {Croft}, S. and {Enriquez}, J. Emilio and {Foster}, G. and {Gizani}, N. and {Hellbourg}, G. and {Isaacson}, H. and {Kaspi}, V.~M. and {Lazio}, T.~J.~W. and {Lebofsky}, M. and {Lynch}, R.~S. and {MacMahon}, D. and {McLaughlin}, M.~A. and {Ransom}, S.~M. and {Scholz}, P. and {Seymour}, A. and {Spitler}, L.~G. and {Tendulkar}, S.~P. and {Werthimer}, D. and {Zhang}, Y.~G.},
        title = "{Highest Frequency Detection of FRB 121102 at 4-8 GHz Using the Breakthrough Listen Digital Backend at the Green Bank Telescope}",
      journal = {\apj},
         year = 2018,
        month = aug,
       volume = {863},
       number = {1},
          eid = {2},
        pages = {2},
          doi = {10.3847/1538-4357/aad005},
archivePrefix = {arXiv},
       eprint = {1804.04101},
 primaryClass = {astro-ph.HE},
       adsurl = {https://ui.adsabs.harvard.edu/abs/2018ApJ...863....2G}
}

@ARTICLE{glowacki25,
       author = {{Glowacki}, Marcin and {Bera}, Apurba and {James}, Clancy and {Patterson}, Jasper and {Deller}, Adam T. and {Gordon}, Alexa and {Marnoch}, Lachlan and {Muller}, August and {Prochaska}, Xavier and {Ryder}, Stuart and {Shannon}, Ryan M. and {Tejos}, Nicolas and {Mannings}, Alexandra G.},
        title = "{An investigation into correlations between FRB and host galaxy properties}",
      journal = {\pasa},
         year = 2025,
        month = nov,
       volume = {42},
          eid = {e157},
        pages = {e157},
          doi = {10.1017/pasa.2025.10122},
archivePrefix = {arXiv},
       eprint = {2506.23403},
 primaryClass = {astro-ph.HE},
       adsurl = {https://ui.adsabs.harvard.edu/abs/2025PASA...42..157G}
}

@ARTICLE{Good20,
       author = {{Good}, Deborah and {Chime/Frb Collaboration}},
        title = "{CHIME/FRB Detection of Three More Radio Bursts from SGR 1935+2154}",
      journal = {The Astronomer's Telegram},
         year = 2020,
        month = oct,
       volume = {14074},
        pages = {1},
       adsurl = {https://ui.adsabs.harvard.edu/abs/2020ATel14074....1G}
}

@ARTICLE{gordon24,
       author = {{Gordon}, Alexa C. and {Fong}, Wen-fai and {Simha}, Sunil and {Dong}, Yuxin and {Kilpatrick}, Charles D. and {Deller}, Adam T. and {Ryder}, Stuart D. and {Eftekhari}, Tarraneh and {Glowacki}, Marcin and {Marnoch}, Lachlan and {Muller}, August R. and {Nugent}, Anya E. and {Palmese}, Antonella and {Prochaska}, J. Xavier and {Rafelski}, Marc and {Shannon}, Ryan M. and {Tejos}, Nicolas},
        title = "{A Fast Radio Burst in a Compact Galaxy Group at z {\ensuremath{\sim}} 1}",
      journal = {\apjl},
         year = 2024,
        month = mar,
       volume = {963},
       number = {2},
          eid = {L34},
        pages = {L34},
          doi = {10.3847/2041-8213/ad2773},
archivePrefix = {arXiv},
       eprint = {2311.10815},
 primaryClass = {astro-ph.GA},
       adsurl = {https://ui.adsabs.harvard.edu/abs/2024ApJ...963L..34G}
}

@ARTICLE{Gronke17,
       author = {{Gronke}, Max and {Dijkstra}, Mark and {McCourt}, Michael and {{Oh}, S. Peng}},
        title = "{Resonant line transfer in a fog: using Lyman-alpha to probe tiny structures in atomic gas}",
      journal = {\aap},
         year = 2017,
        month = nov,
       volume = {607},
          eid = {A71},
        pages = {A71},
          doi = {10.1051/0004-6361/201731013},
archivePrefix = {arXiv},
       eprint = {1704.06278},
 primaryClass = {astro-ph.GA},
       adsurl = {https://ui.adsabs.harvard.edu/abs/2017A&A...607A..71G}
}

@ARTICLE{Gronke_Oh20,
       author = {{Gronke}, Max and {Oh}, S. Peng},
        title = "{Is multiphase gas cloudy or misty?}",
      journal = {\mnras},
         year = 2020,
        month = may,
       volume = {494},
       number = {1},
        pages = {L27-L31},
          doi = {10.1093/mnrasl/slaa033},
archivePrefix = {arXiv},
       eprint = {1912.07808},
 primaryClass = {astro-ph.GA},
       adsurl = {https://ui.adsabs.harvard.edu/abs/2020MNRAS.494L..27G}
}

@ARTICLE{Gronke_Oh23,
       author = {{Gronke}, Max and {Oh}, S. Peng},
        title = "{Cooling-driven coagulation}",
      journal = {\mnras},
         year = 2023,
        month = sep,
       volume = {524},
       number = {1},
        pages = {498-511},
          doi = {10.1093/mnras/stad1874},
archivePrefix = {arXiv},
       eprint = {2209.00732},
 primaryClass = {astro-ph.GA},
       adsurl = {https://ui.adsabs.harvard.edu/abs/2023MNRAS.524..498G}
}

@ARTICLE{haardt_madau96,
       author = {{Haardt}, Francesco and {Madau}, Piero},
        title = "{Radiative Transfer in a Clumpy Universe. II. The Ultraviolet Extragalactic Background}",
      journal = {\apj},
         year = 1996,
        month = apr,
       volume = {461},
        pages = {20},
          doi = {10.1086/177035},
archivePrefix = {arXiv},
       eprint = {astro-ph/9509093},
 primaryClass = {astro-ph},
       adsurl = {https://ui.adsabs.harvard.edu/abs/1996ApJ...461...20H}
}

@ARTICLE{Hennawi06,
   author = {{Hennawi}, J.~F. and {Prochaska}, J.~X. and {Burles}, S. and 
	{Strauss}, M.~A. and {Richards}, G.~T. and {Schlegel}, D.~J. and 
	{Fan}, X. and {Schneider}, D.~P. and {Zakamska}, N.~L. and {Oguri}, M. and 
	{Gunn}, J.~E. and {Lupton}, R.~H. and {Brinkmann}, J.},
    title = "{Quasars Probing Quasars. I. Optically Thick Absorbers near Luminous Quasars}",
  journal = {\apj},
   eprint = {astro-ph/0603742},
     year = 2006,
    month = nov,
   volume = 651,
    pages = {61-83},
      doi = {10.1086/507069},
   adsurl = {http://adsabs.harvard.edu/abs/2006ApJ...651...61H}
}

@article{Heimersheim_2022,
    doi = {10.3847/1538-4357/ac70c9},
    url = {https://dx.doi.org/10.3847/1538-4357/ac70c9},
    year = {2022},
    month = {jul},
    publisher = {The American Astronomical Society},
    volume = {933},
    number = {1},
    pages = {57},
    author = {Stefan Heimersheim and Nina S. Sartorio and Anastasia Fialkov and Duncan R. Lorimer},
    title = {What It Takes to Measure Reionization with Fast Radio Bursts},
    journal = {ApJ}
}

@article{Hashimoto_2021,
    author = {Hashimoto, Tetsuya and Goto, Tomotsugu and Lu, Ting-Yi and On, Alvina Y L and Santos, Daryl Joe D and Kim, Seong Jin and Eser, Ece Kilerci and Ho, Simon C-C and Hsiao, Tiger Y-Y and Lin, Leo Y-W},
    title = "{Revealing the cosmic reionization history with fast radio bursts in the era of Square Kilometre Array}",
    journal = {MNRAS},
    volume = {502},
    number = {2},
    pages = {2346-2355},
    year = {2021},
    month = {01},
    issn = {0035-8711},
    doi = {10.1093/mnras/stab186},
    url = {https://doi.org/10.1093/mnras/stab186},
    eprint = {https://academic.oup.com/mnras/article-pdf/502/2/2346/36218083/stab186.pdf},
}

@INPROCEEDINGS{Huchra05,
       author = {{Huchra}, J. and {Jarrett}, T. and {Skrutskie}, M. and {Cutri}, R. and {Schneider}, S. and {Macri}, L. and {Steining}, R. and {Mader}, J. and {Martimbeau}, N. and {George}, T.},
        title = "{The 2MASS Redshift Survey and Low Galactic Latitude Large-Scale Structure}",
    booktitle = {Nearby Large-Scale Structures and the Zone of Avoidance},
         year = 2005,
       editor = {{Fairall}, Anthony P. and {Woudt}, Patrick A.},
       series = {Astronomical Society of the Pacific Conference Series},
       volume = {329},
        month = jun,
        pages = {135},
       adsurl = {https://ui.adsabs.harvard.edu/abs/2005ASPC..329..135H}
}

@ARTICLE{Hummels19,
       author = {{Hummels}, Cameron B. and {Smith}, Britton D. and {Hopkins}, Philip F. and {O'Shea}, Brian W. and {Silvia}, Devin W. and {Werk}, Jessica K. and {Lehner}, Nicolas and {Wise}, John H. and {Collins}, David C. and {Butsky}, Iryna S.},
        title = "{The Impact of Enhanced Halo Resolution on the Simulated Circumgalactic Medium}",
      journal = {\apj},
         year = 2019,
        month = sep,
       volume = {882},
       number = {2},
          eid = {156},
        pages = {156},
          doi = {10.3847/1538-4357/ab378f},
archivePrefix = {arXiv},
       eprint = {1811.12410},
 primaryClass = {astro-ph.GA},
       adsurl = {https://ui.adsabs.harvard.edu/abs/2019ApJ...882..156H}
}

@ARTICLE{jaroszynski19,
       author = {{Jaroszynski}, M.},
        title = "{Fast radio bursts and cosmological tests}",
      journal = {\mnras},
         year = 2019,
        month = apr,
       volume = {484},
       number = {2},
        pages = {1637-1644},
          doi = {10.1093/mnras/sty3529},
archivePrefix = {arXiv},
       eprint = {1812.11936},
 primaryClass = {astro-ph.CO},
       adsurl = {https://ui.adsabs.harvard.edu/abs/2019MNRAS.484.1637J}
}

@ARTICLE{jow24,
       author = {{Jow}, Dylan L. and {Wu}, Xiaohan and {Pen}, Ue-Li},
        title = "{Refractive lensing of scintillating FRBs by subparsec cloudlets in the multiphase CGM}",
      journal = {Proceedings of the National Academy of Science},
         year = 2024,
        month = sep,
       volume = {121},
       number = {39},
          eid = {e2406783121},
        pages = {e2406783121},
          doi = {10.1073/pnas.2406783121},
archivePrefix = {arXiv},
       eprint = {2309.07256},
 primaryClass = {astro-ph.GA},
       adsurl = {https://ui.adsabs.harvard.edu/abs/2024PNAS..12106783J}
}

@ARTICLE{Kumar+17,
       author = {{Kumar}, Pawan and {Lu}, Wenbin and {Bhattacharya}, Mukul},
        title = "{Fast radio burst source properties and curvature radiation model}",
      journal = {\mnras},
         year = 2017,
        month = jul,
       volume = {468},
       number = {3},
        pages = {2726-2739},
          doi = {10.1093/mnras/stx665},
archivePrefix = {arXiv},
       eprint = {1703.06139},
 primaryClass = {astro-ph.HE},
       adsurl = {https://ui.adsabs.harvard.edu/abs/2017MNRAS.468.2726K}
}

@ARTICLE{kumar24,
       author = {{Kumar}, Pawan and {Beniamini}, Paz and {Gupta}, Om and {Cordes}, James M.},
        title = "{Constraining the FRB mechanism from scintillation in the host galaxy}",
      journal = {\mnras},
         year = 2024,
        month = jan,
       volume = {527},
       number = {1},
        pages = {457-470},
          doi = {10.1093/mnras/stad3010},
archivePrefix = {arXiv},
       eprint = {2307.15294},
 primaryClass = {astro-ph.HE},
       adsurl = {https://ui.adsabs.harvard.edu/abs/2024MNRAS.527..457K}
}

@ARTICLE{lang71,
       author = {{Lang}, Kenneth R.},
        title = "{Interstellar Scintillation of Pulsar Radiation}",
      journal = {\apj},
         year = 1971,
        month = mar,
       volume = {164},
        pages = {249},
          doi = {10.1086/150836},
       adsurl = {https://ui.adsabs.harvard.edu/abs/1971ApJ...164..249L}
}

@ARTICLE{l26a,
       author = {{Lapiner}, Sharon and {Mandelker}, Nir and {Beniamini}, Paz and {Oh}, S. Peng},
        title = "{Probing the cosmic web with fast radio bursts: I. Scattering}",
      journal = {\aap},
         year = 2026,
        month = aug,
       volume = {712},
          eid = {A108},
        pages = {A108},
          doi = {10.1051/0004-6361/202659987},
archivePrefix = {arXiv},
       eprint = {2603.22108},
 primaryClass = {astro-ph.CO},
       adsurl = {https://ui.adsabs.harvard.edu/abs/2026A&A...712A.108L}
}

@ARTICLE{lee22,
       author = {{Lee}, Khee-Gan and {Ata}, Metin and {Khrykin}, Ilya S. and {Huang}, Yuxin and {Prochaska}, J. Xavier and {Cooke}, Jeff and {Zhang}, Jielai and {Batten}, Adam},
        title = "{Constraining the Cosmic Baryon Distribution with Fast Radio Burst Foreground Mapping}",
      journal = {\apj},
         year = 2022,
        month = mar,
       volume = {928},
       number = {1},
          eid = {9},
        pages = {9},
          doi = {10.3847/1538-4357/ac4f62},
archivePrefix = {arXiv},
       eprint = {2109.00386},
 primaryClass = {astro-ph.CO},
       adsurl = {https://ui.adsabs.harvard.edu/abs/2022ApJ...928....9L}
}

@ARTICLE{Lehner22,
       author = {{Lehner}, Nicolas and {Kopenhafer}, Claire and {O'Meara}, John M. and {Howk}, J. Christopher and {Fumagalli}, Michele and {Prochaska}, J. Xavier and {Acharyya}, Ayan and {O'Shea}, Brian W. and {Peeples}, Molly S. and {Tumlinson}, Jason and {Hummels}, Cameron B.},
        title = "{KODIAQ-Z: Metals and Baryons in the Cool Intergalactic and Circumgalactic Gas at 2.2 {\ensuremath{\lesssim}} z {\ensuremath{\lesssim}} 3.6}",
      journal = {\apj},
         year = 2022,
        month = sep,
       volume = {936},
       number = {2},
          eid = {156},
        pages = {156},
          doi = {10.3847/1538-4357/ac7400},
archivePrefix = {arXiv},
       eprint = {2112.03304},
 primaryClass = {astro-ph.GA},
       adsurl = {https://ui.adsabs.harvard.edu/abs/2022ApJ...936..156L}
}

@ARTICLE{Libeskind18,
       author = {{Libeskind}, Noam I. and {van de Weygaert}, Rien and {Cautun}, Marius and {Falck}, Bridget and {Tempel}, Elmo and {Abel}, Tom and {Alpaslan}, Mehmet and {Arag{\'o}n-Calvo}, Miguel A. and {Forero-Romero}, Jaime E. and {Gonzalez}, Roberto and {Gottl{\"o}ber}, Stefan and {Hahn}, Oliver and {Hellwing}, Wojciech A. and {Hoffman}, Yehuda and {Jones}, Bernard J.~T. and {Kitaura}, Francisco and {Knebe}, Alexander and {Manti}, Serena and {Neyrinck}, Mark and {Nuza}, Sebasti{\'a}n E. and {Padilla}, Nelson and {Platen}, Erwin and {Ramachandra}, Nesar and {Robotham}, Aaron and {Saar}, Enn and {Shandarin}, Sergei and {Steinmetz}, Matthias and {Stoica}, Radu S. and {Sousbie}, Thierry and {Yepes}, Gustavo},
        title = "{Tracing the cosmic web}",
      journal = {\mnras},
         year = 2018,
        month = jan,
       volume = {473},
       number = {1},
        pages = {1195-1217},
          doi = {10.1093/mnras/stx1976},
archivePrefix = {arXiv},
       eprint = {1705.03021},
 primaryClass = {astro-ph.CO},
       adsurl = {https://ui.adsabs.harvard.edu/abs/2018MNRAS.473.1195L}
}

@ARTICLE{Lidz_Malloy14,
       author = {{Lidz}, Adam and {Malloy}, Matthew},
        title = "{On Modeling and Measuring the Temperature of the z \raisebox{-0.5ex}\textasciitilde 5 Intergalactic Medium}",
      journal = {\apj},
         year = "2014",
        month = "Jun",
       volume = {788},
       number = {2},
          eid = {175},
        pages = {175},
          doi = {10.1088/0004-637X/788/2/175},
archivePrefix = {arXiv},
       eprint = {1403.6350},
 primaryClass = {astro-ph.CO},
       adsurl = {https://ui.adsabs.harvard.edu/abs/2014ApJ...788..175L}
}

@ARTICLE{Lorimer07,
       author = {{Lorimer}, D.~R. and {Bailes}, M. and {McLaughlin}, M.~A. and {Narkevic}, D.~J. and {Crawford}, F.},
        title = "{A Bright Millisecond Radio Burst of Extragalactic Origin}",
      journal = {Science},
         year = 2007,
        month = nov,
       volume = {318},
       number = {5851},
        pages = {777},
          doi = {10.1126/science.1147532},
archivePrefix = {arXiv},
       eprint = {0709.4301},
 primaryClass = {astro-ph},
       adsurl = {https://ui.adsabs.harvard.edu/abs/2007Sci...318..777L}
}

@ARTICLE{Lorimer24,
       author = {{Lorimer}, Duncan R. and {McLaughlin}, Maura A. and {Bailes}, Matthew},
        title = "{The discovery and significance of fast radio bursts}",
      journal = {\apss},
         year = 2024,
        month = jun,
       volume = {369},
       number = {6},
          eid = {59},
        pages = {59},
          doi = {10.1007/s10509-024-04322-6},
archivePrefix = {arXiv},
       eprint = {2405.19106},
 primaryClass = {astro-ph.HE},
       adsurl = {https://ui.adsabs.harvard.edu/abs/2024Ap&SS.369...59L}
}

@ARTICLE{lu-piro19,
       author = {{Lu}, Wenbin and {Piro}, Anthony L.},
        title = "{Implications from ASKAP Fast Radio Burst Statistics}",
      journal = {\apj},
         year = 2019,
        month = sep,
       volume = {883},
       number = {1},
          eid = {40},
        pages = {40},
          doi = {10.3847/1538-4357/ab3796},
archivePrefix = {arXiv},
       eprint = {1903.00014},
 primaryClass = {astro-ph.HE},
       adsurl = {https://ui.adsabs.harvard.edu/abs/2019ApJ...883...40L}
}

@article{lu23,
    author = {Lu, Yue Samuel and Mandelker, Nir and Oh, S Peng and Dekel, Avishai and van den Bosch, Frank C and Springel, Volker and Nagai, Daisuke and van de Voort, Freeke},
    title = "{The structure and dynamics of massive high-z cosmic-web filaments: Three radial zones in filament cross-sections}",
    journal = {Monthly Notices of the Royal Astronomical Society},
    pages = {stad3779},
    year = {2023},
    month = {12},
    issn = {0035-8711},
    doi = {10.1093/mnras/stad3779},
    url = {https://doi.org/10.1093/mnras/stad3779},
    eprint = {https://academic.oup.com/mnras/advance-article-pdf/doi/10.1093/mnras/stad3779/54093701/stad3779.pdf},
}

@ARTICLE{macquart04,
       author = {{Macquart}, J. -P.},
        title = "{Scattering of gravitational radiation. Second order moments of the wave amplitude}",
      journal = {\aap},
         year = 2004,
        month = aug,
       volume = {422},
        pages = {761-775},
          doi = {10.1051/0004-6361:20034512},
archivePrefix = {arXiv},
       eprint = {astro-ph/0402661},
 primaryClass = {astro-ph},
       adsurl = {https://ui.adsabs.harvard.edu/abs/2004A&A...422..761M}
}

@ARTICLE{macquart13,
       author = {{Macquart}, Jean-Pierre and {Koay}, Jun Yi},
        title = "{Temporal Smearing of Transient Radio Sources by the Intergalactic Medium}",
      journal = {\apj},
         year = 2013,
        month = oct,
       volume = {776},
       number = {2},
          eid = {125},
        pages = {125},
          doi = {10.1088/0004-637X/776/2/125},
archivePrefix = {arXiv},
       eprint = {1308.4459},
 primaryClass = {astro-ph.CO},
       adsurl = {https://ui.adsabs.harvard.edu/abs/2013ApJ...776..125M}
}

@ARTICLE{macquart20,
       author = {{Macquart}, J. -P. and {Prochaska}, J.~X. and {McQuinn}, M. and {Bannister}, K.~W. and {Bhandari}, S. and {Day}, C.~K. and {Deller}, A.~T. and {Ekers}, R.~D. and {James}, C.~W. and {Marnoch}, L. and {Os{\l}owski}, S. and {Phillips}, C. and {Ryder}, S.~D. and {Scott}, D.~R. and {Shannon}, R.~M. and {Tejos}, N.},
        title = "{A census of baryons in the Universe from localized fast radio bursts}",
      journal = {\nat},
         year = 2020,
        month = may,
       volume = {581},
       number = {7809},
        pages = {391-395},
          doi = {10.1038/s41586-020-2300-2},
archivePrefix = {arXiv},
       eprint = {2005.13161},
 primaryClass = {astro-ph.CO},
       adsurl = {https://ui.adsabs.harvard.edu/abs/2020Natur.581..391M}
}

@ARTICLE{m18,
       author = {{Mandelker}, Nir and {van Dokkum}, Pieter G. and {Brodie}, Jean P. and {van den Bosch}, Frank C. and {Ceverino}, Daniel},
        title = "{Cold Filamentary Accretion and the Formation of Metal-poor Globular Clusters and Halo Stars}",
      journal = {\apj},
         year = 2018,
        month = jul,
       volume = {861},
       number = {2},
          eid = {148},
        pages = {148},
          doi = {10.3847/1538-4357/aaca98},
archivePrefix = {arXiv},
       eprint = {1711.09108},
 primaryClass = {astro-ph.GA},
       adsurl = {https://ui.adsabs.harvard.edu/abs/2018ApJ...861..148M}
}

@ARTICLE{m19,
       author = {{Mandelker}, Nir and {van den Bosch}, Frank C. and {Springel}, Volker and {van de Voort}, Freeke},
        title = "{Shattering of Cosmic Sheets due to Thermal Instabilities: A Formation Channel for Metal-free Lyman Limit Systems}",
      journal = {\apjl},
         year = 2019,
        month = aug,
       volume = {881},
       number = {1},
          eid = {L20},
        pages = {L20},
          doi = {10.3847/2041-8213/ab30cb},
archivePrefix = {arXiv},
       eprint = {1906.10693},
 primaryClass = {astro-ph.GA},
       adsurl = {https://ui.adsabs.harvard.edu/abs/2019ApJ...881L..20M}
}

@ARTICLE{m21,
       author = {{Mandelker}, Nir and {van den Bosch}, Frank C. and {Springel}, Volker and {van de Voort}, Freeke and {Burchett}, Joseph N. and {Butsky}, Iryna S. and {Nagai}, Daisuke and {Oh}, S. Peng},
        title = "{Thermal Instabilities and Shattering in the High-redshift WHIM: Convergence Criteria and Implications for Low-metallicity Strong H I Absorbers}",
      journal = {\apj},
         year = 2021,
        month = dec,
       volume = {923},
       number = {1},
          eid = {115},
        pages = {115},
          doi = {10.3847/1538-4357/ac2d29},
archivePrefix = {arXiv},
       eprint = {2107.03395},
 primaryClass = {astro-ph.CO},
       adsurl = {https://ui.adsabs.harvard.edu/abs/2021ApJ...923..115M}
}

@ARTICLE{marcote20,
       author = {{Marcote}, B. and {Nimmo}, K. and {Hessels}, J.~W.~T. and {Tendulkar}, S.~P. and {Bassa}, C.~G. and {Paragi}, Z. and {Keimpema}, A. and {Bhardwaj}, M. and {Karuppusamy}, R. and {Kaspi}, V.~M. and {Law}, C.~J. and {Michilli}, D. and {Aggarwal}, K. and {Andersen}, B. and {Archibald}, A.~M. and {Bandura}, K. and {Bower}, G.~C. and {Boyle}, P.~J. and {Brar}, C. and {Burke-Spolaor}, S. and {Butler}, B.~J. and {Cassanelli}, T. and {Chawla}, P. and {Demorest}, P. and {Dobbs}, M. and {Fonseca}, E. and {Giri}, U. and {Good}, D.~C. and {Gourdji}, K. and {Josephy}, A. and {Kirichenko}, A. Yu. and {Kirsten}, F. and {Landecker}, T.~L. and {Lang}, D. and {Lazio}, T.~J.~W. and {Li}, D.~Z. and {Lin}, H. -H. and {Linford}, J.~D. and {Masui}, K. and {Mena-Parra}, J. and {Naidu}, A. and {Ng}, C. and {Patel}, C. and {Pen}, U. -L. and {Pleunis}, Z. and {Rafiei-Ravandi}, M. and {Rahman}, M. and {Renard}, A. and {Scholz}, P. and {Siegel}, S.~R. and {Smith}, K.~M. and {Stairs}, I.~H. and {Vanderlinde}, K. and {Zwaniga}, A.~V.},
        title = "{A repeating fast radio burst source localized to a nearby spiral galaxy}",
      journal = {\nat},
         year = 2020,
        month = jan,
       volume = {577},
       number = {7789},
        pages = {190-194},
          doi = {10.1038/s41586-019-1866-z},
archivePrefix = {arXiv},
       eprint = {2001.02222},
 primaryClass = {astro-ph.HE},
       adsurl = {https://ui.adsabs.harvard.edu/abs/2020Natur.577..190M}
}

@ARTICLE{Martin14a,
   author = {{Martin}, D.~C. and {Chang}, D. and {Matuszewski}, M. and {Morrissey}, P. and 
	{Rahman}, S. and {Moore}, A. and {Steidel}, C.~C.},
    title = "{Intergalactic Medium Emission Observations with the Cosmic Web Imager. I. The Circum-QSO Medium of QSO 1549+19, and Evidence for a Filamentary Gas Inflow}",
  journal = {\apj},
archivePrefix = "arXiv",
   eprint = {1402.4816},
     year = 2014,
    month = may,
   volume = 786,
      eid = {106},
    pages = {106},
      doi = {10.1088/0004-637X/786/2/106},
   adsurl = {http://adsabs.harvard.edu/abs/2014ApJ...786..106M}
}

@ARTICLE{Martin14b,
   author = {{Martin}, D.~C. and {Chang}, D. and {Matuszewski}, M. and {Morrissey}, P. and 
	{Rahman}, S. and {Moore}, A. and {Steidel}, C.~C. and {Matsuda}, Y.
	},
    title = "{Intergalactic Medium Emission Observations with the Cosmic Web Imager. II. Discovery of Extended, Kinematically Linked Emission around SSA22 Ly{$\alpha$} Blob 2}",
  journal = {\apj},
archivePrefix = "arXiv",
   eprint = {1402.4809},
     year = 2014,
    month = may,
   volume = 786,
      eid = {107},
    pages = {107},
      doi = {10.1088/0004-637X/786/2/107},
   adsurl = {http://adsabs.harvard.edu/abs/2014ApJ...786..107M}
}

@ARTICLE{Martin23,
       author = {{Martin}, D. Christopher and {Darvish}, Behnam and {Lin}, Zeren and {Cen}, Renyue and {Matuszewski}, Mateusz and {Morrissey}, Patrick and {Neill}, James D. and {Moore}, Anna M.},
        title = "{Extensive diffuse Lyman-{\ensuremath{\alpha}} emission correlated with cosmic structure}",
      journal = {Nature Astronomy},
         year = 2023,
        month = nov,
       volume = {7},
        pages = {1390-1401},
          doi = {10.1038/s41550-023-02054-1},
       adsurl = {https://ui.adsabs.harvard.edu/abs/2023NatAs...7.1390M}
}

@ARTICLE{mas-ribas25,
       author = {{Mas-Ribas}, Llu{\'\i}s and {McQuinn}, Matthew and {Prochaska}, J. Xavier},
        title = "{Circumgalactic Medium Cloud Sizes from Refractive Fast Radio Burst Scattering}",
      journal = {\apj},
         year = 2025,
        month = sep,
       volume = {990},
       number = {2},
          eid = {179},
        pages = {179},
          doi = {10.3847/1538-4357/adf43b},
       adsurl = {https://ui.adsabs.harvard.edu/abs/2025ApJ...990..179M}
}

@ARTICLE{masui15,
       author = {{Masui}, Kiyoshi and {Lin}, Hsiu-Hsien and {Sievers}, Jonathan and {Anderson}, Christopher J. and {Chang}, Tzu-Ching and {Chen}, Xuelei and {Ganguly}, Apratim and {Jarvis}, Miranda and {Kuo}, Cheng-Yu and {Li}, Yi-Chao and {Liao}, Yu-Wei and {McLaughlin}, Maura and {Pen}, Ue-Li and {Peterson}, Jeffrey B. and {Roman}, Alexander and {Timbie}, Peter T. and {Voytek}, Tabitha and {Yadav}, Jaswant K.},
        title = "{Dense magnetized plasma associated with a fast radio burst}",
      journal = {\nat},
         year = 2015,
        month = dec,
       volume = {528},
       number = {7583},
        pages = {523-525},
          doi = {10.1038/nature15769},
archivePrefix = {arXiv},
       eprint = {1512.00529},
 primaryClass = {astro-ph.HE},
       adsurl = {https://ui.adsabs.harvard.edu/abs/2015Natur.528..523M}
}

@ARTICLE{mccourt12,
       author = {{McCourt}, Michael and {Sharma}, Prateek and {Quataert}, Eliot and {Parrish}, Ian J.},
        title = "{Thermal instability in gravitationally stratified plasmas: implications for multiphase structure in clusters and galaxy haloes}",
      journal = {\mnras},
         year = 2012,
        month = feb,
       volume = {419},
       number = {4},
        pages = {3319-3337},
          doi = {10.1111/j.1365-2966.2011.19972.x},
archivePrefix = {arXiv},
       eprint = {1105.2563},
 primaryClass = {astro-ph.CO},
       adsurl = {https://ui.adsabs.harvard.edu/abs/2012MNRAS.419.3319M}
}

@ARTICLE{mccourt18,
       author = {{McCourt}, Michael and {Oh}, S. Peng and {O'Leary}, Ryan and {Madigan}, Ann-Marie},
        title = "{A characteristic scale for cold gas}",
      journal = {\mnras},
         year = 2018,
        month = feb,
       volume = {473},
       number = {4},
        pages = {5407-5431},
          doi = {10.1093/mnras/stx2687},
archivePrefix = {arXiv},
       eprint = {1610.01164},
 primaryClass = {astro-ph.GA},
       adsurl = {https://ui.adsabs.harvard.edu/abs/2018MNRAS.473.5407M}
}

@ARTICLE{McQuinn16,
   author = {{McQuinn}, M.},
    title = "{The Evolution of the Intergalactic Medium}",
  journal = {\araa},
archivePrefix = "arXiv",
   eprint = {1512.00086},
     year = 2016,
    month = sep,
   volume = 54,
    pages = {313-362},
      doi = {10.1146/annurev-astro-082214-122355},
   adsurl = {http://adsabs.harvard.edu/abs/2016ARA%26A..54..313M}
}

@ARTICLE{meece15,
       author = {{Meece}, Gregory R. and {O'Shea}, Brian W. and {Voit}, G. Mark},
        title = "{Growth and Evolution of Thermal Instabilities in Idealized Galaxy Cluster Cores}",
      journal = {\apj},
         year = 2015,
        month = jul,
       volume = {808},
       number = {1},
          eid = {43},
        pages = {43},
          doi = {10.1088/0004-637X/808/1/43},
archivePrefix = {arXiv},
       eprint = {1503.02645},
 primaryClass = {astro-ph.CO},
       adsurl = {https://ui.adsabs.harvard.edu/abs/2015ApJ...808...43M}
}

@ARTICLE{narayan92,
       author = {{Narayan}, Ramesh},
        title = "{The Physics of Pulsar Scintillation}",
      journal = {Philosophical Transactions of the Royal Society of London Series A},
         year = 1992,
        month = oct,
       volume = {341},
       number = {1660},
        pages = {151-165},
          doi = {10.1098/rsta.1992.0090},
       adsurl = {https://ui.adsabs.harvard.edu/abs/1992RSPTA.341..151N}
}

@ARTICLE{nimmo25,
       author = {{Nimmo}, Kenzie and {Pleunis}, Ziggy and {Beniamini}, Paz and {Kumar}, Pawan and {Lanman}, Adam E. and {Li}, D.~Z. and {Main}, Robert and {Sammons}, Mawson W. and {Andrew}, Shion and {Bhardwaj}, Mohit and {Chatterjee}, Shami and {Curtin}, Alice P. and {Fonseca}, Emmanuel and {Gaensler}, B.~M. and {Joseph}, Ronniy C. and {Kader}, Zarif and {Kaspi}, Victoria M. and {Lazda}, Mattias and {Leung}, Calvin and {Masui}, Kiyoshi W. and {Mckinven}, Ryan and {Michilli}, Daniele and {Pandhi}, Ayush and {Pearlman}, Aaron B. and {Rafiei-Ravandi}, Masoud and {Sand}, Ketan R. and {Shin}, Kaitlyn and {Smith}, Kendrick and {Stairs}, Ingrid H.},
        title = "{Magnetospheric origin of a fast radio burst constrained using scintillation}",
      journal = {\nat},
         year = 2025,
        month = jan,
       volume = {637},
       number = {8044},
        pages = {48-51},
          doi = {10.1038/s41586-024-08297-w},
archivePrefix = {arXiv},
       eprint = {2406.11053},
 primaryClass = {astro-ph.HE},
       adsurl = {https://ui.adsabs.harvard.edu/abs/2025Natur.637...48N}
}

@ARTICLE{ocker22-20190520,
       author = {{Ocker}, Stella Koch and {Cordes}, James M. and {Chatterjee}, Shami and {Niu}, Chen-Hui and {Li}, Di and {McKee}, James W. and {Law}, Casey J. and {Tsai}, Chao-Wei and {Anna-Thomas}, Reshma and {Yao}, Ju-Mei and {Cruces}, Marilyn},
        title = "{The Large Dispersion and Scattering of FRB 20190520B Are Dominated by the Host Galaxy}",
      journal = {\apj},
         year = 2022,
        month = jun,
       volume = {931},
       number = {2},
          eid = {87},
        pages = {87},
          doi = {10.3847/1538-4357/ac6504},
archivePrefix = {arXiv},
       eprint = {2202.13458},
 primaryClass = {astro-ph.HE},
       adsurl = {https://ui.adsabs.harvard.edu/abs/2022ApJ...931...87O}
}

@ARTICLE{ocker22-horizons,
       author = {{Ocker}, Stella Koch and {Cordes}, James M. and {Chatterjee}, Shami and {Gorsuch}, Miranda R.},
        title = "{Radio Scattering Horizons for Galactic and Extragalactic Transients}",
      journal = {\apj},
         year = 2022,
        month = jul,
       volume = {934},
       number = {1},
          eid = {71},
        pages = {71},
          doi = {10.3847/1538-4357/ac75ba},
archivePrefix = {arXiv},
       eprint = {2203.16716},
 primaryClass = {astro-ph.GA},
       adsurl = {https://ui.adsabs.harvard.edu/abs/2022ApJ...934...71O}
}

@ARTICLE{ocker25,
       author = {{Ocker}, Stella Koch and {Chen}, Mandy C. and {Oh}, S. Peng and {Sharma}, Prateek},
        title = "{Microphysics of Circumgalactic Turbulence Probed by Fast Radio Bursts and Quasars}",
      journal = {\apj},
         year = 2025,
        month = jul,
       volume = {988},
       number = {1},
          eid = {69},
        pages = {69},
          doi = {10.3847/1538-4357/ade0bc},
archivePrefix = {arXiv},
       eprint = {2503.02329},
 primaryClass = {astro-ph.GA},
       adsurl = {https://ui.adsabs.harvard.edu/abs/2025ApJ...988...69O}
}

@ARTICLE{Pasha23,
       author = {{Pasha}, Imad and {Mandelker}, Nir and {van den Bosch}, Frank C. and {Springel}, Volker and {van de Voort}, Freeke},
        title = "{Quenching in cosmic sheets: tracing the impact of large-scale structure collapse on the evolution of dwarf galaxies}",
      journal = {\mnras},
         year = 2023,
        month = apr,
       volume = {520},
       number = {2},
        pages = {2692-2708},
          doi = {10.1093/mnras/stac3776},
archivePrefix = {arXiv},
       eprint = {2204.04097},
 primaryClass = {astro-ph.GA},
       adsurl = {https://ui.adsabs.harvard.edu/abs/2023MNRAS.520.2692P}
}

@ARTICLE{Peeples14,
       author = {{Peeples}, Molly S. and {Werk}, Jessica K. and {Tumlinson}, Jason and {Oppenheimer}, Benjamin D. and {Prochaska}, J. Xavier and {Katz}, Neal and {Weinberg}, David H.},
        title = "{A Budget and Accounting of Metals at z \raisebox{-0.5ex}\textasciitilde 0: Results from the COS-Halos Survey}",
      journal = {\apj},
         year = 2014,
        month = may,
       volume = {786},
       number = {1},
          eid = {54},
        pages = {54},
          doi = {10.1088/0004-637X/786/1/54},
archivePrefix = {arXiv},
       eprint = {1310.2253},
 primaryClass = {astro-ph.CO},
       adsurl = {https://ui.adsabs.harvard.edu/abs/2014ApJ...786...54P}
}

@ARTICLE{Peeples19,
       author = {{Peeples}, Molly S. and {Corlies}, Lauren and {Tumlinson}, Jason and {O'Shea}, Brian W. and {Lehner}, Nicolas and {O'Meara}, John M. and {Howk}, J. Christopher and {Earl}, Nicholas and {Smith}, Britton D. and {Wise}, John H. and {Hummels}, Cameron B.},
        title = "{Figuring Out Gas \& Galaxies in Enzo (FOGGIE). I. Resolving Simulated Circumgalactic Absorption at 2 {\ensuremath{\leq}} z {\ensuremath{\leq}} 2.5}",
      journal = {\apj},
         year = 2019,
        month = mar,
       volume = {873},
       number = {2},
          eid = {129},
        pages = {129},
          doi = {10.3847/1538-4357/ab0654},
archivePrefix = {arXiv},
       eprint = {1810.06566},
 primaryClass = {astro-ph.GA},
       adsurl = {https://ui.adsabs.harvard.edu/abs/2019ApJ...873..129P}
}

@ARTICLE{Pleunis21,
       author = {{Pleunis}, Ziggy and {Good}, Deborah C. and {Kaspi}, Victoria M. and {Mckinven}, Ryan and {Ransom}, Scott M. and {Scholz}, Paul and {Bandura}, Kevin and {Bhardwaj}, Mohit and {Boyle}, P.~J. and {Brar}, Charanjot and {Cassanelli}, Tomas and {Chawla}, Pragya and {(Adam) Dong}, Fengqiu and {Fonseca}, Emmanuel and {Gaensler}, B.~M. and {Josephy}, Alexander and {Kaczmarek}, Jane F. and {Leung}, Calvin and {Lin}, Hsiu-Hsien and {Masui}, Kiyoshi W. and {Mena-Parra}, Juan and {Michilli}, Daniele and {Ng}, Cherry and {Patel}, Chitrang and {Rafiei-Ravandi}, Masoud and {Rahman}, Mubdi and {Sanghavi}, Pranav and {Shin}, Kaitlyn and {Smith}, Kendrick M. and {Stairs}, Ingrid H. and {Tendulkar}, Shriharsh P.},
        title = "{Fast Radio Burst Morphology in the First CHIME/FRB Catalog}",
      journal = {\apj},
         year = 2021,
        month = dec,
       volume = {923},
       number = {1},
          eid = {1},
        pages = {1},
          doi = {10.3847/1538-4357/ac33ac},
archivePrefix = {arXiv},
       eprint = {2106.04356},
 primaryClass = {astro-ph.HE},
       adsurl = {https://ui.adsabs.harvard.edu/abs/2021ApJ...923....1P}
}

@ARTICLE{prochaska17,
       author = {{Prochaska}, J. Xavier and {Werk}, Jessica K. and {Worseck}, G{\'a}bor and {Tripp}, Todd M. and {Tumlinson}, Jason and {Burchett}, Joseph N. and {Fox}, Andrew J. and {Fumagalli}, Michele and {Lehner}, Nicolas and {Peeples}, Molly S. and {Tejos}, Nicolas},
        title = "{The COS-Halos Survey: Metallicities in the Low-redshift Circumgalactic Medium}",
      journal = {\apj},
         year = 2017,
        month = mar,
       volume = {837},
       number = {2},
          eid = {169},
        pages = {169},
          doi = {10.3847/1538-4357/aa6007},
archivePrefix = {arXiv},
       eprint = {1702.02618},
 primaryClass = {astro-ph.GA},
       adsurl = {https://ui.adsabs.harvard.edu/abs/2017ApJ...837..169P}
}

@ARTICLE{prochaska19-sci,
       author = {{Prochaska}, J. Xavier and {Macquart}, Jean-Pierre and {McQuinn}, Matthew and {Simha}, Sunil and {Shannon}, Ryan M. and {Day}, Cherie K. and {Marnoch}, Lachlan and {Ryder}, Stuart and {Deller}, Adam and {Bannister}, Keith W. and {Bhandari}, Shivani and {Bordoloi}, Rongmon and {Bunton}, John and {Cho}, Hyerin and {Flynn}, Chris and {Mahony}, Elizabeth K. and {Phillips}, Chris and {Qiu}, Hao and {Tejos}, Nicolas},
        title = "{The low density and magnetization of a massive galaxy halo exposed by a fast radio burst}",
      journal = {Science},
         year = 2019,
        month = oct,
       volume = {366},
       number = {6462},
        pages = {231-234},
          doi = {10.1126/science.aay0073},
archivePrefix = {arXiv},
       eprint = {1909.11681},
 primaryClass = {astro-ph.GA},
       adsurl = {https://ui.adsabs.harvard.edu/abs/2019Sci...366..231P}
}

@ARTICLE{Putman2012,
       author = {{Putman}, M.~E. and {Peek}, J.~E.~G. and {Joung}, M.~R.},
        title = "{Gaseous Galaxy Halos}",
      journal = {\araa},
         year = 2012,
        month = sep,
       volume = {50},
        pages = {491-529},
          doi = {10.1146/annurev-astro-081811-125612},
archivePrefix = {arXiv},
       eprint = {1207.4837},
 primaryClass = {astro-ph.GA},
       adsurl = {https://ui.adsabs.harvard.edu/abs/2012ARA&A..50..491P}
}

@ARTICLE{rafiei21_dm_z_cf_chime_cat1,
       author = {{Rafiei-Ravandi}, Masoud and {Smith}, Kendrick M. and {Li}, Dongzi and {Masui}, Kiyoshi W. and {Josephy}, Alexander and {Dobbs}, Matt and {Lang}, Dustin and {Bhardwaj}, Mohit and {Patel}, Chitrang and {Bandura}, Kevin and {Berger}, Sabrina and {Boyle}, P.~J. and {Brar}, Charanjot and {Breitman}, Daniela and {Cassanelli}, Tomas and {Chawla}, Pragya and {Adam Dong}, Fengqiu and {Fonseca}, Emmanuel and {Gaensler}, B.~M. and {Giri}, Utkarsh and {Good}, Deborah C. and {Halpern}, Mark and {Kaczmarek}, Jane and {Kaspi}, Victoria M. and {Leung}, Calvin and {Lin}, Hsiu-Hsien and {Mena-Parra}, Juan and {Meyers}, B.~W. and {Michilli}, D. and {M{\"u}nchmeyer}, Moritz and {Ng}, Cherry and {Petroff}, Emily and {Pleunis}, Ziggy and {Rahman}, Mubdi and {Sanghavi}, Pranav and {Scholz}, Paul and {Shin}, Kaitlyn and {Stairs}, Ingrid H. and {Tendulkar}, Shriharsh P. and {Vanderlinde}, Keith and {Zwaniga}, Andrew},
        title = "{CHIME/FRB Catalog 1 Results: Statistical Cross-correlations with Large-scale Structure}",
      journal = {\apj},
         year = 2021,
        month = nov,
       volume = {922},
       number = {1},
          eid = {42},
        pages = {42},
          doi = {10.3847/1538-4357/ac1dab},
archivePrefix = {arXiv},
       eprint = {2106.04354},
 primaryClass = {astro-ph.CO},
       adsurl = {https://ui.adsabs.harvard.edu/abs/2021ApJ...922...42R}
}

@ARTICLE{rahmati13,
       author = {{Rahmati}, Alireza and {Pawlik}, Andreas H. and {Rai{\v{c}}evi{\'c}}, Milan and {Schaye}, Joop},
        title = "{On the evolution of the H I column density distribution in cosmological simulations}",
      journal = {\mnras},
         year = 2013,
        month = apr,
       volume = {430},
       number = {3},
        pages = {2427-2445},
          doi = {10.1093/mnras/stt066},
archivePrefix = {arXiv},
       eprint = {1210.7808},
 primaryClass = {astro-ph.CO},
       adsurl = {https://ui.adsabs.harvard.edu/abs/2013MNRAS.430.2427R}
}

@ARTICLE{Rauch98,
   author = {{Rauch}, M.},
    title = "{The Lyman Alpha Forest in the Spectra of QSOs}",
  journal = {\araa},
   eprint = {astro-ph/9806286},
     year = 1998,
   volume = 36,
    pages = {267-316},
      doi = {10.1146/annurev.astro.36.1.267},
   adsurl = {http://adsabs.harvard.edu/abs/1998ARA%26A..36..267R}
}

@ARTICLE{Rees77,
       author = {{Rees}, M.~J. and {Ostriker}, J.~P.},
        title = "{Cooling, dynamics and fragmentation of massive gas clouds: clues to the masses and radii of galaxies and clusters.}",
      journal = {\mnras},
         year = "1977",
        month = "Jun",
       volume = {179},
        pages = {541-559},
          doi = {10.1093/mnras/179.4.541},
       adsurl = {https://ui.adsabs.harvard.edu/abs/1977MNRAS.179..541R}
}

@ARTICLE{rickett77,
       author = {{Rickett}, B.~J.},
        title = "{Interstellar scattering and scintillation of radio waves.}",
      journal = {\araa},
         year = 1977,
        month = jan,
       volume = {15},
        pages = {479-504},
          doi = {10.1146/annurev.aa.15.090177.002403},
       adsurl = {https://ui.adsabs.harvard.edu/abs/1977ARA&A..15..479R}
}

@ARTICLE{rickett90,
       author = {{Rickett}, B.~J.},
        title = "{Radio propagation through the turbulent interstellar plasma.}",
      journal = {\araa},
         year = 1990,
        month = jan,
       volume = {28},
        pages = {561-605},
          doi = {10.1146/annurev.aa.28.090190.003021},
       adsurl = {https://ui.adsabs.harvard.edu/abs/1990ARA&A..28..561R}
}

@ARTICLE{Robert19,
       author = {{Robert}, P. Fr{\'e}d{\'e}ric and {Murphy}, Michael T. and {O'Meara}, John M. and {Crighton}, Neil H.~M. and {Fumagalli}, Michele},
        title = "{Exploring the origins of a new, apparently metal-free gas cloud at z = 4.4}",
      journal = {\mnras},
         year = 2019,
        month = feb,
       volume = {483},
       number = {2},
        pages = {2736-2747},
          doi = {10.1093/mnras/sty3287},
archivePrefix = {arXiv},
       eprint = {1812.05098},
 primaryClass = {astro-ph.GA},
       adsurl = {https://ui.adsabs.harvard.edu/abs/2019MNRAS.483.2736R}
}

@ARTICLE{Sameer24,
       author = {{Sameer} and {Charlton}, Jane C. and {Wakker}, Bart P. and {Kacprzak}, Glenn G. and {Nielsen}, Nikole M. and {Churchill}, Christopher W. and {Richter}, Philipp and {Muzahid}, Sowgat and {Ho}, Stephanie H. and {Nateghi}, Hasti and {Rosenwasser}, Benjamin and {Narayanan}, Anand and {Ganguly}, Rajib},
        title = "{Cloud-by-cloud multiphase investigation of the circumgalactic medium of low-redshift galaxies}",
      journal = {\mnras},
         year = 2024,
        month = jun,
       volume = {530},
       number = {4},
        pages = {3827-3854},
          doi = {10.1093/mnras/stae962},
archivePrefix = {arXiv},
       eprint = {2403.05617},
 primaryClass = {astro-ph.GA},
       adsurl = {https://ui.adsabs.harvard.edu/abs/2024MNRAS.530.3827S}
}

@ARTICLE{sammons23,
       author = {{Sammons}, Mawson W. and {Deller}, Adam T. and {Glowacki}, Marcin and {Gourdji}, Kelly and {James}, C.~W. and {Prochaska}, J. Xavier and {Qiu}, Hao and {Scott}, Danica R. and {Shannon}, R.~M. and {Trott}, C.~M.},
        title = "{Two-screen scattering in CRAFT FRBs}",
      journal = {\mnras},
         year = 2023,
        month = nov,
       volume = {525},
       number = {4},
        pages = {5653-5668},
          doi = {10.1093/mnras/stad2631},
archivePrefix = {arXiv},
       eprint = {2305.11477},
 primaryClass = {astro-ph.HE},
       adsurl = {https://ui.adsabs.harvard.edu/abs/2023MNRAS.525.5653S}
}

@BOOK{schneider92,
       author = {{Schneider}, Peter and {Ehlers}, J{\"u}rgen and {Falco}, Emilio E.},
        title = "{Gravitational Lenses}",
         year = 1992,
    publisher = {Springer Berlin, Heidelberg },
          doi = {10.1007/978-3-662-03758-4},
       adsurl = {https://ui.adsabs.harvard.edu/abs/1992grle.book.....S}
}

@ARTICLE{scott25,
       author = {{Scott}, Danica R. and {Dial}, Tyson and {Bera}, Apurba and {Deller}, Adam T. and {Glowacki}, Marcin and {Gourdji}, Kelly and {James}, Clancy W. and {Shannon}, Ryan M. and {Bannister}, Keith W. and {Ekers}, Ron D. and {Paterson}, Jasper and {Sammons}, Mawson and {Sutinjo}, Adrian T. and {Uttarkar}, Pavan A.},
        title = "{High-time-resolution properties of 35 fast radio bursts detected by the Commensal Real-time ASKAP Fast Transients survey}",
      journal = {\pasa},
         year = 2025,
        month = oct,
       volume = {42},
          eid = {e133},
        pages = {e133},
          doi = {10.1017/pasa.2025.10103},
archivePrefix = {arXiv},
       eprint = {2505.17497},
 primaryClass = {astro-ph.HE},
       adsurl = {https://ui.adsabs.harvard.edu/abs/2025PASA...42..133S}
}

@ARTICLE{sharma12,
       author = {{Sharma}, Prateek and {McCourt}, Michael and {Quataert}, Eliot and {Parrish}, Ian J.},
        title = "{Thermal instability and the feedback regulation of hot haloes in clusters, groups and galaxies}",
      journal = {\mnras},
         year = 2012,
        month = mar,
       volume = {420},
       number = {4},
        pages = {3174-3194},
          doi = {10.1111/j.1365-2966.2011.20246.x},
archivePrefix = {arXiv},
       eprint = {1106.4816},
 primaryClass = {astro-ph.CO},
       adsurl = {https://ui.adsabs.harvard.edu/abs/2012MNRAS.420.3174S}
}

@ARTICLE{shaw25,
       author = {{Shaw}, Abinash Kumar and {Ghara}, Raghunath and {Beniamini}, Paz and {Zaroubi}, Saleem and {Kumar}, Pawan},
        title = "{Asking Fast Radio Bursts for More than Reionization History}",
      journal = {\apj},
         year = 2025,
        month = nov,
       volume = {993},
       number = {2},
          eid = {209},
        pages = {209},
          doi = {10.3847/1538-4357/ae07c5},
archivePrefix = {arXiv},
       eprint = {2409.03255},
 primaryClass = {astro-ph.CO},
       adsurl = {https://ui.adsabs.harvard.edu/abs/2025ApJ...993..209S}
}

@ARTICLE{shen06,
       author = {{Shen}, Jiajian and {Abel}, Tom and {Mo}, H.~J. and {Sheth}, Ravi K.},
        title = "{An Excursion Set Model of the Cosmic Web: The Abundance of Sheets, Filaments, and Halos}",
      journal = {\apj},
         year = 2006,
        month = jul,
       volume = {645},
       number = {2},
        pages = {783-791},
          doi = {10.1086/504513},
archivePrefix = {arXiv},
       eprint = {astro-ph/0511365},
 primaryClass = {astro-ph},
       adsurl = {https://ui.adsabs.harvard.edu/abs/2006ApJ...645..783S}
}

@ARTICLE{shin25,
       author = {{Shin}, Kaitlyn and {Leung}, Calvin and {Simha}, Sunil and {Andersen}, Bridget C. and {Fonseca}, Emmanuel and {Nimmo}, Kenzie and {Bhardwaj}, Mohit and {Brar}, Charanjot and {Chatterjee}, Shami and {Cook}, Amanda M. and {Gaensler}, B.~M. and {Joseph}, Ronniy C. and {Jow}, Dylan and {Kaczmarek}, Jane and {Kahinga}, Lordrick and {Kaspi}, Victoria M. and {Kharel}, Bikash and {Lanman}, Adam E. and {Lazda}, Mattias and {Main}, Robert A. and {Mas-Ribas}, Lluis and {Masui}, Kiyoshi W. and {Mena-Parra}, Juan and {Michilli}, Daniele and {Pandhi}, Ayush and {Patil}, Swarali Shivraj and {Pearlman}, Aaron B. and {Pleunis}, Ziggy and {Prochaska}, J. Xavier and {Rafiei-Ravandi}, Masoud and {Sammons}, Mawson W. and {Sand}, Ketan R. and {Smith}, Kendrick and {Stairs}, Ingrid},
        title = "{Investigating the Sightline of a Highly Scattered Fast Radio Burst through a Cosmic Sheet Structure in the Local Universe}",
      journal = {\apj},
         year = 2025,
        month = nov,
       volume = {993},
       number = {2},
          eid = {208},
        pages = {208},
          doi = {10.3847/1538-4357/ae093b},
archivePrefix = {arXiv},
       eprint = {2410.07307},
 primaryClass = {astro-ph.HE},
       adsurl = {https://ui.adsabs.harvard.edu/abs/2025ApJ...993..208S}
}

@ARTICLE{Springel05,
       author = {{Springel}, Volker and {White}, Simon D.~M. and {Jenkins}, Adrian and {Frenk}, Carlos S. and {Yoshida}, Naoki and {Gao}, Liang and {Navarro}, Julio and {Thacker}, Robert and {Croton}, Darren and {Helly}, John and {Peacock}, John A. and {Cole}, Shaun and {Thomas}, Peter and {Couchman}, Hugh and {Evrard}, August and {Colberg}, J{\"o}rg and {Pearce}, Frazer},
        title = "{Simulations of the formation, evolution and clustering of galaxies and quasars}",
      journal = {\nat},
         year = 2005,
        month = jun,
       volume = {435},
       number = {7042},
        pages = {629-636},
          doi = {10.1038/nature03597},
archivePrefix = {arXiv},
       eprint = {astro-ph/0504097},
 primaryClass = {astro-ph},
       adsurl = {https://ui.adsabs.harvard.edu/abs/2005Natur.435..629S}
}

@ARTICLE{Stanimirovic.Zweibel.2018,
       author = {{Stanimirovi{\'c}}, Sne{\v{z}}ana and {Zweibel}, Ellen G.},
        title = "{Atomic and Ionized Microstructures in the Diffuse Interstellar Medium}",
      journal = {\araa},
         year = 2018,
        month = sep,
       volume = {56},
        pages = {489-540},
          doi = {10.1146/annurev-astro-081817-051810},
archivePrefix = {arXiv},
       eprint = {1810.00933},
 primaryClass = {astro-ph.GA},
       adsurl = {https://ui.adsabs.harvard.edu/abs/2018ARA&A..56..489S}
}

@ARTICLE{Steidel00,
       author = {{Steidel}, Charles C. and {Adelberger}, Kurt L. and {Shapley}, Alice E. and {Pettini}, Max and {Dickinson}, Mark and {Giavalisco}, Mauro},
        title = "{Ly{\ensuremath{\alpha}} Imaging of a Proto-Cluster Region at <z>=3.09}",
      journal = {\apj},
         year = 2000,
        month = mar,
       volume = {532},
       number = {1},
        pages = {170-182},
          doi = {10.1086/308568},
archivePrefix = {arXiv},
       eprint = {astro-ph/9910144},
 primaryClass = {astro-ph},
       adsurl = {https://ui.adsabs.harvard.edu/abs/2000ApJ...532..170S}
}

@ARTICLE{Steidel10,
   author = {{Steidel}, C.~C. and {Erb}, D.~K. and {Shapley}, A.~E. and {Pettini}, M. and 
	{Reddy}, N. and {Bogosavljevi{\'c}}, M. and {Rudie}, G.~C. and 
	{Rakic}, O.},
    title = "{The Structure and Kinematics of the Circumgalactic Medium from Far-ultraviolet Spectra of z \~{}= 2-3 Galaxies}",
  journal = {\apj},
archivePrefix = "arXiv",
   eprint = {1003.0679},
 primaryClass = "astro-ph.CO",
     year = 2010,
    month = jul,
   volume = 717,
    pages = {289-322},
      doi = {10.1088/0004-637X/717/1/289},
   adsurl = {http://adsabs.harvard.edu/abs/2010ApJ...717..289S}
}

@ARTICLE{Stern21,
       author = {{Stern}, Jonathan and {Faucher-Gigu{\`e}re}, Claude-Andr{\'e} and {Fielding}, Drummond and {Quataert}, Eliot and {Hafen}, Zachary and {Gurvich}, Alexander B. and {Ma}, Xiangcheng and {Byrne}, Lindsey and {El-Badry}, Kareem and {Angl{\'e}s-Alc{\'a}zar}, Daniel and {Chan}, T.~K. and {Feldmann}, Robert and {Kere{\v{s}}}, Du{\v{s}}an and {Wetzel}, Andrew and {Murray}, Norman and {Hopkins}, Philip F.},
        title = "{Virialization of the Inner CGM in the FIRE Simulations and Implications for Galaxy Disks, Star Formation, and Feedback}",
      journal = {\apj},
         year = 2021,
        month = apr,
       volume = {911},
       number = {2},
          eid = {88},
        pages = {88},
          doi = {10.3847/1538-4357/abd776},
archivePrefix = {arXiv},
       eprint = {2006.13976},
 primaryClass = {astro-ph.GA},
       adsurl = {https://ui.adsabs.harvard.edu/abs/2021ApJ...911...88S}
}

@ARTICLE{Suresh19,
       author = {{Suresh}, Joshua and {Nelson}, Dylan and {Genel}, Shy and {Rubin}, Kate H.~R. and {Hernquist}, Lars},
        title = "{Zooming in on accretion - II. Cold circumgalactic gas simulated with a super-Lagrangian refinement scheme}",
      journal = {\mnras},
         year = 2019,
        month = mar,
       volume = {483},
       number = {3},
        pages = {4040-4059},
          doi = {10.1093/mnras/sty3402},
archivePrefix = {arXiv},
       eprint = {1811.01949},
 primaryClass = {astro-ph.GA},
       adsurl = {https://ui.adsabs.harvard.edu/abs/2019MNRAS.483.4040S}
}

@ARTICLE{tegmark04,
       author = {{Tegmark}, Max and {Strauss}, Michael A. and {Blanton}, Michael R. and {Abazajian}, Kevork and {Dodelson}, Scott and {Sandvik}, Havard and {Wang}, Xiaomin and {Weinberg}, David H. and {Zehavi}, Idit and {Bahcall}, Neta A. and {Hoyle}, Fiona and {Schlegel}, David and {Scoccimarro}, Roman and {Vogeley}, Michael S. and {Berlind}, Andreas and {Budavari}, Tam{\'a}s and {Connolly}, Andrew and {Eisenstein}, Daniel J. and {Finkbeiner}, Douglas and {Frieman}, Joshua A. and {Gunn}, James E. and {Hui}, Lam and {Jain}, Bhuvnesh and {Johnston}, David and {Kent}, Stephen and {Lin}, Huan and {Nakajima}, Reiko and {Nichol}, Robert C. and {Ostriker}, Jeremiah P. and {Pope}, Adrian and {Scranton}, Ryan and {Seljak}, Uro{\v{s}} and {Sheth}, Ravi K. and {Stebbins}, Albert and {Szalay}, Alexander S. and {Szapudi}, Istv{\'a}n and {Xu}, Yongzhong and {Annis}, James and {Brinkmann}, J. and {Burles}, Scott and {Castander}, Francisco J. and {Csabai}, Istvan and {Loveday}, Jon and {Doi}, Mamoru and {Fukugita}, Masataka and {Gillespie}, Bruce and {Hennessy}, Greg and {Hogg}, David W. and {Ivezi{\'c}}, {\v{Z}}eljko and {Knapp}, Gillian R. and {Lamb}, Don Q. and {Lee}, Brian C. and {Lupton}, Robert H. and {McKay}, Timothy A. and {Kunszt}, Peter and {Munn}, Jeffrey A. and {O'Connell}, Liam and {Peoples}, John and {Pier}, Jeffrey R. and {Richmond}, Michael and {Rockosi}, Constance and {Schneider}, Donald P. and {Stoughton}, Christopher and {Tucker}, Douglas L. and {vanden Berk}, Daniel E. and {Yanny}, Brian and {York}, Donald G.},
        title = "{Cosmological parameters from SDSS and WMAP}",
      journal = {\prd},
         year = 2004,
        month = may,
       volume = {69},
       number = {10},
          eid = {103501},
        pages = {103501},
          doi = {10.1103/PhysRevD.69.103501},
archivePrefix = {arXiv},
       eprint = {astro-ph/0310723},
 primaryClass = {astro-ph},
       adsurl = {https://ui.adsabs.harvard.edu/abs/2004PhRvD..69j3501T}
}

@ARTICLE{Teyssier.02,
       author = {{Teyssier}, R.},
        title = "{Cosmological hydrodynamics with adaptive mesh refinement. A new high resolution code called RAMSES}",
      journal = {\aap},
         year = 2002,
        month = apr,
       volume = {385},
        pages = {337-364},
          doi = {10.1051/0004-6361:20011817},
archivePrefix = {arXiv},
       eprint = {astro-ph/0111367},
 primaryClass = {astro-ph},
       adsurl = {https://ui.adsabs.harvard.edu/abs/2002A&A...385..337T}
}

@ARTICLE{tornotti25a,
       author = {{Tornotti}, Davide and {Fumagalli}, Michele and {Fossati}, Matteo and {Arrigoni Battaia}, Fabrizio and {Benitez-Llambay}, Alejandro and {Dayal}, Pratika and {Dutta}, Rajeshwari and {Peroux}, Celine and {Rafelski}, Marc and {Revalski}, Mitchell},
        title = "{The MUSE Ultra Deep Field: A 5 Mpc Stretch of the z ≍ 4 Cosmic Web Revealed in Emission}",
      journal = {\apjl},
         year = 2025,
        month = feb,
       volume = {980},
       number = {2},
          eid = {L43},
        pages = {L43},
          doi = {10.3847/2041-8213/adb0ba},
archivePrefix = {arXiv},
       eprint = {2412.06895},
 primaryClass = {astro-ph.GA},
       adsurl = {https://ui.adsabs.harvard.edu/abs/2025ApJ...980L..43T}
}

@ARTICLE{tornotti25b,
       author = {{Tornotti}, Davide and {Fumagalli}, Michele and {Fossati}, Matteo and {Benitez-Llambay}, Alejandro and {Izquierdo-Villalba}, David and {Travascio}, Andrea and {Arrigoni Battaia}, Fabrizio and {Cantalupo}, Sebastiano and {Beckett}, Alexander and {Bonoli}, Silvia and {Dayal}, Pratika and {D'Odorico}, Valentina and {Dutta}, Rajeshwari and {Lusso}, Elisabeta and {Peroux}, Celine and {Rafelski}, Marc and {Revalski}, Mitchell and {Spinoso}, Daniele and {Swinbank}, Mark},
        title = "{High-definition imaging of a filamentary connection between a close quasar pair at z = 3}",
      journal = {Nature Astronomy},
         year = 2025,
        month = apr,
       volume = {9},
        pages = {577-588},
          doi = {10.1038/s41550-024-02463-w},
archivePrefix = {arXiv},
       eprint = {2406.17035},
 primaryClass = {astro-ph.CO},
       adsurl = {https://ui.adsabs.harvard.edu/abs/2025NatAs...9..577T}
}

@ARTICLE{Tumlinson2017,
       author = {{Tumlinson}, Jason and {Peeples}, Molly S. and {Werk}, Jessica K.},
        title = "{The Circumgalactic Medium}",
      journal = {\araa},
         year = 2017,
        month = aug,
       volume = {55},
       number = {1},
        pages = {389-432},
          doi = {10.1146/annurev-astro-091916-055240},
archivePrefix = {arXiv},
       eprint = {1709.09180},
 primaryClass = {astro-ph.GA},
       adsurl = {https://ui.adsabs.harvard.edu/abs/2017ARA&A..55..389T}
}

@ARTICLE{Umehata19,
       author = {{Umehata}, H. and {Fumagalli}, M. and {Smail}, I. and {Matsuda}, Y. and
         {Swinbank}, A.~M. and {Cantalupo}, S. and {Sykes}, C. and
         {Ivison}, R.~J. and {Steidel}, C.~C. and {Shapley}, A.~E. and
         {Vernet}, J. and {Yamada}, T. and {Tamura}, Y. and {Kubo}, M. and
         {Nakanishi}, K. and {Kajisawa}, M. and {Hatsukade}, B. and {Kohno}, K.},
        title = "{Gas filaments of the cosmic web located around active galaxies in a protocluster}",
      journal = {Science},
         year = 2019,
        month = oct,
       volume = {366},
       number = {6461},
        pages = {97-100},
          doi = {10.1126/science.aaw5949},
archivePrefix = {arXiv},
       eprint = {1910.01324},
 primaryClass = {astro-ph.GA},
       adsurl = {https://ui.adsabs.harvard.edu/abs/2019Sci...366...97U}
}

@ARTICLE{vandeVoort19,
       author = {{van de Voort}, Freeke and {Springel}, Volker and {Mandelker}, Nir and {van den Bosch}, Frank C. and {Pakmor}, R{\"u}diger},
        title = "{Cosmological simulations of the circumgalactic medium with 1 kpc resolution: enhanced H I column densities}",
      journal = {\mnras},
         year = 2019,
        month = jan,
       volume = {482},
       number = {1},
        pages = {L85-L89},
          doi = {10.1093/mnrasl/sly190},
archivePrefix = {arXiv},
       eprint = {1808.04369},
 primaryClass = {astro-ph.GA},
       adsurl = {https://ui.adsabs.harvard.edu/abs/2019MNRAS.482L..85V}
}

@ARTICLE{vedantham19,
       author = {{Vedantham}, H.~K. and {Phinney}, E.~S.},
        title = "{Radio wave scattering by circumgalactic cool gas clumps}",
      journal = {\mnras},
         year = 2019,
        month = feb,
       volume = {483},
       number = {1},
        pages = {971-984},
          doi = {10.1093/mnras/sty2948},
archivePrefix = {arXiv},
       eprint = {1811.10876},
 primaryClass = {astro-ph.GA},
       adsurl = {https://ui.adsabs.harvard.edu/abs/2019MNRAS.483..971V}
}

@ARTICLE{Viel13,
       author = {{Viel}, Matteo and {Becker}, George D. and {Bolton}, James S. and
         {Haehnelt}, Martin G.},
        title = "{Warm dark matter as a solution to the small scale crisis: New constraints from high redshift Lyman-{\ensuremath{\alpha}} forest data}",
      journal = {\prd},
         year = 2013,
        month = aug,
       volume = {88},
       number = {4},
          eid = {043502},
        pages = {043502},
          doi = {10.1103/PhysRevD.88.043502},
archivePrefix = {arXiv},
       eprint = {1306.2314},
 primaryClass = {astro-ph.CO},
       adsurl = {https://ui.adsabs.harvard.edu/abs/2013PhRvD..88d3502V}
}

@ARTICLE{voit-donahue15,
       author = {{Voit}, G. Mark and {Donahue}, Megan},
        title = "{Cooling Time, Freefall Time, and Precipitation in the Cores of ACCEPT Galaxy Clusters}",
      journal = {\apjl},
         year = 2015,
        month = jan,
       volume = {799},
       number = {1},
          eid = {L1},
        pages = {L1},
          doi = {10.1088/2041-8205/799/1/L1},
archivePrefix = {arXiv},
       eprint = {1409.1601},
 primaryClass = {astro-ph.GA},
       adsurl = {https://ui.adsabs.harvard.edu/abs/2015ApJ...799L...1V}
}

@ARTICLE{Wadiasingh2019,
       author = {{Wadiasingh}, Zorawar and {Timokhin}, Andrey},
        title = "{Repeating Fast Radio Bursts from Magnetars with Low Magnetospheric Twist}",
      journal = {\apj},
         year = 2019,
        month = jul,
       volume = {879},
       number = {1},
          eid = {4},
        pages = {4},
          doi = {10.3847/1538-4357/ab2240},
archivePrefix = {arXiv},
       eprint = {1904.12036},
 primaryClass = {astro-ph.HE},
       adsurl = {https://ui.adsabs.harvard.edu/abs/2019ApJ...879....4W}
}

@ARTICLE{Wang12,
       author = {{Wang}, Huiyuan and {Mo}, H.~J. and {Yang}, Xiaohu and {van den Bosch}, Frank C.},
        title = "{Reconstructing the cosmic velocity and tidal fields with galaxy groups selected from the Sloan Digital Sky Survey}",
      journal = {\mnras},
         year = 2012,
        month = feb,
       volume = {420},
       number = {2},
        pages = {1809-1824},
          doi = {10.1111/j.1365-2966.2011.20174.x},
archivePrefix = {arXiv},
       eprint = {1108.1008},
 primaryClass = {astro-ph.CO},
       adsurl = {https://ui.adsabs.harvard.edu/abs/2012MNRAS.420.1809W}
}

@ARTICLE{wang25_dm_z_cf_chime_cat2,
       author = {{Wang}, Haochen and {Masui}, Kiyoshi and {Andrew}, Shion and {Fonseca}, Emmanuel and {Gaensler}, B.~M. and {Joseph}, R.~C. and {Kaspi}, Victoria M. and {Kharel}, Bikash and {Lanman}, Adam E. and {Leung}, Calvin and {Mas-Ribas}, Lluis and {Mena-Parra}, Juan and {Nimmo}, Kenzie and {Pearlman}, Aaron B. and {Pen}, Ue-Li and {Prochaska}, J. Xavier and {Raikman}, Ryan and {Shin}, Kaitlyn and {Siegel}, Seth R. and {Smith}, Kendrick M. and {Stairs}, Ingrid H.},
        title = "{Measurement of the Dispersion-Galaxy Cross-Power Spectrum with the Second CHIME/FRB Catalog}",
      journal = {arXiv e-prints},
         year = 2025,
        month = jun,
          eid = {arXiv:2506.08932},
        pages = {arXiv:2506.08932},
          doi = {10.48550/arXiv.2506.08932},
archivePrefix = {arXiv},
       eprint = {2506.08932},
 primaryClass = {astro-ph.CO},
       adsurl = {https://ui.adsabs.harvard.edu/abs/2025arXiv250608932W}
}

@ARTICLE{Wechsler_Tinker18,
   author = {{Wechsler}, R.~H. and {Tinker}, J.~L.},
    title = "{The Connection Between Galaxies and Their Dark Matter Halos}",
  journal = {\araa},
archivePrefix = "arXiv",
   eprint = {1804.03097},
     year = 2018,
    month = sep,
   volume = 56,
    pages = {435-487},
      doi = {10.1146/annurev-astro-081817-051756},
   adsurl = {http://adsabs.harvard.edu/abs/2018ARA%26A..56..435W}
}

@ARTICLE{werk14,
       author = {{Werk}, Jessica K. and {Prochaska}, J. Xavier and {Tumlinson}, Jason and {Peeples}, Molly S. and {Tripp}, Todd M. and {Fox}, Andrew J. and {Lehner}, Nicolas and {Thom}, Christopher and {O'Meara}, John M. and {Ford}, Amanda Brady and {Bordoloi}, Rongmon and {Katz}, Neal and {Tejos}, Nicolas and {Oppenheimer}, Benjamin D. and {Dav{\'e}}, Romeel and {Weinberg}, David H.},
        title = "{The COS-Halos Survey: Physical Conditions and Baryonic Mass in the Low-redshift Circumgalactic Medium}",
      journal = {\apj},
         year = 2014,
        month = sep,
       volume = {792},
       number = {1},
          eid = {8},
        pages = {8},
          doi = {10.1088/0004-637X/792/1/8},
archivePrefix = {arXiv},
       eprint = {1403.0947},
 primaryClass = {astro-ph.CO},
       adsurl = {https://ui.adsabs.harvard.edu/abs/2014ApJ...792....8W}
}

@ARTICLE{White78,
       author = {{White}, S.~D.~M. and {Rees}, M.~J.},
        title = "{Core condensation in heavy halos: a two-stage theory for galaxy formation and clustering.}",
      journal = {\mnras},
         year = "1978",
        month = "May",
       volume = {183},
        pages = {341-358},
          doi = {10.1093/mnras/183.3.341},
       adsurl = {https://ui.adsabs.harvard.edu/abs/1978MNRAS.183..341W}
}

@ARTICLE{yao17,
       author = {{Yao}, J.~M. and {Manchester}, R.~N. and {Wang}, N.},
        title = "{A New Electron-density Model for Estimation of Pulsar and FRB Distances}",
      journal = {\apj},
         year = 2017,
        month = jan,
       volume = {835},
       number = {1},
          eid = {29},
        pages = {29},
          doi = {10.3847/1538-4357/835/1/29},
archivePrefix = {arXiv},
       eprint = {1610.09448},
 primaryClass = {astro-ph.GA},
       adsurl = {https://ui.adsabs.harvard.edu/abs/2017ApJ...835...29Y}
}

@ARTICLE{Yao25,
       author = {{Yao}, Zhiyuan and {Mandelker}, Nir and {Oh}, S. Peng and {Aung}, Han and {Dekel}, Avishai},
        title = "{Effects of cloud geometry and metallicity on shattering and coagulation of cold gas, and implications for cold streams penetrating virial shocks}",
      journal = {\mnras},
         year = 2025,
        month = jan,
       volume = {536},
       number = {3},
        pages = {3053-3089},
          doi = {10.1093/mnras/stae2771},
archivePrefix = {arXiv},
       eprint = {2410.12914},
 primaryClass = {astro-ph.GA},
       adsurl = {https://ui.adsabs.harvard.edu/abs/2025MNRAS.536.3053Y}
}

@ARTICLE{zahedy19,
       author = {{Zahedy}, Fakhri S. and {Chen}, Hsiao-Wen and {Johnson}, Sean D. and {Pierce}, Rebecca M. and {Rauch}, Michael and {Huang}, Yun-Hsin and {Weiner}, Benjamin J. and {Gauthier}, Jean-Ren{\'e}},
        title = "{Characterizing circumgalactic gas around massive ellipticals at z {\ensuremath{\sim}} 0.4 - II. Physical properties and elemental abundances}",
      journal = {\mnras},
         year = 2019,
        month = apr,
       volume = {484},
       number = {2},
        pages = {2257-2280},
          doi = {10.1093/mnras/sty3482},
archivePrefix = {arXiv},
       eprint = {1809.05115},
 primaryClass = {astro-ph.GA},
       adsurl = {https://ui.adsabs.harvard.edu/abs/2019MNRAS.484.2257Z}
}

@ARTICLE{Zeldovich1970,
       author = {{Zel'dovich}, Ya. B.},
        title = "{Gravitational instability: An approximate theory for large density perturbations.}",
      journal = {\aap},
         year = 1970,
        month = mar,
       volume = {5},
        pages = {84-89},
       adsurl = {https://ui.adsabs.harvard.edu/abs/1970A&A.....5...84Z}
}

@ARTICLE{zhang18,
       author = {{Zhang}, Bing},
        title = "{Fast Radio Burst Energetics and Detectability from High Redshifts}",
      journal = {\apjl},
         year = 2018,
        month = nov,
       volume = {867},
       number = {2},
          eid = {L21},
        pages = {L21},
          doi = {10.3847/2041-8213/aae8e3},
archivePrefix = {arXiv},
       eprint = {1808.05277},
 primaryClass = {astro-ph.HE},
       adsurl = {https://ui.adsabs.harvard.edu/abs/2018ApJ...867L..21Z}
}

@ARTICLE{ziegler25,
       author = {{Ziegler}, Joshua J. and {Shapiro}, Paul R. and {Dawoodbhoy}, Taha and {Beniamini}, Paz and {Kumar}, Pawan and {Freese}, Katherine and {Ocvirk}, Pierre and {Aubert}, Dominique and {Lewis}, Joseph S.~W. and {Teyssier}, Romain and {Park}, Hyunbae and {Ahn}, Kyungjin and {Sorce}, Jenny G. and {Iliev}, Ilian T. and {Yepes}, Gustavo and {Gottl{\"o}ber}, Stefan},
        title = "{Predictions for dispersion measures of fast radio bursts through the epoch of reionization using CoDa II}",
      journal = {\mnras},
         year = 2025,
        month = sep,
       volume = {542},
       number = {2},
        pages = {1518-1531},
          doi = {10.1093/mnras/staf1308},
archivePrefix = {arXiv},
       eprint = {2411.02699},
 primaryClass = {astro-ph.CO},
       adsurl = {https://ui.adsabs.harvard.edu/abs/2025MNRAS.542.1518Z}
}

\appendix

{
\section{Symbols}
\label{se:sym}
The symbols used in this paper are listed in \tab{sym}. 

\rowcolors{3}{TabCol}{}
\begin{table*}
\caption{List of Symbols}\label{tab:sym}
\rowcolors{2}{TabCol}{}
\begin{tabular}{ m{2.2cm} m{10cm} m{4.3cm} }
  \hhline{===}
  \textbf{Symbol} & \textbf{Description} & \textbf{Expr.} \\
  \hhline{===}
    
    $\n$ & Density of a cloudlet (\fig{clump_prop}). & 
    \\*
    $\lc$ & Length scale of a cloudlet (\fig{clump_prop}). & 
    \\*
    $\lcmin$ & Minimum cooling length (\eqnp{lshatter}, \fig{clump_prop}. Fiducial $\lc=\lcmin$). & $\lcmin=\min(\lcool)$
    \\*
    $\fv$ & Volume-filling fraction of cloudlets. & 
    \\*
    $\DL$ & Width of a screen (\se{vir}). & 
    \\*
    $\dL$ & Effective width of a screen. & $\dL=\fv\DL$ 
    \\*
    $\fa$ & Area-covering fraction of cloudlets. & $\fa=\fv\DL/\lc$ 
    \\*
    \hline
    $\nuo$ $(\lamo)$ & Observed frequency (wavelength). & 
    \\*
    $\dmsc$ & DM due to density fluctuations. & $\dmsc=\fa\lc\n$
    \\*
    $\lpi$ & Coherence or diffractive scale (\eqsitinp{lpi_0}{lpi_dm}). & 
    \\*
    $\dsl, \dlo, \dso$ & Angular diameter distance between: source and screen, screen and observer, source and observer. & 
    \\*
    $\Deff$ & Effective angular diameter distance. & $\Deff={\dsl\dlo}/{\dso}\approx\min(\dsl,\dlo)$ 
    \\*
    $\rf$ & Fresnel scale (\eqnp{rF}). & $\rf\propto({\lamo\Deff}/{\opzo[]})^{1/2}$ 
    \\*
    $\taus$ & Scattering time or temporal broadening (\eqsnp{taus,taus_dm}). & $\taus\propto \lamo({\rf}/{\lpi})^2$
    \\*
    $\tausin$ & $\taus$ of a screen at $z$ in its own rest frame (=intrinsic). & $\tausin=\taus\opzo[]^{17/5}$
    \\*
    $\zs$ & Redshift of an FRB source ($\approx$ redshift of the host screen). & 
    \\*
    \hline
    $\dnus$ & Scintillation/decoherence bandwidth  (\eqsnp{dnus,dnus_dm}). & $\dnus\propto\taus^{-1}$ 
    \\*
    $\rref$ & Refractive scale. Valid only in the strong scintillation regime. & $\rref=\rf^2/\lpi$ 
    \\*
    $u$ & Scintillation strength, $u>1$ in the strong regime (\eqnp{u}). & $u=\frac{\rf}{\lpi}=\pfrac{\nuo}{\dnus}^{1/2}=\pfrac{\rref}{\lpi}^{1/2}$ 
    \\*
    $\nuou$ & Transition frequency between the weak and strong scintillation regimes (\eqsnp{nuou, nuou_dm}. Strong regime for $\nuo<\nuou$). & $\nuou=\nuo\,[u(\nuo)]^{10/17}$ 
    \\*
    $\dffu$ & Flux modulation of the 2nd screen due to its scintillation regime (\eqnp{dffu}. The 1st screen is closer to the source). & $\dffu\sim1$ in the strong regime. 
    \\*
    \hline
    $\dnuo$ & Spectral resolution of the detector (total bandwidth of the detector divided by the number of channels). & 
    \\*
    $\dffob$ & Flux modulation due to the detector spectral resolution and scint. bandwidth of the second screen (\eqnp{dffob_1}). & $\dffob\sim1$ when $\dnuo\lsim\dnus$
    \\*
    \hline
    $\lpiso$ & Coherence scale of the CW screen, projected onto the FRB host screen plane (\eqnp{lpip}, \fig{lpiso}). & $\lpiso=\lpi\dso/\dlo$
    \\*
    $\lpisl$ & Coherence scale of the CW screen, projected onto the MW screen plane (\eqnp{lpip}). & $\lpisl=\lpi\dso/\dsl$
    \\*
    $\lcone$ & Scattering cone scale (\eqnp{Rcone}, \fig{Rcone}. Used for the host and MW screens). & $\lcone=\rref\opzo[]\propto\lamo\Deff/\lpi$ 
    \\*
    $\nuoRs$ & Transition frequency between an extended source ($\nuo<\nuoRs$) and a point source ($\nuo\geq\nuoRs$). \Eqs{nuoRs_base, nuoRs, nuoRsmw}, \figsiti{dff_cw}{dff_mw_zS2}. & 
    \\*
    $\dnusRs$ & Scintillation bandwidth of the 2nd screen at the transition frequency $\nuoRs$. & $\dnusRs=\dnus(\nuoRs)$  
    \\*
    $\dffRs$ & Flux modulation of the 2nd screen due to a broadened source size (\eqnp{dffRs}). & $\dffRs\sim1$ for a point source. 
    \\*
    $\dff$ & Total flux modulation (=modulation index) of the 2nd screen (\eqsnp{dff_def,dff}). & $\dff=\dffu\ \dffob\ \dffRs$
    \\*
    \hline
    $\dffmw$ & Flux modulation of a MW screen. & 
    \\*
    $\dnusmw$ & Scintillation bandwidth of a MW screen (\eqnp{dnus_dm}). & $\dnusmw=\dnus[,\mmw]$ 
    \\*
    $\dnusmwghz$ & Scintillation bandwidth of a MW screen, measured at $1\ghz$. & $\dnusmwghz=\dnusmw(\nuo=1\ghz)$
    \\*
    $\nuoRsmw$ & Transition frequency below which scintillation of the MW screen is suppressed by an intervening CW screen (\eqnp{nuoRsmw}). & $\nuoRsmw=\nuo\pfrac{\lpisl[,cw]}{\Rmw}^{-5/17}$ 
    \\*
    $\dnusRsmw$ & Scintillation bandwidth of a MW screen at $\nuo=\nuoRsmw$. & $\dnusRsmw=\dnusmw(\nuoRsmw)$ 
    \\*
    $\dmw$ & Distance of a MW screen from the observer. & $\dmw=\dlo[,\mmw]\approx\Deff[,\mmw]$
    \\*
    $\dmsmw$ & DM of a MW screen, assumed to be dominated by density fluctuations. & $\dmsmw{\simeq}\dmsc[,\mmw]{=}[\fa\lc\n]_{{\mmw}}$
    \\*
    $\dmigm$ & Extragalactic DM along the path from the host to the observer (\eqnp{dmigm}). & $\dmigm 
    = c\int_{0}^{\zs} dz \frac{\nt(z)}{\opzo[]^2 H(z)}$
    \\*
    $\nmw$ & Cumulative number of CWOs along an average los that can cause suppression of $\dffmw$ (\eqnp{nmw}). & 
    \\*
    $\ffrbmw$ & Fraction of FRBs whose $\dffmw$ (\eqnp{ffrbmw}) is suppressed due to interception with CWOs. & $\ffrbmw=1-\e{\nmw}$
    \\*
    $\nuoRs[\hg-\mmw]$ & Transition frequency below which a MW screen is suppressed by the host screen. In most cases, $\nuoRs[\hg-\mmw]$ is low (\eqnp{nuoRshgmw}, \fig{dff_hgmw_2d}). & $\nuoRs[\hg-\mmw]=\nuo\pfrac{\lpisl[,\hg]}{\Rmw}^{-5/17}$ 
    \\*
    \hline
    $\nuoRscw$ & Transition frequency above which the first CW screen following the host can be detected with $\dff[cw]\sim1$ in spite of broadening by the host screen (\eqnp{nuoRs}). & $\nuoRscw=\nuo\pfrac{\lpiso[,cw]}{\Rhg}^{-5/17}$ 
    \\*
    $\dnusRscw$ & Scintillation bandwidth of a CW screen at $\nuo=\nuoRscw$ (\eqnp{dnus}). & $\dnusRscw=\dnus(\nuoRscw)$ 
    \\*
    $\dhg$ & Distance of the host screen from the source. & $\dhg=\dsl[,\hg]\approx\Deff[,\hg]$ 
    \\*
    $\dmshg$ & DM of the host screen, assumed to be dominated by density fluctuations. & $\dmshg\simeq\dmsc[,\hg]=[\fa\lc\n]_{{\hg}}$ 
    \\*
    $\tausinhgghz$ & The scattering time of the host screen in its own rest frame (=intrinsic), measured at $1\ghz$. & $\tausinhg\approx\taus\opzs^{17/5}$
    \\*
    \hhline{===}
\end{tabular}
\end{table*}
\rowcolors{0}{}{}
}

\section{Virial properties of CWOs}
\label{se:vir}

In this section, we briefly summarise the main results from our model for the virial temperature and scale-length of different CWOs. For a complete description of the virial properties of CW filaments (sheets), see \citeta{m18} (\dmpaper). 

\smallskip
The virial temperature of a halo is given by \citep[e.g.][]{dekel13},
\be
    \label{eq:Tvir_halo}
    \Tvh \sim 2.4 \tm 10^6 \Kel\:\Dvhf^{1/3}\:\Mhf^{2/3}\opzo,
\ee
where $\Mhf\sim\Mh/10^{12}\msun$, $\opzo=\opzo[]/5$, and $\Dvhf$ is the average overdensity in the halo normalized to a value of $180$.

\smallskip
For a filament in the Einstein--de Sitter (EdS) regime at $z>1$,\footnote{See \dmpaper for corresponding expressions valid at $z<1$ for both sheets and filaments.} the virial temperature as a function of the halo mass it feeds is given 
by (\citeta{m18}; \citealt{lu23, Aung24}),
\be 
    \label{eq:Tvir_fil}
    \Tvf \sim 2.55\tm 10^5\Kel\;\fff\,\Machf^{-1} \:\Mhf^{0.77}\,\opzo^{2},
\ee
where $\fff$ is the fraction of total halo accretion flowing along a given filament normalized to $1/3$ \citep{Dekel09a,danovich12}, and $\Machf$ is the filament `effective Mach number', defined as the velocity of filament gas normalized to the halo virial velocity.

\smallskip
In our companion paper (\dmpaper), we derive the virial properties of CW sheets, yielding in the EdS regime 
\begin{equation}
    \label{eq:Tvir_sheet}
    \Tvsh
    \sim 2.4 \tm10^5\Kel \ \fff^{}\fshf^{2}\Machff^{-1}\Machshf^{-2} \Dvshf^{-1} \:\Mhf^{0.77} \opzo^{3}.
\end{equation}
Analogously to how \citeta{m18} derived the virial properties of filaments using the accretion onto haloes, we constrain the virial properties of sheets using the accretion rate onto filaments. We define $\fshf$ to be the fraction of total accretion onto the filament flowing along the sheet normalized to a fiducial value of $1/4$ \citep{lu23}. The virial overdensity of the sheet, $\Dvshf$, is normalized to a value of $6$ \citep{shen06}, and $\Machsh$ is the `effective Mach number' of the sheet, defined as the velocity of sheet gas normalized to the filament virial velocity \citep{lu23}.

\smallskip
The virial radius of a halo as a function of halo mass is 
\be
    \Rvh\sim 64\kpc \,\Dvhf^{-1/3} \: \Mhf^{1/3} \opzo^{-1}.
    \label{eq:Rv_h}
\ee
The virial radii of filaments and sheets as a function of their virial temperature are estimated by
\be
    \Rvf  
    \sim 63\kpc\ \Dvff^{-0.5} \: \Tvff^{0.5} \opzo^{-1.5},
    \label{eq:Rv_f}
\ee
where $\Dvf$ is the virial overdensity of the filament, with $\Dvff=\Dvf/36$ (\citealt{shen06}; \citeta{m18}), and  
\be
    \Xv 
    \sim 170 \kpc \ 
    \Dvshf^{-0.5} \: \Tvshf^{0.5}\opzo^{-1.5}.
    \label{eq:Xv_sh}
\ee

\smallskip
Finally, for haloes and filaments, the path length through the CWO, $\DL$, corresponding to a los with an impact parameter $b$, is given by $\DL=2\Rv\sqrt{1-b^2}$. For sheets, we assume an interception perpendicular to the sheet plane, and therefore, the path length does not depend on $b$ (can be treated as $b=0$) and is estimated by $\DL=2\Xv$.

\subsection{Pressure in CWOs}
\label{se:P}

In order to model the properties of cloudlets in CWOs, we must first confront the question of \emph{which CWOs are expected to experience shattering in the first place?} We expect this to occur when the virial temperature exceeds $\gsim10^5\Kel$, roughly $(5-10)$ times higher than the equilibrium temperature of $(1-2)\tm 10^4\Kel$ \citep[see also][]{Gronke_Oh20, Gronke_Oh23, Yao25}. In these cases, an accretion shock forms at the CWO outer boundary, with a post-shock temperature of roughly $\Tv$ (\citeta{m19,m21}; \citealp{lu23}). Thermal fragmentation (`shattering') can occur as this gas cools. We assume the external pressure remains constant at roughly the virial pressure (estimated below) over the time-scale of fragmentation, sustained by the ram pressure from infalling gas and non-thermal pressure from turbulent motions in the post-shock gas. The properties of the resulting shattered clouds are then set by the ambient hot gas pressure.

\smallskip
In our companion paper (\scattpaper), we model the post-shock pressure of CWOs.\footnote{Our fiducial model assumes that the total post-shock pressure of CWOs consists of thermal pressure alone; however, we explore the effect of substantial non-thermal contributions in \se{model_var}.}
For simplicity, we find $P_\mr{shock}$ of each CWO using the mean baryon density and virial temperature rather than introducing a radial profile.\footnote{See \scattpaper for the full derivation.} 
We assume a mean molecular weight of $\mu=0.59$ for the hot phase, typical for fully ionized gas with primordial composition. 
The post-shock gas pressure for haloes, filaments, and sheets, respectively, is
\begin{align}
    &P_\mr{shock,h} \sim 1.9\tm 10^3 \Kel \cmc \Dvhf \: \Tvhf \opzo^3,
    \label{eq:Ph_Tv}
    \\
    &P_\mr{shock,f} \sim 650 \Kel \cmc \Dvff \: \Tvff \opzo^3,
    \label{eq:Pf}
    \\
    &P_\mr{shock,sh} \sim 110 \Kel \cmc \Dvshf \: \Tvshf \opzo^3,
    \label{eq:Psh}
\end{align}
where $\Tv[,5.5]=\Tv/(\Tvfid\Kel)$ is the virial temperature of the corresponding CWO, 
normalized to $\Tvfid\Kel$. 
For haloes, it is convenient to follow the evolution in time as a function of halo mass rather than virial temperature,
\be
    P_\mr{shock,h} \sim 1.5\tm 10^4 \Kel \cmc \Dvhf^{4/3} \: \Mh[,12]^{2/3} \opzo^4.
    \label{eq:Ph}
\ee

\section{Scintillating CW screen with a single cloud}
\label{se:dff_cw_app}

\begin{figure*}
    \centering
    \includegraphics[width=0.85\linewidth]{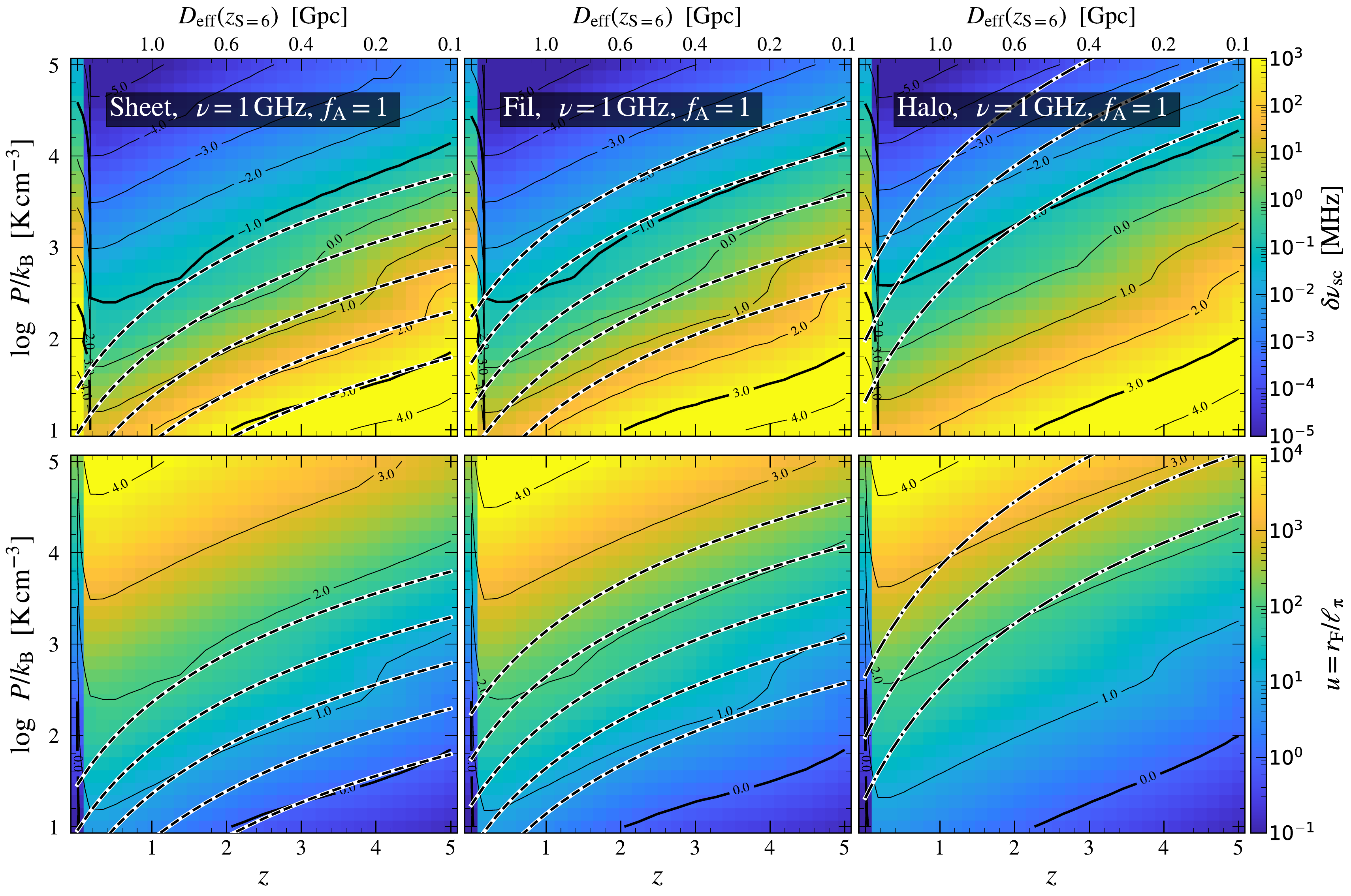}
    \caption{The scintillation bandwidth (top) and scintillation strength (bottom) of CW scattering screens with a single scattering clump, $\fa=1$, at an observed frequency of $\nuo=1\ghz$. The bottom x-axis indicates the redshift of the screen, and the source is assumed to be at redshift $\zs=6$. The corresponding $\Deff(z, \zs=6)$ is marked on the top x-axis. Most CWOs under these conditions are in the strong regime, with the exception of sheets with $\Tvsh\sim 10^{5}\Kel$, which have a strength of roughly $u=1$ at $\nuo=\nuou\sim1\ghz$. Such sheets are in the strong regime when observed at $\nuo<\nuou\sim1\ghz$.}
    \label{fig:dnus_u}
\end{figure*}

In this section, we demonstrate a few basic properties of a scintillating CW screen, which can either affect the detection of flux modulation caused by the CW screen itself, or the flux modulation of a subsequent screen (in the MW). We assume here a CW screen with a single cloud, $\fa=1$.

In the top row of \fig{dnus_u}, we show $\dnus$ (\eqnp{dnus}) in the $P-z$ plane for the three types of CWOs (sheets, filaments and haloes from left to right), assuming a frequency of $\nuo=1\ghz$, a source redshift of $\zs=6$, and only one cloud along the los, $\fa=1$. 
In each case, we estimate the inner scale of dissipation, $\li$, following \citet{beniamini20} assuming a fixed magnetic field, $B$, which determines if $\lpi<\li$ or $\lpi>\li$ in \eq{lpi_0}. This can change the expression for $\dnus$ in \eq{dnus}, which assumes $\lpi>\li$. The effect of $B$ on relevant CWOs is small overall, and we rarely have $\lpi<\li$ (\scattpaper). The electron density, $\n$, and outer scale, $\lo=\lc$, are taken from \fig{clump_prop}. 
In the top x-axis, we show the effective angular diameter distance, $\Deff$, corresponding to the redshift of the CWO in the bottom x-axis, assuming an FRB source at $\zs=6$. However, we note that $\Deff$ is not very sensitive to the choice of $\zs$.\footnote{See fig.~B.1 in \scattpaper and fig.~5 in \citet{macquart13}.}
We obtain $\dnus\ll 1\ghz$ for all relevant CWOs except for sheets with $\Tvsh=10^{5}\Kel$, for which $\dnus\sim1\ghz$.

\smallskip
In the bottom row of \fig{dnus_u}, we show the scintillation strength, $u$ (\eqnp{u}), in the $P-z$ plane for the three types of CWOs. As before, we assume a frequency of $\nuo=1\ghz$ and only one cloud along the los through the CWO, $\fa=1$. Under these conditions, $u>1$ for most of the CWOs considered in this work. The only exception is sheets with $\Tvsh=10^{5}\Kel$, which lie close to the $u=1$ line, meaning that the maximal frequency for which they are in the strong regime is $\nuou\sim 1\ghz$. These sheets would be in the strong regime if observed at lower frequencies of $\nuo<\nuou$, or if their $\fa$ were larger. This is consistent with our estimate of $\dnus$ discussed above.

\section{Extended source vs point source}
\label{se:ExtPoint}

\begin{figure}
    \centering
    \includegraphics[width=0.85\linewidth]{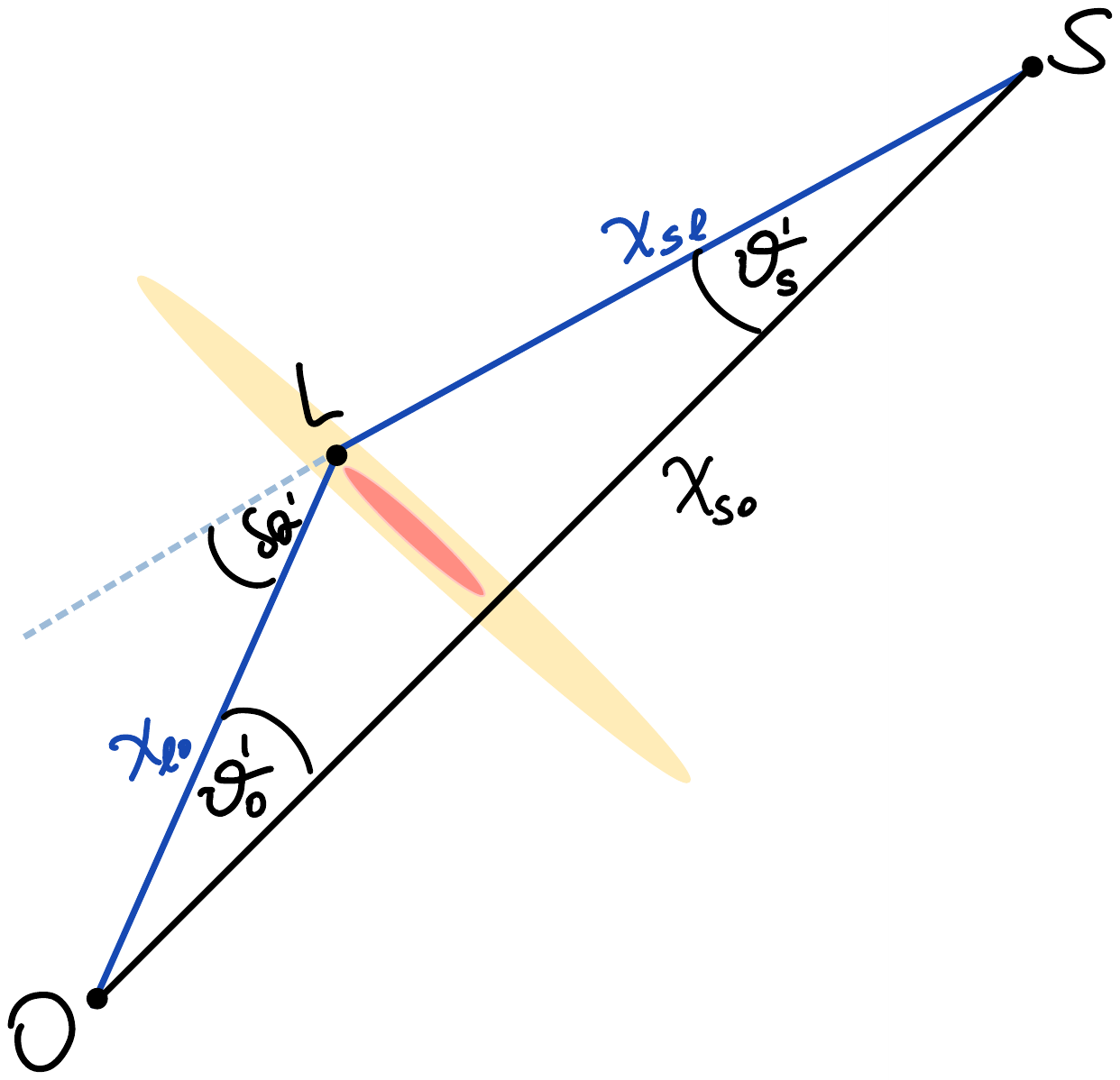}
    \caption{The geometric path of a ray (blue lines) due to a single scattering screen (yellow) at location $L$, between the source, $S$, and an observer, $O$. Comoving distances are marked with $\chi$ and angular diameter distances are denoted with $d$, such that $\chilo=\dlo(1+z_\mr{L}), \chisl=\dsl\opzs$, and $\chiso=\dso\opzs$. The scattering angle caused by the screen is $\dth\propto[\lpi(1+z_\mr{L})]^{-1}$, and $\tho=\dth\chisl/\chiso$, $\ths=\dth\chilo/\chiso$. The scattering cone (or disc) on the screen is indicated by the red ellipse, and is given by $\lcone\approx\tho\chilo=\ths\chisl$.}
    \label{fig:time_delay}
\end{figure}

\begin{figure*}
    \centering
    \includegraphics[width=0.85\linewidth]{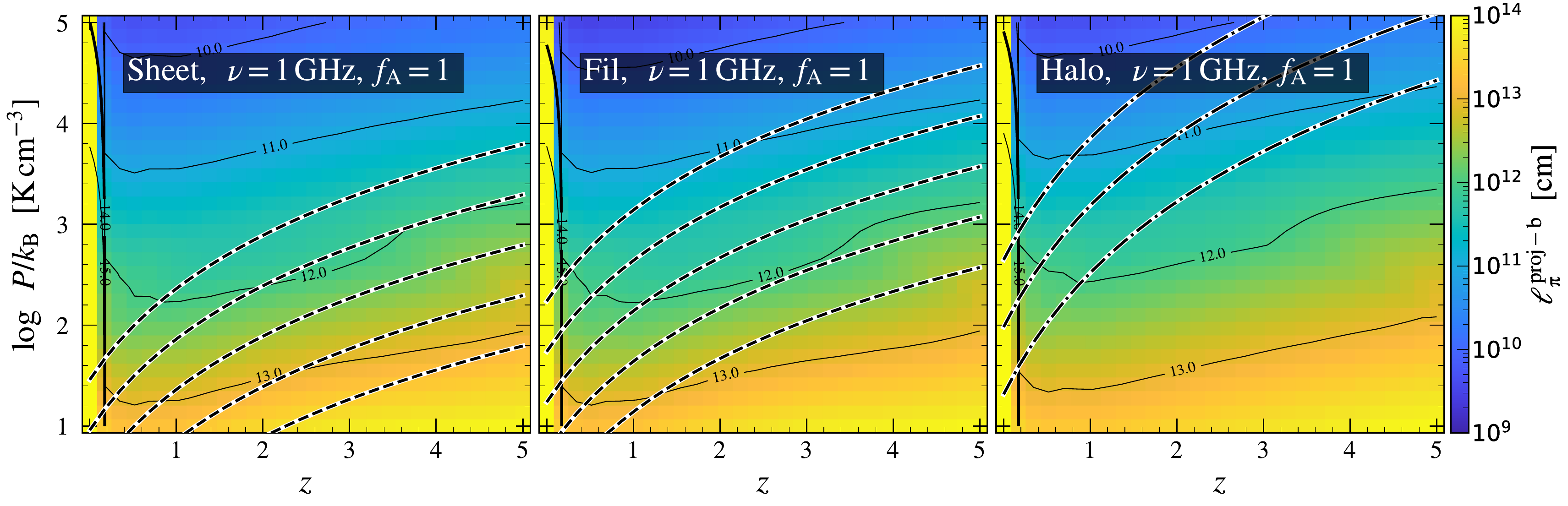}
    \caption{The diffractive scale of a CW screen projected back onto the source plane, $\lpiso=\lpi\dso/\dlo$, assuming $\nuo=1\ghz$, $\fa=1$, and a source at $\zs=6$ (to find $\dso$). This can be compared to the scattering cone of the host screen, $\Rhg$, shown in \fig{Rcone}, to determine the effect of source broadening on the observed flux modulation. See text for details.}
    \label{fig:lpiso}
\end{figure*}

\begin{figure}
    \centering
    \includegraphics[width=0.85\linewidth]{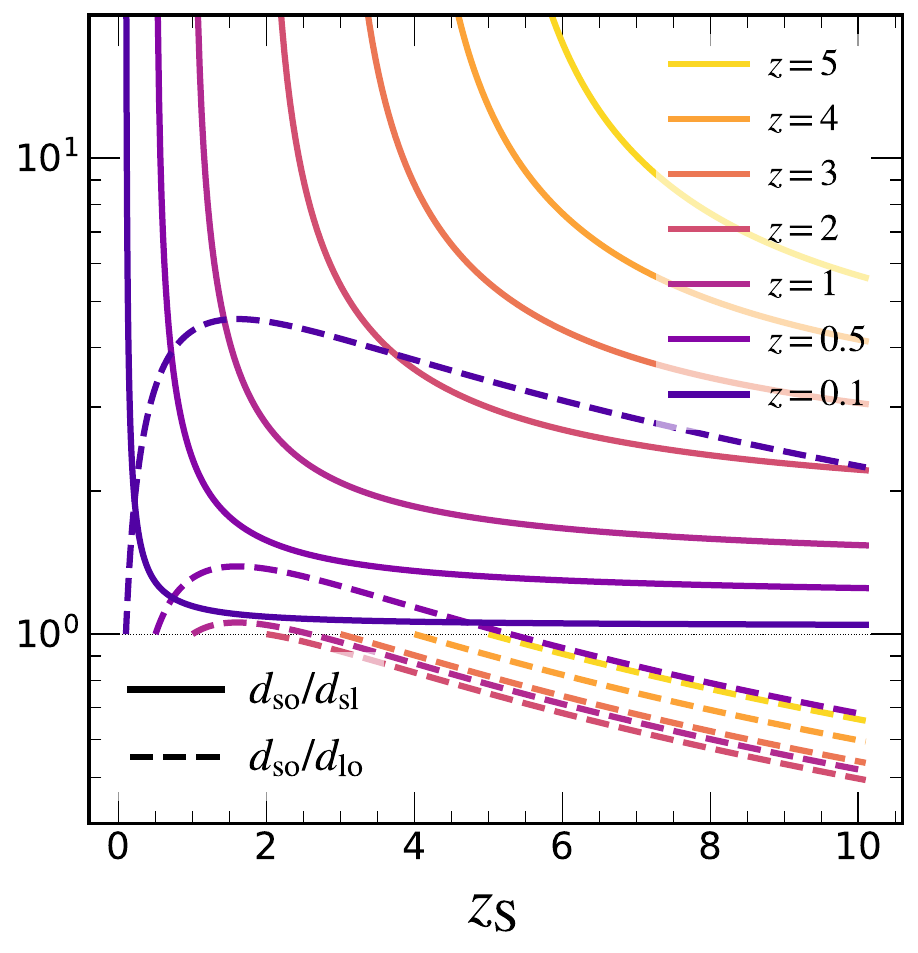}
    \caption{The ratios of angular diameter distances $\dso/\dsl$ (solid, relevant for $\lpisl$, \eqnp{nuoRsmw}), and $\dso/\dlo$ (dash-dotted lines, see $\lpiso$ in \eqnp{nuoRs}), both shown as a function of source redshift, $\zs$. The different coloured lines represent screens at different redshifts, $z$.}
    \label{fig:dso/dsl}
\end{figure}

In this section, we provide a detailed description of the distinction between an extended source and a point source, expanding on the discussion in \se{dffRs}. 
We illustrate some of the notations to follow in \fig{time_delay} (see also appendix B in \scattpaper), where we show a single screen ($L$) between the source ($S$) and the observer ($O$). We denote the comoving distances between the three points as $\chiso$, $\chilo$, and $\chisl$. 
For a screen with a deflection angle $\dth\propto[\lpi(1+z_\mr{L})]^{-1}$, the angles marked in the cartoons are $\tho\simeq\dth\chisl/\chiso$ and $\ths\simeq\dth\chilo/\chiso$.
The observed image of the broadened source projected back onto the screen, namely, the scattering cone scale (red ellipse), is given by $\lcone\approx\tho\chilo\simeq\ths\chisl$.\footnote{Unlike our standard notation, the angle $\ths$ is in the rest frame of the source, rather than that of the screen.}
For $z_1>z_2$, the angular diameter distance between two points is $d_{12}=\chi_{12}/(1+z_1)=(\chi_1-\chi_2)/(1+z_1)$. We can therefore write $\lcone\approx\thoo\dlo=\thso\dsl$, where $\thoo=\tho(1+z_\mr{L})$ and $\thso=\ths\opzs$ are the angular sizes of the observed image with respect to the observer and the source.

\smallskip
To find the condition for strong flux modulation from a given scintillating screen (screen 2) in the presence of an extended source due to the los passing through a previous screen (screen 1, closer to the source), we follow \citeta{narayan92}, and compare the diffractive angle of the second screen, $\thd[2]=\lpi[2]/\dlo[2]$, to the apparent angular size of the broadened source in the observer frame, $\thr[1]=\lcone[,1]/\dlo[1]=(1+z_1)\rref[1]/\dlo[1]$. 
The ratio of these angles determines whether the broadening by the first screen is perceived by the second screen as a point source, $\thd[2]/\thr[1]\geq1$ (unresolved), or an extended source, $\thd[2]/\thr[1]<1$ (resolved),
\be
    \frac{\thd[2]}{\thr[1]}
    = \frac{\lpi[2]}{\lcone[1]}\frac{\dlo[1]}{\dlo[2]}
    = \frac{\lpib[2]}{\lcone[1]},
    \label{eq:proj_b}
\ee
where we define $\lpib[2]=\lpi[2]\dlo[1]/\dlo[2]$. 
\Eq{proj_b} is equivalent to the ratio of the second screen coherence scale and the apparent broadened size of the source, when both are projected back onto the plane of the first screen.

\smallskip
It is often convenient to distinguish between an extended and a point source by projecting the relevant scales forward, onto the plane of the second screen.
Since the result of both projections must be equal, we can find the expression for a forward projection by inserting the relation $\rf^2=\rref\lpi$, where $\rf^2\propto\Deff\lamz$ and $\rref=\lcone\opzo[]$, into \eq{proj_b} for each screen. This gives $\lpi[2](\dlo[1]/\dlo[2])/\lcone[1]=\lpi[1](\dsl[2]/\dsl[1])/\lcone[2]$, where $\lpif[1]=\lpi[1]\dsl[2]/\dsl[1]$ is the coherence scale of the first screen, projected forward onto the second screen, and $\lcone[2]$ is the scattering cone scale of the second screen. Namely, the broadening by the first screen is unresolved by the second screen if $\lpif[1]/\lcone[2]\ge1$. 
Analogous to \eq{proj_b}, for a forward projection, the condition for a point source can also be expressed by the angles, but from the perspective of the source (i.e. the ``apparent'' angle, $\thrf[2]$, of the second screen, subtended by the distance $\dsl[2]$ from the source location). In this case, $\thdf[1]=\lpi[1]/\dsl[1]$, and $\thrf[2]=\lcone[2]/\dsl[2]$, giving ${\thdf[1]}/{\thrf[2]}={\lpif[1]}/{\lcone[2]}$.
To summarise the condition for both projections, the broadening of the source by the first screen is unresolved by the second screen if 
\be
    \frac{\lpib[2]}{\lcone[1]} = \frac{\lpif[1]}{\lcone[2]} \ge1.
\ee

\smallskip
In the main text, we make use of both versions of this condition. 
In \se{mdff_mw}, we explore the effect of broadening by a CW screen on the observed scintillation of a MW screen. 
For this set-up, the proximity of the MW screen (2nd screen) to the observer allows us to use the approximation $\dsl[2]=\dsl[,\mmw]\approx\dso$.\footnote{Projection onto the MW screen is roughly equivalent to a projection onto the observer plane.}
The forward projection is therefore more convenient in this case since $\lpif[1]=\lpisl[,cw]\approx\lpi[,cw]\frac{\dso}{\dsl[,cw]}$ can be calculated independently of the second MW screen location. 
In \se{mdff_cw}, we explore the conditions for strong flux modulation by the CW screen, where a source is broadened by a previous screen in the host galaxy ($\hg$).
Similarly, using the proximity of the host screen to the source, $\dlo[1]=\dlo[,\hg]\approx\dso$,\footnote{For this set-up, projection onto the host screen is roughly equivalent to a projection onto the source plane.}
the backward projection of $\lpi[,cw]$ onto the host screen gives $\lpib[2]=\lpiso[,cw]\approx\lpi[,cw]\frac{\dso}{\dlo[,cw]}$ again, independent of the first host screen location.

\smallskip
In \fig{lpiso}, we show the coherence scale of CWOs when projected back onto the plane of the host screen, $\lpiso$ (\eqnp{lpip}), at $1\ghz$ for $\fa=1$. As before, we assume a source at $\zs=6$. As described in \se{dff}, this scale can be compared to $\Rhg$ (\eqnp{Rcone}, \fig{Rcone}) to assess whether the broadened source size will be perceived by the CW screen as a point source. For example, for a host screen with $\dmsfahg=10^{2.5}\dmunits$, the broadened size of the source is $\Rhg(\zs=6)\sim5\tm10^{11}\cm$, which is roughly comparable to $\lpiso$ of a $10^{6}\Kel$ filament at $z\sim4$ with $\fa=1$, with $\lpiso$ of a few times $10^{11}\cm$.

\smallskip
For quick estimates of our results assuming different $\zs$, we provide in \fig{dso/dsl} the ratios (y-axis) of $\dso/\dsl$ (solid) and $\dso/\dlo$ (dashed), which are used in the projected coherence scales $\lpisl$ and $\lpiso$ of CW screens (\eqnp{lpip}. See also \eqsnp{nuoRsmw, nuoRs}).
The x-axis indicates the redshift of the source, $\zs$, and the different colours represent screens at different redshifts, $z$ (see legend).

\section[Fraction of FRBs with suppressed dffmw due to CW screens]{Fraction of FRBs with suppressed $\dffmw$ due to CW screens}
\label{se:mdff_mw_app}

In our companion paper (\scattpaper), we showed that if turbulent shattered cloudlets exist with $\fv\gsim10^{-3}$ in massive haloes, the observed scattering time is expected to show a correlation with the redshift of the source for $\zs\lsim1$ (or $\dmigm\lsim10^{3}\dmunits$). We further showed that this effect was dominated by massive $\sim10^{12.5-13}\msun$ haloes (fig.~9 in \scattpaper). Here, we explore whether our results in \se{mdff_mw} are also dominated by such masses. 

\begin{figure}
    \centering
    \includegraphics[width=0.85\linewidth]{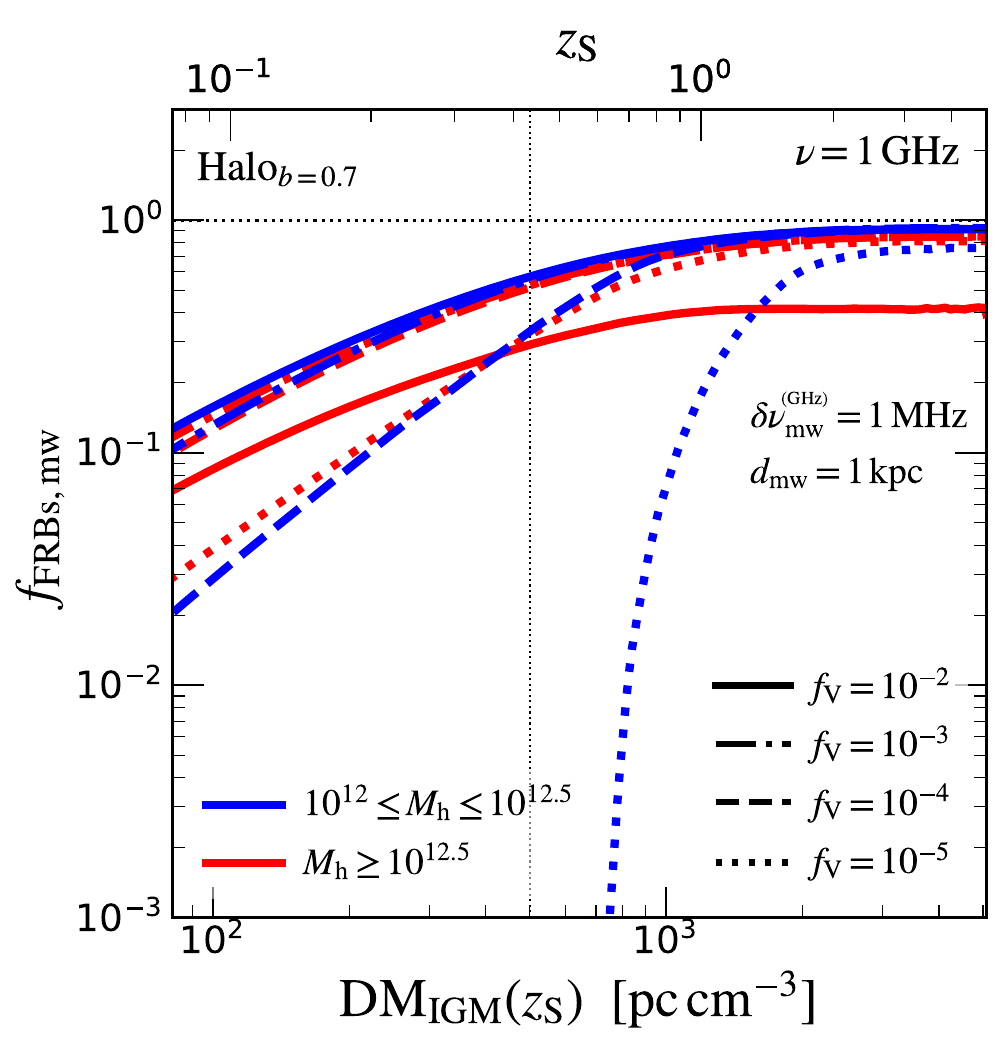}
    \caption{Similar to \fig{ffrbmw_h}, we show $\ffrbmw$ as a function of the extragalactic DM. Here, we divide the results into two mass bins of $10^{12}\leq\Mh\leq10^{12.5}\msun$ (blue) and $\Mh\geq10^{12.5}\msun$ (red). Different line styles show different values of $\fv$. For simplicity, we show only one frequency, $\nuo=1\ghz$. For $\fv\sim10^{-3}$ (dash-dotted lines), an encounter with a halo in either mass bin yields a similar $\ffrbmw\gsim0.5$ for an FRB with $\dmigm\geq500\dmunits$. 
    However, for higher or lower volume-filling fractions, the results differ between the halo mass bins, with the largest difference at the lowest $\fv$ considered in this example. 
    For $\fv=10^{-2}$ (solid lines), massive haloes around $10^{13}\msun$ have $\fvmax<10^{-2}$, and are therefore excluded, which results in $\ffrbmw$ lower by a factor of $\sim2$ compared to the low-mass bin. 
    For lower $\fv\sim10^{-4}$, we obtain $\ffrbmw\gsim0.5$ in the massive bin, and a more modest $\ffrbmw\gsim0.3$ for the lower-mass bin.}
    \label{fig:ffrbmw_h_mbins}
\end{figure}

\smallskip
Similar to \fig{ffrbmw_h}, we show in \fig{ffrbmw_h_mbins} $\ffrbmw$ for haloes as a function of $\dmigm$, here divided into two mass bins, $\Mh=10^{12-12.5}\msun$ (blue) and $\Mh\geq10^{12.5}\msun$ (red), for different volume-filling fractions (line styles). For clarity, we show only one frequency, $\nuo=1\ghz$. 
The difference between the two mass bins is most evident at the lowest volume-filling fraction, and the smallest difference is for $\fv=10^{-3}$, where we see similar results across the lower- and higher-mass bins. 
For instance, for FRBs with $\dmigm\gsim500\dmunits$ and intervening haloes with $\fv=10^{-4}$, lower-mass haloes yield $\ffrbmw\gsim0.3$ while higher-mass haloes yield $\ffrbmw\gsim0.5$. 
Unlike the results in \scattpaper, for $\fv=10^{-2}$ the low-mass bin is dominant (due to the exclusion of the most massive haloes in this bin with $\fvmax<10^{-2}$), with $\ffrbmw$ for the high-mass bin lower by a factor of $\sim2$ than for the low-mass bin. 
The main reason for the different behaviour seen here for haloes with $\fv=10^{-2}$ (or $\fv=10^{-3}$) compared to scattering is that here we are quantifying a cut-off ($\nuoRsmw$) rather than the cumulative strength of a specific quantity (for instance, $\taus$ in \scattpaper).
In other words, if $\nuoRsmw$ exceeds the observed frequency for either a $10^{12}\msun$ or a $10^{13}\msun$ halo, we count each of them as being able to suppress $\dffmw$.\footnote{Strictly speaking, the extent of suppression (i.e. the value of $\dffmw$) would differ if caused by a stronger (weaker) scattering screen with a higher (lower) $\nuoRsmw$. However, given the strong dependence with a power of $17/5$ in \eq{dffRs}, a relatively mild difference between $\nuo$ and $\nuoRsmw$ will severely reduce $\dffmw$.}

\subsection[ffrbmw by filaments and sheets]{$\ffrbmw$ by filaments and sheets}
\label{se:ffrbmw_fsh}

Similar to \se{ffrbmw}, where we estimate $\ffrbmw$ for haloes, we repeat the exercise here for sheets and filaments. In this case, the mass in \eqsiti{dnmw_dzdm}{ffrbmw} is replaced by the shock temperature, giving
\be
    \frac{d^2\nmw}{dz\,dT} =
    \begin{cases}
    \frac{d^2\num}{dzdT} \min(1, \fa), & \nuo\leq\nuoRsmw,
    \\[5pt]
    0, & \nuo>\nuoRsmw,
    \end{cases}
    \label{eq:dnmw_dzdm_fsh}
\ee
\be
    \nmw(\zs,\nuo,\geq\Tv) =
    \int_{0}^{\zs}\int_{\Tv}
    \frac{d^2\nmw}{dz\,dT} \ dz\,dT,
    \label{eq:nmw_fsh}
\ee
and
\be
    \ffrbmw(\zs,\nuo,\geq\Tv) = 1-\e{-\nmw}.
    \label{eq:ffrbmw_fsh}
\ee

\begin{figure}
    \centering
    \includegraphics[width=1\linewidth]{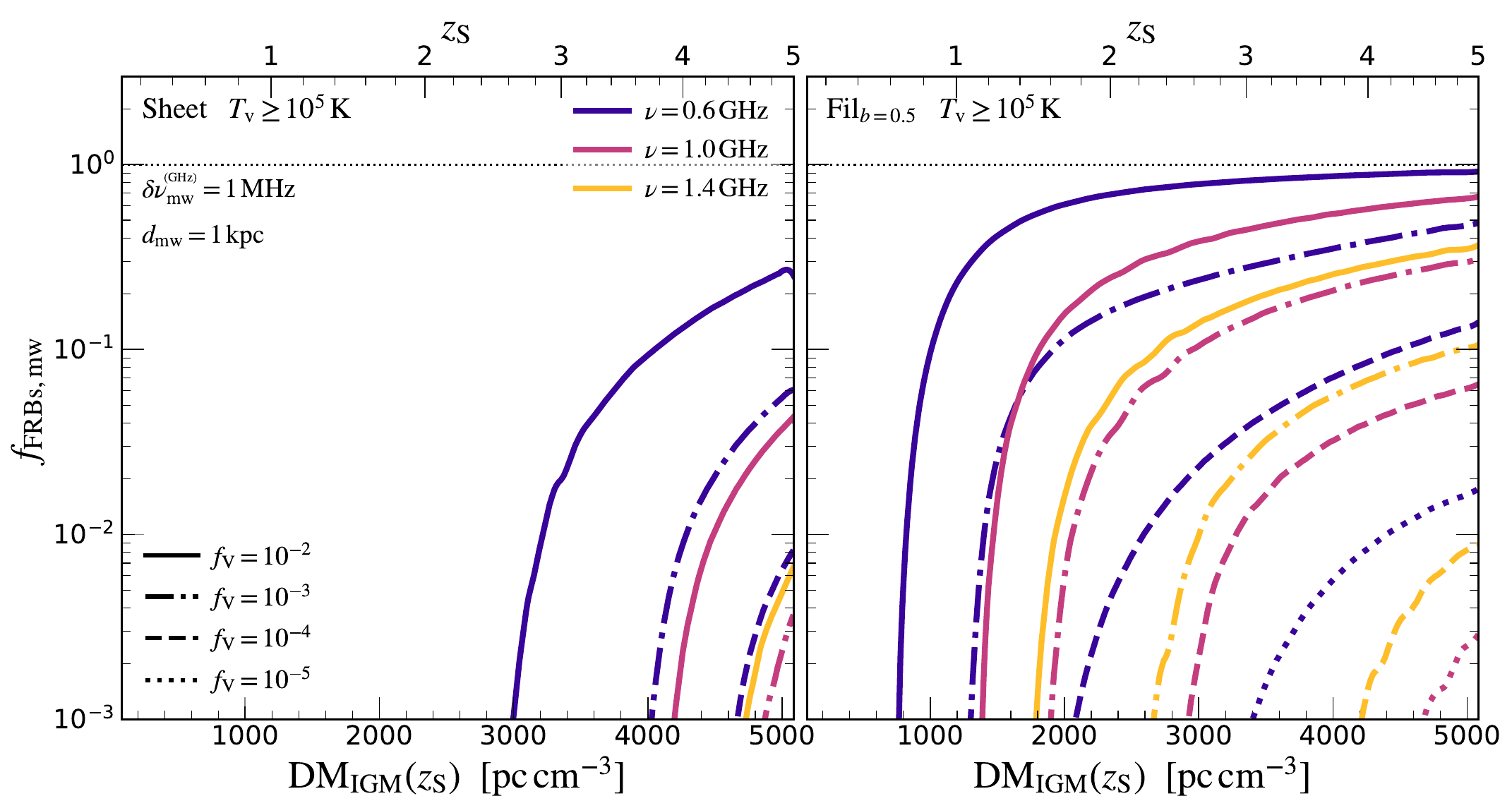}
    \caption{Fraction of FRBs ($\ffrbmw$, y-axis) whose MW scintillation is suppressed by a $\Tv\geq10^{5}\Kel$ sheets (left) and filaments (right).
    To compare with non-localised observations, we express $\ffrbmw$ as a function of the average extragalactic $\dmigm$ (bottom x-axis). The DM is calculated using \eq{dmigm} for a given $\zs$ (top x-axis).
    As before, we assume $\dnusmwghz=1\mhz$, $\dmw=1\kpc$ for the MW screen, and an impact parameter of $b=0.5$ for filaments. The different colours indicate the observed frequency, and the line styles represent the volume-filling fractions.
    As discussed in \se{dff_mw_sig}, these CWOs are expected to shatter at higher redshifts. This, combined with the low abundance of the higher-temperature CWOs (especially for sheets), yields $\ffrbmw$ values that are considerably lower than for haloes.
    For example, for filaments with $\fv\sim10^{-2}$ ($\fv\gsim10^{-3}$), $\gsim40\%$ ($\gsim10\%$) of FRBs with $\dmigm\gsim3000\dmunits$ ($\zs\gsim2.6$) are expected to have a suppressed $\dffmw$ below $\nuo=1\ghz$ due to these systems.
}
    \label{fig:ffrbmw_fsh}
\end{figure}

\smallskip
In \fig{ffrbmw_fsh}, we show the fraction of FRBs ($\ffrbmw$) whose $\dffmw$ is suppressed below a given frequency by interception with $\Tv\geq10^{5}\Kel$ sheets (left) and filaments (right).
We assume $\dnusmwghz=1\mhz$ and $\dmw=1\kpc$ for the MW screen (as in \fig{ffrbmw_h}), and an impact parameter of $b=0.5$ for filaments. The colours indicate different observed frequencies and the line styles represent volume-filling fractions. The fraction $\ffrbmw$ is expressed as a function of the average extragalactic DM (bottom x-axis), which is derived from the source redshift (top x-axis) using \eq{dmigm}.\footnote{Unlike \fig{ffrbmw_h}, the bottom and top x-axes are shown here in linear scale to better highlight higher redshifts where sheets and filaments are expected to shatter according to our fiducial model.}
As discussed in \se{dff_mw_sig}, an average los to an FRB is unlikely to encounter a sheet (though the expected high rate of FRBs offers some chance of this occurring), while for filaments the odds of an encounter are better. 
For instance, $\gsim40\%$ ($\gsim10\%$) of the FRBs with $\dmigm\gsim3000\dmunits$ are expected to have suppressed $\dffmw$ below $\nuo=1\ghz$ due to filaments with $\fv\sim10^{-2}$ ($\fv\sim10^{-3}$).
For such $\dmigm$, all FRBs are expected to encounter haloes with $\geq10^{12}\msun$ along their los. However, the first intercepted CW screen that is able to cause suppression is the one that will, in practice, suppress $\dffmw$ (and will likely suppress the flux modulation of following CW screens as well). Therefore, if filaments have turbulent cloudlets with high enough volume-filling fraction, they may suppress volume-filling fraction along some sightlines to high $\zs$ sources regardless of whether they encounter a halo. 
However, distinguishing between halo-suppression and filament-suppression will likely be very challenging (see \se{screen_ambiguity}).

\section{Scintillation due to CWOs}
\label{se:mdff_cw_app}

In \se{dff}, we discussed the different components that can affect the flux modulation of a CW screen, and showed that the total flux modulation of a CWO is expected to be of order unity if the observed frequency $\nuo$ is within the range $\nuoRs<\nuo<\nuou$, assuming a detector with spectral resolution $\dnuo\lsim\dnus$. 
In \fig{nuou}, we use the same fiducial values as in \fig{dff_cw}, and show the maximal frequency $\nuou$ (y-axis) below which the scintillation due to the CW screen is in the strong regime. 
The colours in \fig{nuou} represent the ratio $\nuoRs/\nuou$. As can be seen, $\nuoRs<\nuou$ for all the systems in this example. Therefore, $\dffu$ is generally much less limiting compared to $\dffRs$ and $\dffob$.

\begin{figure}
    \centering
    \includegraphics[width=1\linewidth]{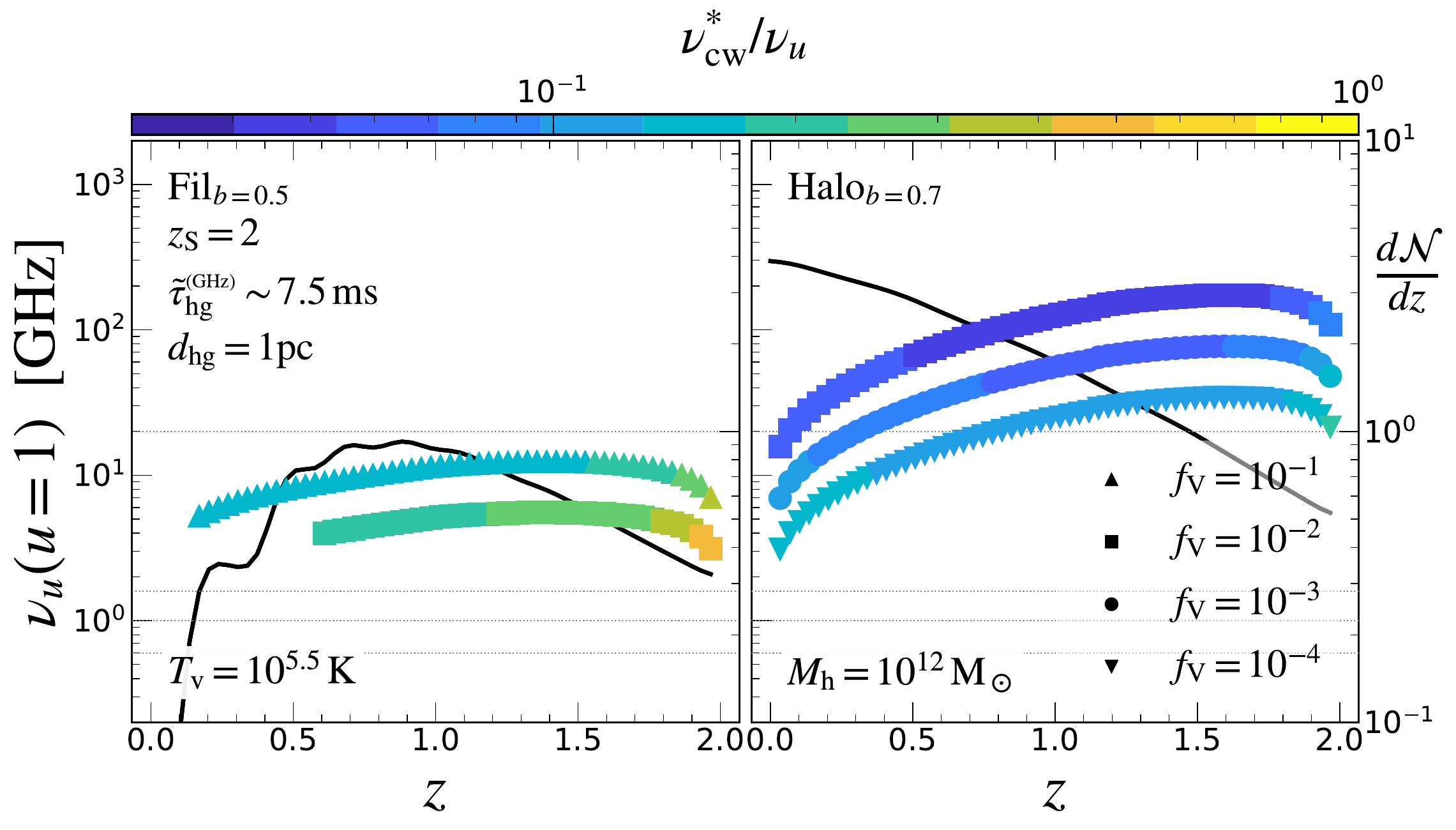}
    \caption{The maximal frequency, $\nuou$, below which the scintillation is in the strong regime, $u=(\nuo/\dnus)^{1/2}>1$.}
    \label{fig:nuou}
\end{figure}

\bsp 
\label{lastpage}
\end{document}